\documentclass[aps,prb,twocolumn,floatfix,superscriptaddress]{revtex4-2}
\usepackage{physics}
\usepackage{graphicx}
\usepackage{lineno,hyperref}
\usepackage{amsmath,amssymb}
\usepackage{array}

\hypersetup{colorlinks=true,linkcolor=blue, filecolor=magenta, citecolor=blue,urlcolor=magenta}

\makeatletter

\usepackage{bm}\usepackage{times}\usepackage{amsfonts}

\makeatother

\begin{document}
\title{Ironing out the details of unconventional superconductivity}
\author{Rafael M. Fernandes}
\affiliation{School of Physics and Astronomy, University of Minnesota, Minneapolis,
MN 55455}
\author{Amalia I. Coldea}
\affiliation{Clarendon Laboratory, Department of Physics, University of Oxford,
Parks Road, Oxford OX1 3PU, UK}
\author{Hong Ding}
\affiliation{Beijing National Laboratory for Condensed Matter Physics and Institute
of Physics, Chinese Academy of Sciences, Beijing 100190, China}
\affiliation{CAS Center for Excellence in Topological Quantum Computation, University
of Chinese Academy of Sciences, Beijing 100190, China}
\author{Ian R. Fisher}
\affiliation{Geballe Laboratory for Advanced Materials and Department of Applied
Physics, Stanford University, Stanford, CA 94305, USA}
\affiliation{Stanford Institute for Materials and Energy Science, SLAC National
Accelerator Laboratory, 2575 Sand Hill Road, Menlo Park, California
94025, USA}
\author{P. J. Hirschfeld}
\affiliation{Department of Physics, University of Florida, 2001 Museum Rd, Gainesville,
FL 32611, USA}
\author{Gabriel Kotliar}
\affiliation{Physics and Astronomy Department, Rutgers University, Piscataway,
New Jersey 08854, USA}
\affiliation{Condensed Matter Physics and Materials Science Department, Brookhaven
National Laboratory, Upton, New York 11973, USA}
\begin{abstract}
Superconductivity is a remarkably widespread phenomenon observed in most metals cooled down to very low temperatures. The ubiquity
of such conventional superconductors, and the wide range of associated
critical temperatures, is readily understood in terms of the celebrated
Bardeen-Cooper-Schrieffer (BCS) theory. Occasionally, however, unconventional superconductors
are found, such as the iron-based materials, which extend and defy this understanding
in new and unexpected ways. In the case of the iron-based superconductors,
this includes a new appreciation of the ways in which the presence
of multiple atomic orbitals can manifest in unconventional superconductivity, giving rise to a rich landscape of gap structures that share the same dominant pairing mechanism.
Besides superconductivity, these materials have also led to new insights into the unusual metallic
state governed by the Hund\textquoteright s interaction, the control
and mechanisms of electronic nematicity, the impact of magnetic fluctuations
and quantum criticality, and the significance of topology in correlated
states. Over the thirteen years since their discovery, they have
proven to be an incredibly fruitful testing ground for the development of new experimental tools and theoretical approaches, both of which have extensively influenced the wider field of
quantum materials.
\end{abstract}
\date{\today}

\maketitle

\section{Introduction}

A comprehensive understanding of conventional superconductors, in
which lattice vibrations bind electrons in Cooper pairs, is provided
by the BCS-Eliashberg theory. Several families of unconventional superconductors,
however, defy explanation within this paradigm, presenting a series
of rich intellectual challenges. For many years, attention was split
between the cuprate superconductors \cite{Keimer15}, with critical temperatures ($T_{c}$)
up to $165$K, and the heavy-fermion \cite{Maple15} and organic superconductors \cite{Singleton02},
which typically have lower $T_{c}$ values. In 2008, a new family
of superconductors was discovered based on iron (Fe) \cite{Kamihara08}. The discovery
was noteworthy given that Fe is generally seen as a strongly magnetic
ion, and magnetism is typically considered to be antithetical to superconductivity.
It rapidly became more remarkable as new members of the family were
discovered with progressively higher $T_{c}$ values -- high enough that the materials were soon referred to as ``high-$T_{c}$'' (for an early review, see \cite{Johnston10}). 

A large body of evidence now indicates that these Fe-based superconductors
(FeSC) are unconventional, i.e. that pairing is not driven by lattice
vibrations (phonons) \cite{Mazin_review,Chubukov_review,DHLee_review}. They have provided a fascinating array of new insights into the conditions of occurrence and nature of unconventional
superconductivity, particularly in systems where the electrons can
occupy multiple orbitals. Prior to their discovery, unconventional
pairing was synonymous with Cooper pairs with non-zero angular momentum
and gap nodes, exemplified, for instance, by the $d$-wave superconducting
state realized in cuprates \cite{Keimer15}. In iron-based materials, however, the Cooper
pairs are widely believed to have zero angular momentum, their unconventional
nature arising from the different phases they take on different bands \cite{Mazin08}.
A variety of pairing structures have been observed, attributed
to the same dominant pairing mechanism acting on distinct types of multi-orbital
electronic structures. 

Besides superconductivity, the normal state of the FeSC is also unusual.
Similar to many other quantum materials, electron-electron interactions
play an important role in shaping their phase diagrams. However, due
to the multi-orbital character of these compounds, it is the Hund's
interaction that is believed to play the most prominent role \cite{Haule09}. The
resulting \textquotedblleft Hund metal\textquotedblright{} interpolates
between a description of incoherent atomic states at high temperatures
and coherent Fermi-surface states at low temperatures \cite{Georges13}. This is in
contrast with the cuprates, where the onsite Hubbard repulsion dominates,
or with the heavy-fermion materials, where the Kondo coupling between
localized and itinerant electrons is the relevant interaction. A related
concept that has emerged in studies of FeSC is that of orbital differentiation,
by which distinct orbitals subjected to the same microscopic interactions
experience different degrees of correlation \cite{Kotliar11,Medici15,Sprau17}. 

\begin{figure*}
\begin{centering}
\includegraphics[width=2\columnwidth]{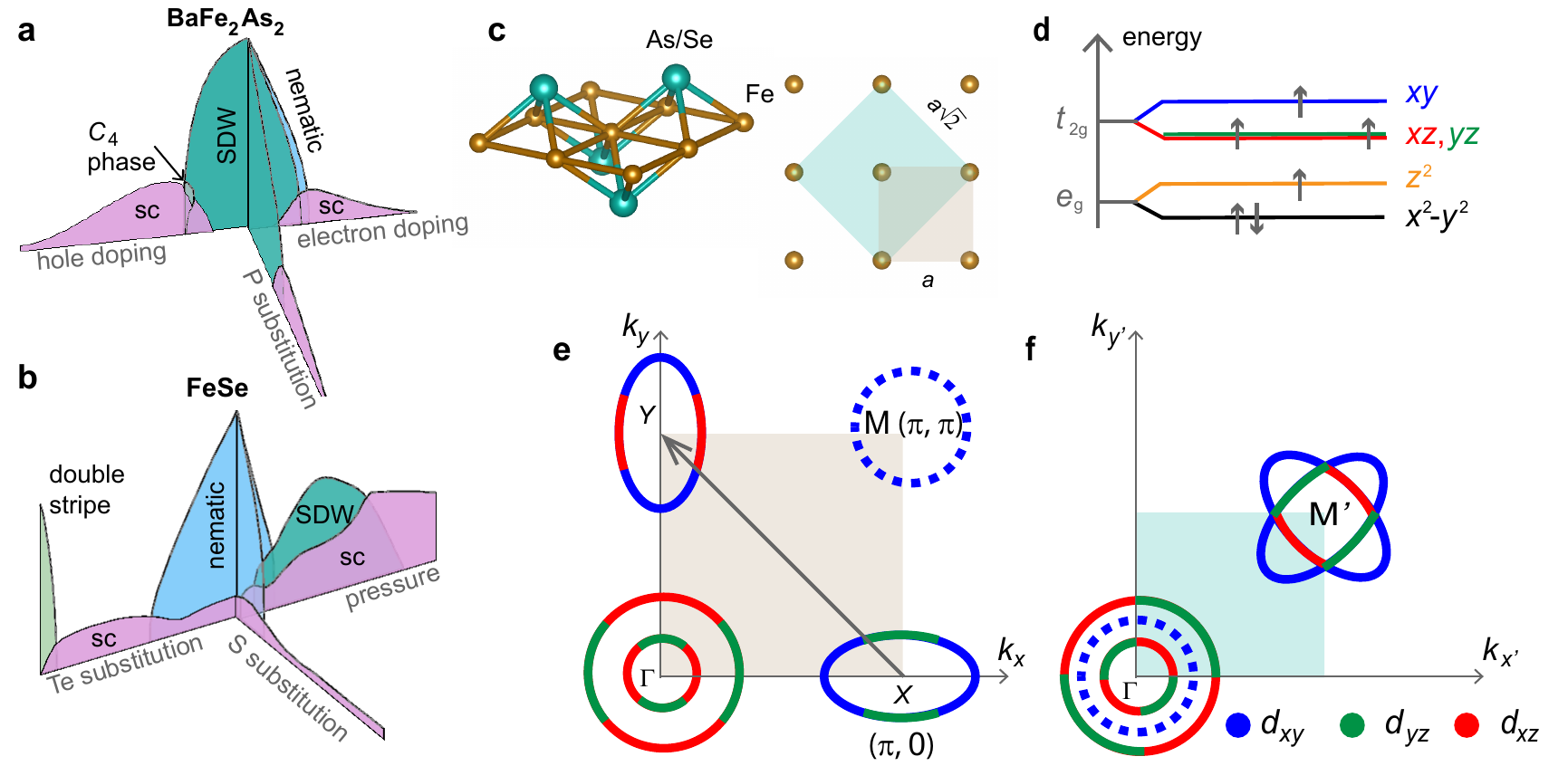}
\par\end{centering}
\caption{\textbf{General structural and electronic properties}. Three-dimensional
phase diagrams of two families of FeSC: (a) BaFe$_{2}$As$_{2}$ \cite{Shibauchi14} and
(b) FeSe \cite{Coldea18,Kreisel20}. The color indicates schematically the presence of different
competing electronic phases -- nematic, spin-density wave (SDW),
double-stripe, $C_{4}$ magnetic phase, and superconductivity (SC).
The tuning parameter can be electron or hole doping, isoelectronic
substitution (As/P or Se/S, Se/Te) or applied pressure. (c) The common
structure responsible for the electronic properties of FeSC consists
of Fe planes and pnictogens (As) or chalcogens (Se) outside the plane.
A simplified representation considering a single Fe per unit cell
is shown in gray, whereas the crystallographic unit cell containing
2 Fe per unit cell (which take into accounts the glide symmetry of
the lattice) is shown in green. (d) A schematic representation of
the crystal field levels of an isolated Fe$^{2+}$ ion ($d^{6}$) inside of
a distorted FeAs$_{4}$ tetrahedron \cite{Haule09}. The alignment of spins indicates the high
spin state to fulfill Hund's rule, but other spin states are possible. (e) The Fermi surface in the tetragonal phase (shown schematically here) consists
of hole pockets at the center and of electron pockets at the corner
of the (e) 1-Fe Brillouin zone and (f) 2-Fe Brillouin zone. In the
latter, the two elliptical electron pockets fold along the diagonal
wave-vector indicated in (e). The colors indicate the dominant orbital
character of each band \cite{Mazin_review}. An additional $d_{xy}$-dominated hole-pocket is shown centered at $(\pi,\pi)$ in the 1-Fe zone [$(0,0)$ in the 2-Fe zone] (dashed circles). The size of this pocket varies widely across materials; in some cases, such a pocket is not present. The momenta in panel (e) are in units of the
inverse lattice constant $1/a$.}
\label{fig_phase_diagrams}
\end{figure*}

It is from this correlated normal state that not only superconductivity
emerges, but also other electronic ordered states. The majority of
FeSC order magnetically \cite{Dai15}; for example, BaFe$_{2}$As$_{2}$ exhibits
magnetic order with a stripe pattern below a critical temperature
of $134$~K, although more unusual non-collinear and non-uniform
spin configurations are found under hole doping [Fig.~\ref{fig_phase_diagrams}(a)].
Other compounds, such as FeSe, exhibit only magnetic fluctuations
at ambient pressure, but no magnetic order [Fig.~\ref{fig_phase_diagrams}(b)].
More ubiquitously, magnetic fluctuations at the stripe-order wave-vectors
are commonly observed for superconducting compositions. The observation,
by neutron scattering, of an associated resonance in the magnetic spectrum at this specific
wave-vector \cite{Lumsden10,Inosov2016} has been widely interpreted as evidence both for the presence
of a sign-changing superconducting gap and also for magnetic fluctuations
playing a key role in the pairing interaction \cite{Scalapino_review}.

Another common feature found in a large number of FeSC phase diagrams
is a tetragonal-to-orthorhombic structural phase transition. It often
occurs either concurrently or at a higher temperature than the magnetic
transition [Fig.~\ref{fig_phase_diagrams}(a)], although in the specific
case of FeSe it occurs in the absence of magnetic order at ambient pressure  [Fig.~\ref{fig_phase_diagrams}(b)].
A variety of experiments have revealed that lattice strain is not
the primary order parameter for this phase transition, although it
has the same symmetry \cite{Fernandes14}. Borrowing language from the field of liquid
crystals, the state is referred to as an electronic nematic \cite{Fradkin10}, in which
electronic degrees of freedom drive the breaking of (discrete) rotational
symmetry, while translational symmetry is not affected. Experiments
indicate that nematic fluctuations extend far across the phase diagram \cite{Chu12,Bohmer16,Gallais_review},
motivating the question of what role, if any, nematicity plays in these materials.

The most recent surprise is the realization that several representative
FeSC compounds can display topologically non-trivial band structures \cite{Zhang18}.
They have been proposed to promote various topological phenomena,
such as spin-momentum-locked surface states and semi-metallic Dirac
bulk states. Due to the intrinsic fully gapped unconventional superconductivity
of these topological materials, they have become prime candidates
in the search for robust topological superconducting states and their
associated Majorana fermions.

The above brief overview brings us to an important feature of the
FeSC. After 13 years of research there is a wide consensus as to the
nature of the various states found in the phase diagrams. In the Landau
paradigm, these phases are characterized by the symmetries that they
break, and there has been little -- if any -- disagreement for any
of the given materials as to which symmetries are broken. Yet, knowing
what these states are is different from understanding how they arise
and inter-relate with each other and with the emergent superconducting
state. This enables a series of well-posed questions that are, in
some sense, much crisper than what can currently be asked for the
other family of unconventional high-$T_{c}$ superconductors, the cuprates \cite{Keimer15}. In this
review, we outline what is known for sure about FeSC, and pose a series
of open questions that we believe are central to understanding the
origins of their unconventional superconductivity.

\section{Electronic structure and Correlations}

All FeSC are characterized by a common structural motif comprising
tetrahedrally coordinated Fe atoms arranged on a square lattice [Fig.~\ref{fig_phase_diagrams}(c)]. 
The coordinating ligands
are typically from group V (the pnictogens P and As) or group VI (the
chalcogens S, Se and Te). Parent compounds have a formal valence Fe$^{2+}$.
In a description in terms of isolated atomic states, this corresponds
to a $3d^{6}$ electronic configuration displaying three possible
spin states. Bond angles vary
somewhat between compounds, differing from the perfect tetrahedral
angle of $109.5^{\circ}$, thus leading to an additional splitting
between the $d_{xy}$ and $d_{xz}/d_{yz}$ orbitals [Fig.~\ref{fig_phase_diagrams}(d)].

From a band theory perspective, the FeSC can be thought of as compensated
semimetals, having the same number of electron-like and hole-like
carriers \cite{Singh08}. A widely used simplified model features a Brillouin zone
corresponding to the unit cell of the square Fe lattice [shaded gray
area in Fig.~\ref{fig_phase_diagrams}(c)]. The low lying bands form
the electron and hole Fermi-surface pockets displayed in Fig.~\ref{fig_phase_diagrams}(e)
and colored according to the orbitals that contribute the largest
spectral weight for that Fermi momentum
\cite{Mazin_review}. More realistic models include
important effects \cite{Vafek13}, such as the puckering of the As/Se atoms above
and below the Fe plane, which introduces a glide plane symmetry and
implies a crystallographic unit cell (and corresponding Brillouin
zone) containing two Fe atoms [blue shaded areas in Figs.~\ref{fig_phase_diagrams}(c)
and (f)] \cite{Eschrig09}. Additional effects include the spin-orbit coupling 
\cite{Borisenko16}, which
gaps out the intersections of the electron pockets in Fig.~\ref{fig_phase_diagrams}(f),
the three-dimensional dispersion of the bands,
\cite{Eschrig09}, and the hybridization
between the As/Se $p$ and Fe $d$ bands \cite{Fang15,SCZhang16}, which is the root of several
topological phenomena.

In the FeSC the charge and orbital degrees of freedom appear to be
itinerant, as most compounds are metallic at all temperatures and the distinctive charge-gap and satellite  features seen in oxides near a Mott transition are absent in x-ray spectroscopy \cite{XR1}. At low temperatures, the normal state of the FeSC is
well described by the Fermi liquid theory. The qualitative features
of the quasiparticle dispersions, predicted by DFT (density functional
theory) calculations and sketched in Fig.~\ref{fig_phase_diagrams}(e),
are often similar to those detected experimentally using ARPES (angle-resolved
photo-emission spectroscopy) \cite{Richard15,Coldea18,Yi_review} and quantum oscillation measurements \cite{Carrington11}. However, the quasiparticle dispersions are generally reduced
relative to the DFT calculations. Such mass enhancements, also observed in optical conductivity measurements \cite{Qazilbash09}, are attributed to electronic correlations, and were anticipated by DFT+DMFT (dynamical mean-field theory) calculations \cite{Haule08_PRL,Vollhardt09,Werner12,Valenti12,Biermann16}. Moreover, the size of the Fermi pockets are smaller in experiments \cite{Coldea08,Kordyuk13} as compared to the DFT predictions, and in some cases this discrepancy is temperature-dependent \cite{Kaminski13,Brouet13}. Although different mechanisms have been proposed to explain this effect \cite{Ortenzi09,Fernandes_Chubukov,Zantout19,Bhattacharyya20,Kim20}, its origin remains under debate.

\begin{figure}[htbp]
\begin{centering}
\includegraphics[width=0.95\columnwidth]{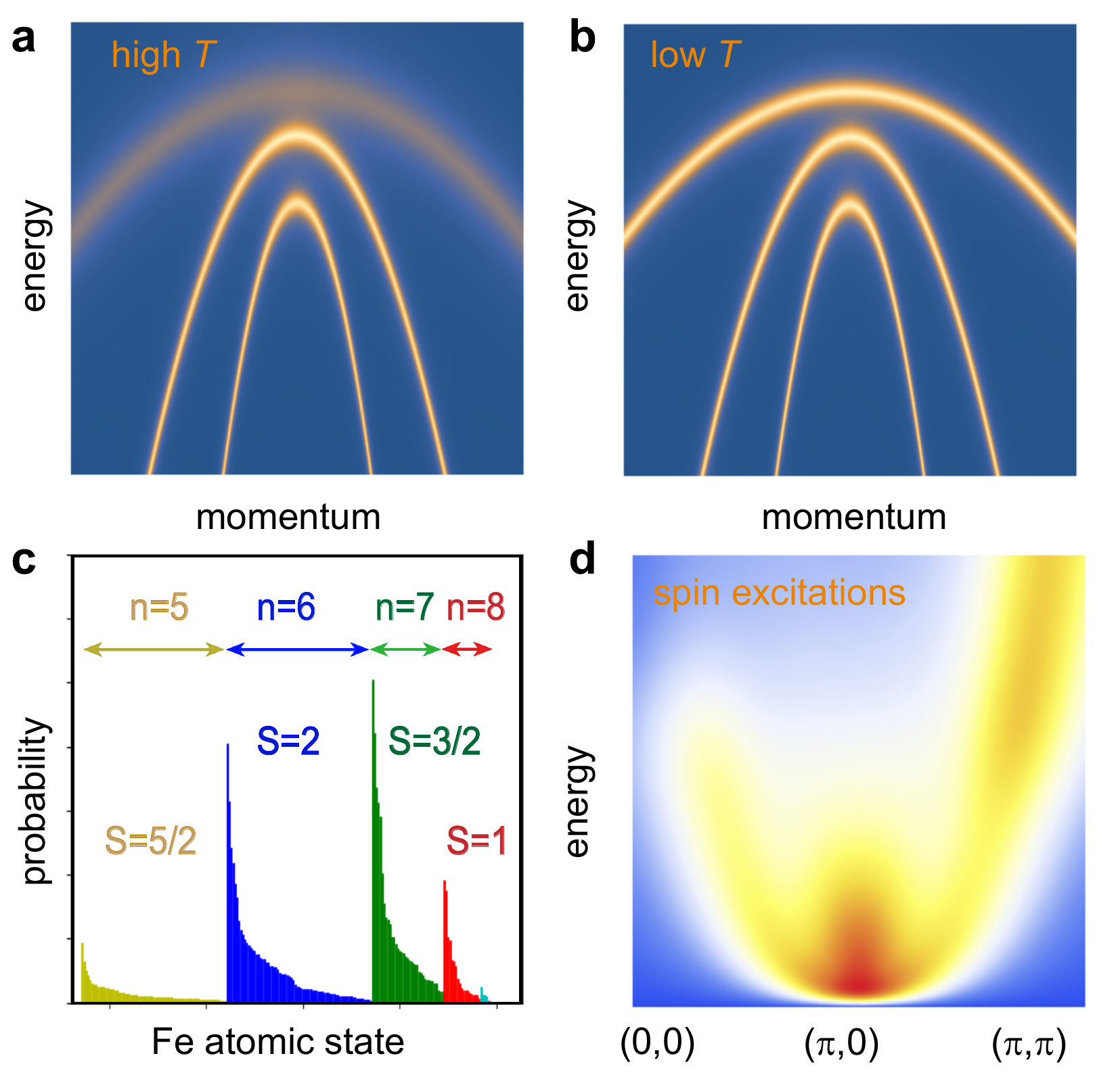} 
\par\end{centering}
\caption{\textbf{Electronic correlations and orbital differentiation}. The
energy dispersions of the three hole bands along a high-symmetry direction
of the 2-Fe Brillouin zone {[}Fig. \ref{fig_phase_diagrams}(f){]}
at (a) high temperatures and (b) low temperatures. The topmost hole
band, which has $d_{xy}$ character, becomes incoherent at high temperatures,
as represented by the faint line. At low temperatures, this band can
reestablish its coherence, but its effective mass can remain sizable,
as indicated by its flatness. The two other hole bands have $d_{xz}$
and $d_{yz}$ character. (c) Histogram of the Fe atomic states in
a parent FeSC, as obtained from DMFT calculations \cite{Haule09}. There are $2^{10}$ possible states involving the $3d$ Fe orbitals. The atomic states are distributed in different colors in the histogram according to their electronic occupation $n$. Within a given sector $n$ of the histogram, the states are ordered by decreasing probability; in all cases, the higher probability corresponds to the high-spin configuration for that occupation. (d) The typical momentum-resolved
spin-excitation spectrum is peaked at different wave-vectors at low
energies {[}in this case, the stripe state $(\pi,0)${]} and high
energies {[}$(\pi,\pi)${]}. \label{fig_correlations}}
\end{figure}

The correlations arise from the Coulomb repulsion between electrons,
which is short-ranged due to screening. This results in an on-site
Hubbard repulsion $U$, which penalizes the system when two electrons
occupy the same site and suppresses spin fluctuations. However, because
multiple orbitals are available in the FeSC, other on-site terms are
also generated by the Coulomb repulsion. Among them is the Hund's
interaction $J_H$, which is unscreened in a solid \cite{Sawatzky88} and favors the alignment of the spins of electrons
in different orbitals. This leads to a correlated metallic state called a Hund
metal, which is different from 
a Mott insulator, in that charge and orbital degrees of freedom are
delocalized, while the spin degrees of freedom remain nearly localized
down to low temperatures. This is illustrated schematically in Fig.~\ref{fig_correlations}(c), which depicts the histogram of all possible $3d$-Fe atomic
states in a Hund metal. While the histogram extends over a wide range of electronic occupations,
showcasing the itinerant nature of the charge carriers, it also displays sharp peaks
at high-spin configurations, illustrating the local nature of the spins.

Fingerprints of the Hund metal phase can be seen in the FeSC upon increasing temperature ($T$), where a coherence-incoherence crossover onsets \cite{Haule09}. In very clean
materials, this manifests in the resistivity behavior, which crosses
over from the characteristic Fermi-liquid $T^{2}$ dependence at low
temperatures to values of the order of several $100$ $\mu \mathrm{\Omega\cdot cm}$
at high temperatures \cite{Schmalian_Kotliar}. 
In a semiclassical treatment, these values imply
that the mean-free-path is of the order of the inverse of the Fermi
momentum, which is not consistent with a picture of propagating Bloch
waves to describe the transport properties.

The coherence-incoherence crossover can also manifest in the electronic spectrum. An example is illustrated in Fig.~\ref{fig_correlations}(a)-(b): at high temperatures, the $d_{xy}$ hole-band is much fainter and flatter than the other two $d_{xz}/d_{yz}$ hole-bands, reflecting the small coherence factor and large effective mass of the former. Upon decreasing the temperature, this $d_{xy}$-band, which only crosses the Fermi level in some compounds [Figs. \ref{fig_phase_diagrams}(e)-(f)], becomes much sharper and thus more coherent. Such an effect, predicted theoretically \cite{Kotliar12,Si13,Medici14}, has been observed in FeSe$_{1-x}$Te$_x$ and LiFeAs \cite{Yi_review}.
In extreme cases, the $d_{xy}$ orbital could remain completely localized down to zero temperature, while the $d_{xz}/d_{yz}$ orbitals remain coherent,
giving rise to an orbital-selective Mott transition \cite{Medici14,Si16}. The fact that the $d_{xy}$ orbital is less coherent than the others is an example of a  phenomenon called orbital differentiation \cite{Kotliar11,Bascones_review,Kreisel20}, by which different orbitals are affected by correlations in distinct ways. This phenomenon is not restricted to the normal state: pronounced orbital differentiation was observed inside the superconducting phase of FeSe \cite{Sprau17}, but its origin is unsettled \cite{Kreisel20,Rhodes18,Zhou18}.  

Correlations also affect the spin-excitation
spectrum probed by neutron scattering, which is rather different at
low and high energies \cite{Dai15}. In momentum space, as sketched in Fig.~\ref{fig_correlations}(d), at low energies
the magnetic spectral weight is strongly peaked near the wave-vector
of the magnetic ground state -- usually, the in-plane stripe ordering
vectors $(\pi,0)$ and $(0,\pi)$.
As one moves towards higher energies, the magnetic spectral weight generally moves towards
$(\pi,\pi)$ \cite{Kotliar_antiphase}, even though Néel order is not observed in the FeSC.

\begin{figure}
\begin{centering}
\includegraphics[width=1\columnwidth]{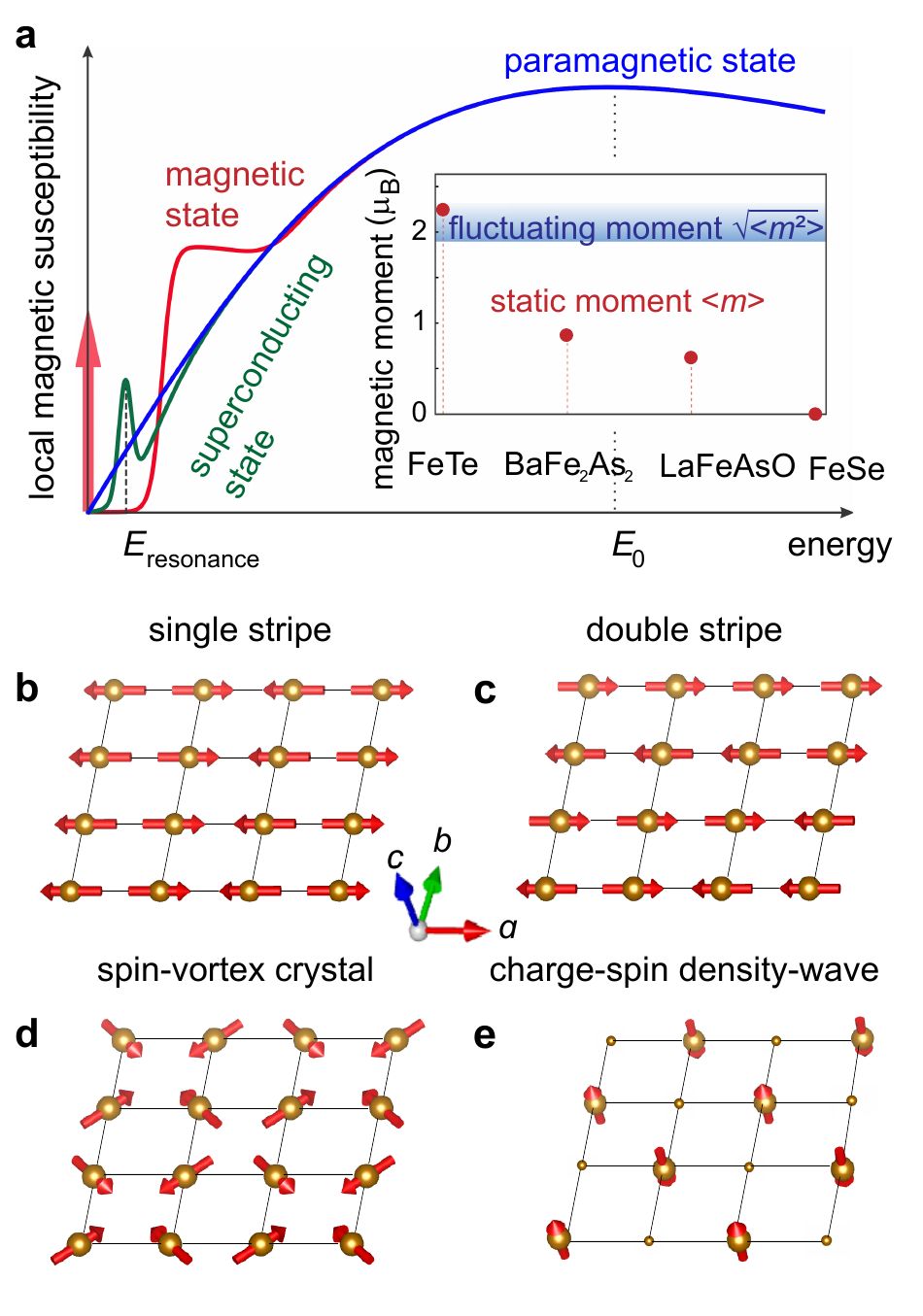} 
\par\end{centering}
\caption{\textbf{Dual local-itinerant nature of magnetism}. (a) Imaginary part
of local susceptibility versus energy in a typical FeSC \cite{Dai15}. All FeSC
show very similar high energy behavior, but differ at low energies
depending on the occurrence of magnetic order (red curve and red arrow,
denoting the Bragg peak) or superconductivity (green). The inset shows
the variation of the static ordered moment $\langle m\rangle$ across
materials, and of the fluctuating local moment, given by the energy-integrated
susceptibility $\sqrt{\langle m^{2}\rangle}$. (b)-(c) Single-stripe and double-stripe configurations
of the Fe spins. The first is realized in most FeSC, whereas the latter,
in FeTe. In momentum space, they correspond respectively to Bragg
peaks at $\left(\pi,0\right)$ {[}or $\left(0,\pi\right)${]} and
$\left(\pi/2,\,\pi/2\right)$ in the 1-Fe Brillouin zone. (d)-(e)
$C_{4}$-symmetric spin configurations observed in electron-doped CaKFe$_{4}$As$_{4}$ \cite{Meier18} and hole-doped SrFe$_{2}$As$_{2}$ \cite{Allred16}, respectively. They correspond to a superposition
of $\left(\pi,0\right)$ and $\left(0,\pi\right)$ wave-vectors resulting
in either a non-collinear spin-vortex phase (d), characterized by
a staggered spin-vorticity across the Fe square plaquettes, or a charge-spin
density-wave phase (e), a non-uniform state with out-of-plane moments
in which half of the Fe atoms display vanishing magnetization and a smaller charge density than the average (smaller yellow spheres). \label{fig_magnetism}}
\end{figure}

This dichotomy between low and high energies is clearly seen  in
the local magnetic susceptibility \cite{Dai15}, whose imaginary part is schematically plotted in Fig.~\ref{fig_magnetism}(a).
At energies $E_{0}$ of the order of $100$ meV, regardless of the
chemical composition, it displays a broad peak that implies a large
local fluctuating magnetic moment -- to be contrasted with the delocalized
nature of the orbital and charge degrees of freedom. Experimental
estimates give a fluctuating moment more or less uniform across different
parent compounds, of about $2$-$3\mu_{B}$ [inset of Fig.~\ref{fig_magnetism}(a)].
In contrast, at energy scales of the order of $10$~meV, the imaginary part of the local
susceptibility in the paramagnetic state increases
with energy \cite{Wang13}, indicative of Landau damping generated by the
decay of spin fluctuations into particle-hole excitations -- a hallmark
of itinerant magnets. Indeed, the system remains metallic inside the magnetically ordered state. 

Thus, while charge/orbital degrees of freedom are itinerant, the spin degrees of freedom display properties typical of strongly-correlated local-spin systems at high energies and of itinerant-spin systems at low energies. This ``orbital-spin" separation \cite{von_delft_PhysRevLett.115.136401} is the most striking feature of the Hund metal. As the temperature is lowered, this correlated metallic state displays Fermi liquid behavior and an ordered phase emerges -- magnetic,
nematic, or superconducting. To understand the low-energy electronic states and the resulting ordered states, it is therefore important to consider both the Fermi surface details [Fig.~\ref{fig_phase_diagrams}(e)], as  revealed by quantum oscillation measurements \cite{Carrington11}, and the magnetic spectrum [Fig. \ref{fig_magnetism}(a)], as revealed by neutron scattering.

\section{Magnetism at the crossroads between itinerancy and localization}

The vast majority of FeSC parent compounds, as is the case of BaFe$_{2}$As$_{2}$
in Fig.~\ref{fig_phase_diagrams}(a), undergo a magnetic transition
to a stripe-like configuration \cite{Dai15}. As shown in Fig.~\ref{fig_magnetism}(b), it
consists of parallel spins along one in-plane
Fe-Fe direction and antiparallel along the other.
There are thus two energetically equivalent stripe states, related
by an in-plane $90^{\circ}$ rotation in both real-space and spin-space.
The spin-orbit coupling generates magnetic anisotropies that force
the spins to point parallel to the selected ordering vector \cite{Christensen15}. As a
result, a spin gap appears in the low-energy part of the local magnetic
susceptibility [Fig.~\ref{fig_magnetism}(a)], while
the high-energy part is not affected.
In contrast to the fluctuating moment, the ordered moment can be rather
small, and it changes considerably across different compounds {[}inset
of Fig.~\ref{fig_magnetism}(a){]} \cite{Lumsden10}. Although some parent compounds -- such as LiFeAs and FeSe -- do not undergo a magnetic transition, they still display low-energy
fluctuations associated with the stripe magnetic state \cite{Braden12,Wang2016}. FeTe is one of the few parent compounds that display
a different magnetic configuration -- the double-stripe state of
Fig.~\ref{fig_magnetism}(c). Even in this case, upon modest substitution of Se for Te, magnetic fluctuations at the single-stripe wave-vectors emerge \cite{Mandrus10,Broholm10}. 

Perturbations such as doping, isovalent
chemical substitutions, and pressure not only tend to reduce the magnetic transition
temperature of the pristine compositions [Fig.~\ref{fig_phase_diagrams}(a)], but they can
also give rise to new magnetic ground states. Locally, impurities
can promote puddles of Néel and other orders \cite{Gastiasoro14}. Globally, doping BaFe$_{2}$As$_{2}$
with electrons stabilizes an incommensurate stripe order \cite{Pratt11}, whereas hole-doping
promotes the so-called $C_{4}$
magnetic phases  \cite{Sheveleva20}. The latter are linear combinations of the magnetic configurations with different stripe
ordering wave-vectors that preserve the tetragonal (i.e. $C_{4}$) symmetry of the lattice \cite{Lorenzana08,RMF_Berg}. They can be either the non-collinear spin-vortex
crystal {[}Fig.~\ref{fig_magnetism}(d){]}, as observed in electron-doped CaKFe$_{4}$As$_{4}$ \cite{Meier18}, or the non-uniform charge-spin
density-wave {[}Fig.~\ref{fig_magnetism}(e){]}, as observed in hole-doped SrFe$_{2}$As$_{2}$ \cite{Allred16} (see Fig.~\ref{fig_phase_diagrams}(a)). 

Overall, the simultaneous presence of features commonly associated with localized and itinerant magnetism has motivated theoretical models adopting both a strong-coupling perspective \cite{Si16,JPHu08,Dagotto_review}, usually based on significant exchange interactions beyond nearest-neighbor spins, and a weak-coupling approach \cite{Bascones_review,Eremin11,Fernandes_Chubukov}, often associated with Fermi-surface nesting. Nesting refers to the situation when the hole and electron pockets in Fig. \ref{fig_phase_diagrams}(e) have comparable shapes and sizes. The deterioration of the nesting conditions upon doping was invoked to explain and anticipate the onset of $C_{4}$ magnetic phases and of incommensurability with doping \cite{Vorontsov10,Fernandes_Chubukov}. DFT has also been widely employed to investigate magnetism in FeSC. While DFT successfully captures the magnetic ground state configuration of most compounds \cite{Mazin08,Yildirim08}, it has problems in explaining the size of the ordered magnetic moment or the absence of magnetism in FeSe \cite{Glasbrenner15}. Advanced, beyond-DFT \textit{ab initio} methods, have been able to address some of these problems \cite{Kotliar11,Hirayama15}.

FeSC also provide a new arena to explore quantum criticality \cite{Abrahams11},
a phenomenon associated with a zero-temperature second-order phase transition tuned by pressure, composition, or
strain (also called a quantum critical point, QCP). Indeed, the extrapolation
of the stripe magnetic transition temperature to zero near the point
where the superconducting dome is peaked [Fig.~\ref{fig_phase_diagrams}(a)]
is reminiscent of certain heavy-fermion materials \cite{Maple15}. Quantum criticality
in those compounds is often empirically associated with non-Fermi-liquid
behavior, such as a metallic resistivity whose temperature dependence deviates
from the $T^{2}$ Fermi-liquid  behavior at low temperatures.
In the FeSC, the clearest evidence for strange metal behavior associated
with a putative QCP is found for BaFe$_{2}$(As$_{1-x}$P$_{x}$)$_{2}$
[Fig.~\ref{fig_phase_diagrams}(a)]. There, a linear-in-$T$
resistivity and an unusual scaling of the magneto-resistance are observed above $T_{{\rm c}}$ near optimal doping \cite{Shibauchi14,Hayes16}, and accompanied by an enhancement of the effective electron mass. Inside the
superconducting dome, a sharp peak of the $T=0$ superconducting
penetration depth is observed \cite{Shibauchi14}, whose origin remains unsettled \cite{Chowdhury13,Levchenko13}. More
generally, whether magnetic quantum criticality is a central ingredient
to the phase diagram of the FeSC remains an unresolved issue, requiring
further experimental and theoretical analyses.

\section{Electronic nematicity and vestigial orders}

\begin{figure}
\begin{centering}
\includegraphics[width=1\columnwidth]{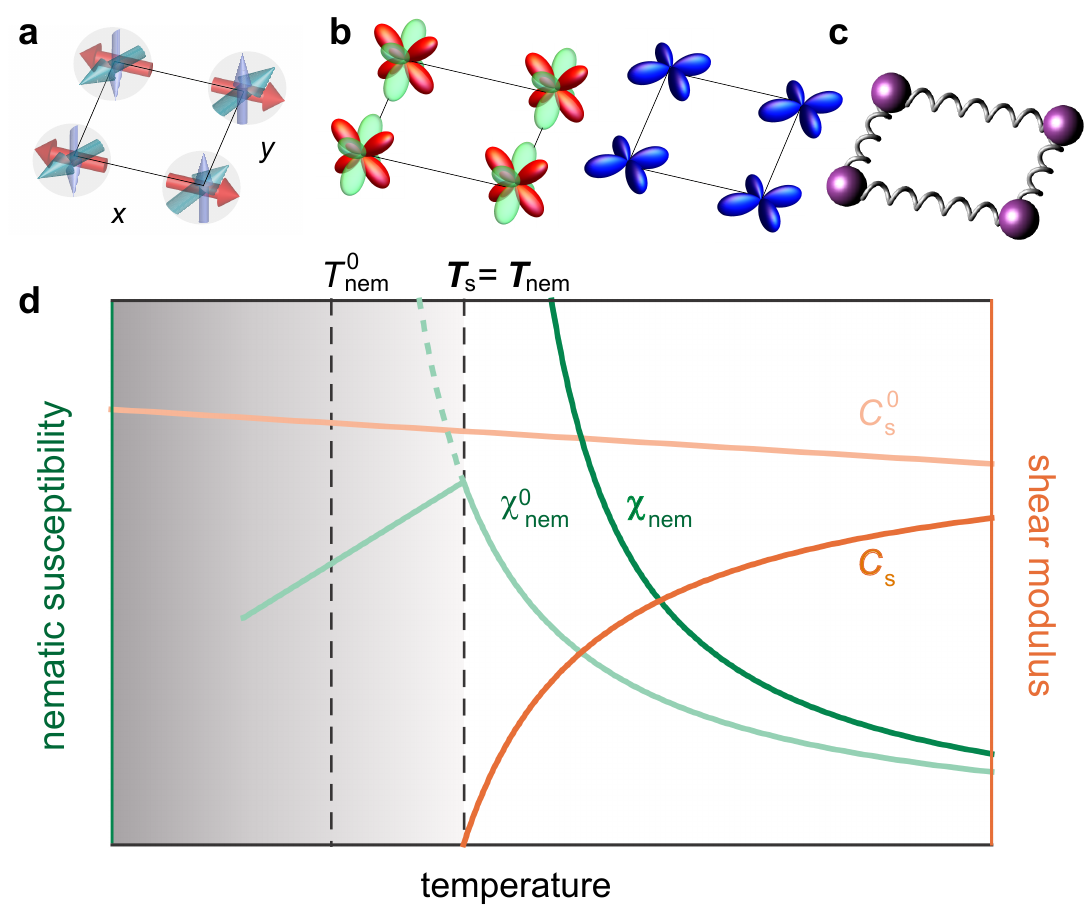} 
\par\end{centering}
\caption{\textbf{Electronic nematic order and its coupling to the lattice}.
The nature of the electronic nematic transition remains a matter of
investigation, and might involve one or a combination of the following
mechanisms \cite{Fernandes14}. (a) In the case of spin-driven nematic order, partial
melting of the stripe magnetic phase results in a state for which
$\left\langle \mathbf{S}_{i}\right\rangle =0$, where $\mathbf{S}_{i}$
denotes the spin of a specific site $i$, but for which $\left\langle \mathbf{S}_{i}\cdot\mathbf{S}_{i+x}\right\rangle =-\left\langle \mathbf{S}_{i}\cdot\mathbf{S}_{i+y}\right\rangle $.
(b) In the case of orbitally-driven nematic order, interactions lead
to a finite difference in the on-site occupancy of orbitals $d_{xz}$
and $d_{yz}$ (denoted by red and green on the left panel) and/or
in the $d_{xy}$ orbital hopping (blue, right panel). Symmetry ensures that all these order parameters take on a finite
value in the nematic state. In all cases, an associated bare (unrenormalized)
nematic susceptibility $\chi_{\mathrm{nem}}^{0}$ can be measured
via a number of experimental techniques \cite{Chu12,Gallais_review,Bohmer16}. Coupling to the lattice results
in a spontaneous strain with the same symmetry (i.e. a concomitant
ferroelastic structural phase transition) at $T_{{\rm s}}=T_{{\rm nem}}$
{[}panel (c){]}, occurring at a slightly higher temperature than the
bare nematic transition temperature $T_{\mathrm{nem}}^{0}$. It also
leads to a renormalization of the nematic susceptibility $\chi_{\mathrm{nem}}$
for temperatures above $T_{{\rm s}}$, and a softening of the elastic
modulus $C_{{\rm s}}$ in the same symmetry channel from its bare
value $C_{{\rm s}}^{0}$ {[}panel (d){]}. Grey shading indicates the
magnitude of the nematic order parameter.}
\label{fig_nematicity} 
\end{figure}

While on symmetry grounds the nematic transition observed in most
phase diagrams of FeSC is no different than a tetragonal-to-orthorhombic
transition, the driving force can arise from a number of possible
physical mechanisms. Quite generally, one can define order parameters
that break the tetragonal symmetry of the system in different channels
-- spin, orbital, and lattice {[}Figs. \ref{fig_nematicity}(a)-(c){]} \cite{Fernandes14}.
Symmetry requires that all of these are simultaneously non-zero or
zero, but cannot answer the question of which is the primary one. Indeed, direct experimental manifestations of nematic order have been reported in orbital \cite{Yi_review}, magnetic \cite{Kasahara10,PCDai14}, and elastic \cite{Bohmer16} degrees of freedom, with associated anisotropies in transport \cite{Chu10,Tanatar10}, optical \cite{Degiorgi15} and local electronic properties \cite{JCDavis10,Rosenthal14}. A crucial insight came from the realization that the strain is either
the primary nematic order parameter -- in which case the nematic
transition would be a simple structural instability -- or a conjugate
field to it -- in which case the instability would be electronically
driven. Measurements of the elasto-resistivity \cite{Chu12}, as well as of the Raman spectrum \cite{Gallais_review} and elastic stiffness \cite{Bohmer16}, settled this issue, establishing the dominant low-energy electronic character of the nematic state. Nevertheless, coupling to the lattice raises the critical temperature by a small amount from $T_{\mathrm{nem}}^0$ to $T_{\mathrm{nem}}$ [Fig. \ref{fig_nematicity}(d)].

To explain the electronic mechanism behind the nematic transition,
two general scenarios have been explored, attributing it primarily to either
spin or orbital degrees of freedom. This distinction, however, can become subtle, since they work in tandem \cite{Dagotto13,Fanfarillo15}, possibly to different extents in different materials.
In the simplest manifestation of the latter case, interactions spontaneously
lift the degeneracy between the $d_{xz}$ and $d_{yz}$ orbitals \cite{Ku09,Phillips10,Kontani16},
resulting in deformations of the Fermi surfaces in Fig. \ref{fig_phase_diagrams}(e). In contrast, the former scenario relies
on the proximity to the stripe magnetic instability, which breaks
both the (discrete) rotational and translational symmetries of the
lattice \cite{Fernandes14,Kivelson08,Sachdev08}. The idea is that the stripe magnetic phase melts in two stages,
first restoring the broken translational symmetry and then the four-fold
rotational symmetry. The intermediate paramagnetic-orthorhombic phase that onsets between the magnetic-orthorhombic and paramagnetic-tetragonal phases is the electronic nematic. Being a partially-melted magnetic
phase, this has been dubbed a ``vestigial'' phase of the stripe
magnetic state \cite{Fernandes19}. Theoretically, because it is stabilized by magnetic fluctuations, vestigial nematicity can be captured by phenomenological Ginzburg-Landau models that go beyond the mean-field approximation. Microscopically, it has been found in both localized spin \cite{Kivelson08,Sachdev08,Wang15,Si16} and itinerant magnetic \cite{Fernandes12,Fanfarillo15} approaches.  

The appeal of the spin-driven mechanism is that it naturally accounts
for the close proximity between the stripe-magnetic and nematic phase
boundaries in the phase diagrams of most FeSC. The nature of these coupled transitions -- split or simultaneous, second-order or first-order -- can be controlled by doping or pressure \cite{Gati19}. This mechanism also explains the
absence of nematic order when the magnetic ground state is not the
stripe one. Experimentally, the strongest evidence in favor of this
scenario is the scaling between the shear modulus $C_{s}$ and the
NMR spin-lattice relaxation rate $1/T_{1}$, suggestive that the lattice
softening is caused by magnetic fluctuations \cite{Fernandes13}. Application of this
mechanism to FeSe is problematic \cite{Baek15}, however, since stripe-magnetic order
only appears when pressure is applied \cite{Bohmer19}, and is completely absent in
the phase diagram obtained by sulfur substitution {[}Fig.~\ref{fig_phase_diagrams}(b){]}.
The orbital-order scenario also faces challenges, at least in its
simplest form, since ARPES measurements indicate the inadequacy of
simple on-site ferro-orbital order \cite{Ding15,Pfau19}. As a result, the origin of nematic order in iron chalcogenides remains an open question.

From a purely phenomenological
perspective, the presence of a doping-dependent nematic phase
transition implies the possibility of a nematic QCP. A variety of
theoretical studies point to possible exotic non-Fermi-liquid effects
proximate to such a QCP, with implications for the description of
the normal state from which the superconductor emerges \cite{Lederer17,Klein18}. 
Probing this,
however, is challenging because of its proximity to a putative magnetic
QCP for the vast majority of materials. The very nature of the coupled
nematic-magnetic quantum phase transitions remains unsettled both
experimentally and theoretically. Nevertheless, recent data unveiling
power-law scaling of the nematic critical temperature as it is suppressed
by doping and externally-induced strains provides strong
evidence for a nematic QCP in the phase diagram of BaFe$_{2}$As$_{2}$,
with an associated quantum critical regime that spans a large part
of the phase diagram \cite{Worasaran20}. Another promising arena to study nematic quantum
criticality is FeSe$_{1-x}$S$_{x}$ \cite{Shibauchi20,Kreisel20,Coldea20} {[}Fig.~\ref{fig_phase_diagrams}(b){]},
where magnetic order is absent -- although experimental evidence
for possible non-Fermi liquid behavior in the vicinity of the nematic
QCP remains controversial \cite{Shibauchi20,Coldea20}.

\begin{figure*}
\begin{centering}
\includegraphics[width=1.7\columnwidth]{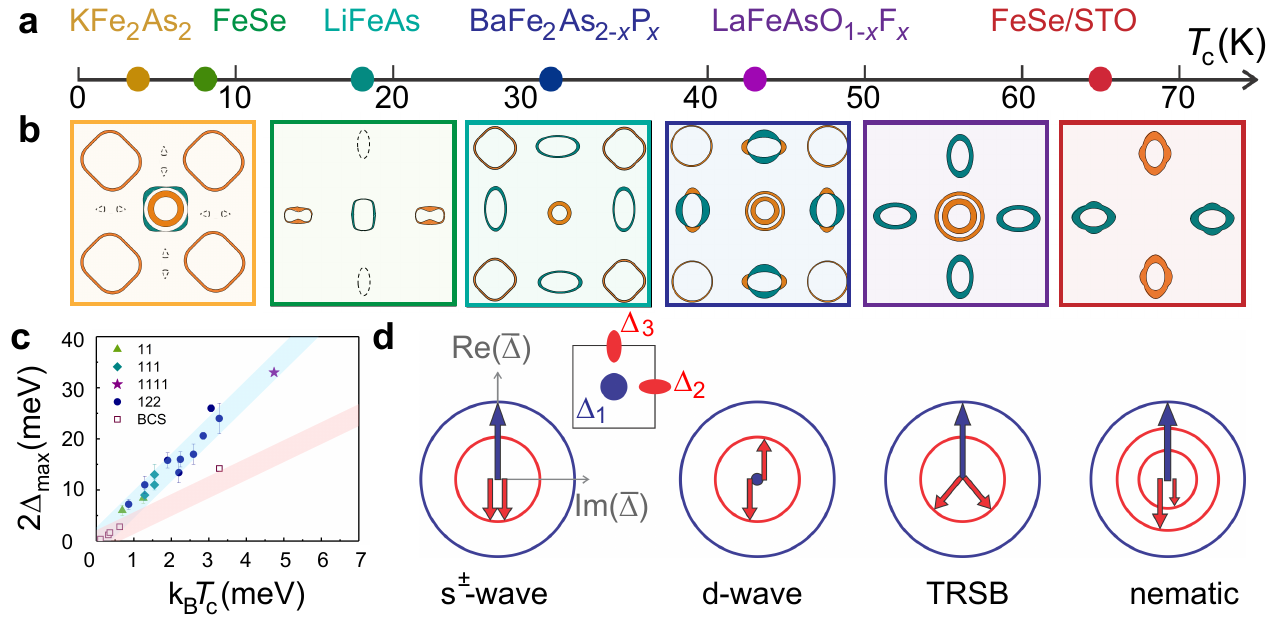}
\par\end{centering}
\caption{\textbf{Superconducting gap structures and gap symmetries}. (a) Superconducting
critical temperatures, $T_{{\rm c}}$, of six canonical Fe-based superconductors.
(b) Schematic gap structures for these materials in the 1-Fe Brillouin
zone (borders colored according to (a)) based on weak-coupling calculations, 
ARPES, and STM experiments (see \cite{Okazaki12,Mazin_review,Chubukov_review,Hirschfeld_review} and references therein).
The line thickness represents
the magnitude of the gap, whereas the green and orange colors denote
different signs. (c) Ratio between twice the maximum gap ($2\Delta_{\mathrm{max}}$,
based on ARPES data) and $k_{B}T_{{\rm c}}$ of FeSC, compared to
that of conventional superconductors \cite{Lee18}. Here, $k_{B}$ is the Boltzmann
constant. Colored symbols correspond to some of the materials labeled
in (a). The open square symbols correspond to conventional BCS superconductors. (d) Possible superconducting ground states realized
in a three-band toy model with repulsive inter-band interactions \cite{Stanev10,Grinenko20,Fernandes_Chubukov}.
The red (blue) arrows are associated with the complex value of the
gap averaged around the electron (hole) pockets, $\bar{\Delta}$ (see
inset). TRSB denotes time-reversal symmetry breaking.}
\label{fig_superconductivity}
\end{figure*}

Theoretically, an understanding of the nematic QCP requires incorporating
two important ingredients. The first is the inevitable coupling to
the lattice, which induces long-range dipolar-like nematic interactions \cite{Karahasanovic16}.
They not only suppress critical nematic fluctuations, promoting a
mean-field transition, but also change the properties of the QCP,
favoring a Fermi-liquid behavior at low enough temperatures \cite{Paul17,Reiss20}. The second
ingredient is the ubiquitous presence of random local strains in the
sample, caused by dopants and other defects \cite{Curro15,Frandsen19,Billinge19}. They couple to the nematic order parameter as a random conjugate field,
promoting effects typical of the random-field Ising-model. It has
been argued that these effects are responsible for a deviation from
Curie-Weiss behavior of the nematic susceptibility \cite{Kuo16}. Significant progress
will require models that combine long-range phonon-mediated interactions,
random strain, and quantum criticality.

\section{Unconventional superconducting states}

The FeSC display a wide range of superconducting transition temperatures \cite{Johnston10},
as illustrated in Fig.~\ref{fig_superconductivity}(a). The largest
$T_{{\rm c}}\approx65$ K is observed in monolayer FeSe grown on SrTiO$_{3}$,
but the precise temperature where phase-coherent superconductivity
sets in is still under dispute (see \cite{Hoffman} for a review). In contrast to cuprates, many pristine
(i.e. unsubstituted) compounds display superconductivity, such as
bulk FeSe, LiFeAs, and CaKFe$_{4}$As$_{4}$. In others, such as BaFe$_{2}$As$_{2}$
and LaFeAsO, the competing magnetic and nematic orders need to be
suppressed, e.g. via doping, chemical substitution, or pressure, to obtain superconductivity
{[}Figs.~\ref{fig_phase_diagrams}(a) and (b){]}. In some compounds, a second superconducting dome can be accessed by pressure or doping \cite{Hoson18}. In all cases, NMR measurements support a singlet pairing state \cite{Johnston10}.

Because the DFT-calculated electron-phonon coupling
cannot account for the $T_c$ of the FeSC \cite{Haule08_PRL,Boeri08}, an electronic mechanism has been proposed \cite{Mazin_review,Chubukov_review,DHLee_review}. However, this does not preclude electron-phonon interaction, which can be enhanced by correlations \cite{Mandal14}, 
from playing a role in superconductivity, as it has been proposed in monolayer FeSe \cite{Lee2014}. Quite generally, electronic repulsion forces the gap function to change sign in real/momentum space. For a large Fermi surface, such as in the cuprates, this can be accomplished by an anisotropic gap (e.g. $d$-wave). For multiple small Fermi pockets, such as in the FeSC, the gap can remain nearly isotropic around each Fermi surface, as long as it acquires different signs (i.e. phases) on different pockets. We refer to any gap structure that satisfies this criterion as $s^{+-}$-wave. In the FeSC, a strong repulsive pairing interaction is believed to be promoted by magnetic correlations associated with the nearby stripe magnetic state {[}Fig. \ref{fig_phase_diagrams}(a){]} \cite{Mazin08}. In a weak-coupling approach, which can be implemented via RPA (random phase approximation) or (f)RG [(functional) renormalization group] calculations, the inter-pocket interaction is boosted by spin fluctuations peaked at the stripe wave-vectors $(\pi,0)$ and $(0,\pi)$, which connect the hole and electron pockets, thus overcoming the intra-pocket repulsion \cite{Mazin_review,Chubukov_review,Platt13}. In a strong-coupling approach, pairing between next-nearest-neighbor sites is promoted by the dominant next-nearest-neighbor antiferromagnetic exchange interaction \cite{JPHu08,Richard15,Si16}. Although there are important differences between the two approaches, both generally find an $s^{+-}$ gap with opposite signs on the electron and hole pockets.

Phase-sensitive interference experiments to distinguish the
$s^{+-}$ state from the more conventional $s^{++}$ state, which
has also been proposed as mediated by orbital fluctuations \cite{Kontani10}, are challenging because the Cooper pairs have
zero angular momentum in both cases. Nevertheless, there is very strong
indirect evidence that an $s^{+-}$-wave state is realized, most notably the observation of a resonance mode in
the magnetic spectrum below $T_{c}$ \cite{Lumsden10,Inosov2016}. This manifests as a sharp peak in the magnetic susceptibility
measured at the stripe wave-vectors and at an energy $E_{\mathrm{resonance}}$
below twice the gap value, $2\Delta$ [see schematic illustration in Fig.~\ref{fig_magnetism}(a)]. Such a feature is naturally
explained if the gaps at momenta separated by the stripe wave-vectors
have opposite signs \cite{Scalapino_review}. Additional indirect evidence includes the observation
of half-integer flux-quantum transitions in loops of polycrystalline
FeSC \cite{Chen10} and the behavior of the momentum-integrated quasi-particle interference \cite{Hirschfeld_review}.
Moreover, the introduction of controlled disorder via irradiation has been employed to indirectly probe the gap structure. Features in agreement with an $s^{+-}$ state have been observed, such as the lifting of accidental nodes by disorder and the rate of suppression of $T_c$ with impurity scattering \cite{Cho18,Hirschfeld_review}. The observation of in-gap bound states at nonmagnetic impurities in doped compounds is also evidence for a sign-changing gap \cite{Yang13}.

Various gap structures can be realized under the $s^{+-}$-wave umbrella, depending on details of the Fermi surface and on the orbital degrees of freedom \cite{Mazin_review,Chubukov_review,Platt13}. While the gap function generally has opposite signs on electron and hole pockets, additional sign changes between same-character pockets may occur \cite{Kotliar_antiphase}. Moreover, while ARPES observes nearly isotropic gaps in many compounds \cite{Richard15}, accidental nodes may occur as well \cite{Okazaki12}, which are well described by weak-coupling models \cite{Kuroki09,Mazin_review}. Some of these gap structures are illustrated in Fig.~\ref{fig_superconductivity}(b), in the 1-Fe Brillouin zone.  They represent
the leading current candidates for the gap structure of the materials in Fig. \ref{fig_superconductivity}(a), partly motivated
by theoretical considerations, but consistent with ARPES, STM, and/or neutron scattering measurements.

The variety of gap structures in Fig.~\ref{fig_superconductivity}(b)
and the wide range of $T_{{\rm c}}$ values in Fig.~\ref{fig_superconductivity}(a)
raise the question of whether there is really a common, dominant pairing
mechanism in the FeSC. Evidence in favor of this comes from
the dimensionless ratio $2\Delta_{\mathrm{max}}/\left(k_{B}T_{{\rm c}}\right)$,
where $\Delta_{\mathrm{max}}$ is the zero-temperature value of the
largest gap. As shown schematically in Fig. \ref{fig_superconductivity}(c),
this ratio usually falls between $6.0$ and $8.5$ for the vast majority
of FeSC (blue shaded region in the figure) \cite{Lee18}. This is to be contrasted with the $3.5$-$4.5$
range observed in canonical electron-phonon superconductors (red shaded
region). Despite significant theoretical progress, particularly in multi-orbital weak-coupling approaches, several important questions about the superconductivity of FeSC remain open. They include establishing how the gap structure and $T_c$ depend on specific materials parameters, such as the FeAs$_{4}$ tetrahedral angle \cite{Johnston10}, and explaining the seemingly universal $2\Delta_{\mathrm{max}}/\left(k_{B}T_{{\rm c}}\right)$ ratio. Another issue is the case of compounds with only hole pockets (such as KFe$_2$As$_2$) or only electron pockets (such as monolayer FeSe), which do not fall within the standard weak-coupling $s^{+-}$ paradigm. Yet, both types of systems display superconductivity, and some of the electron-pocket-only compounds are among the highest $T_{{\rm c}}$'s in all FeSC. Moreover, the relevance of magnetic fluctuations in these compounds is not well established. This begs for new approaches that can elucidate the pairing mechanism in these compounds (see e.g. \cite{Vafek_Chubukov}) and its relationship with the other FeSC.

The phase diagrams in Figs.~\ref{fig_phase_diagrams}(a)-(b) show
that, besides stripe magnetism, nematic order is also suppressed down
to zero temperature near the doping composition where $T_{c}$ is
the largest. This has led to an important question that remains unresolved:
what role do nematic fluctuations play for the pairing state of the
FeSC \cite{Fernandes14}? Theoretically, nematic fluctuations generate an attractive pairing
interaction peaked at zero momentum. Hence, they can boost $T_{c}$
in any symmetry channel promoted by a more dominant pairing interaction
(e.g. due to spin fluctuations). Nematic fluctuations can plausibly
promote superconducting order on their own, particularly near a QCP \cite{Lederer17,Klein18}.
However, in the clearest case of FeSe$_{1-x}$S$_{x}$ {[}Fig.~\ref{fig_phase_diagrams}(b){]},
no strong change in $T_{{\rm c}}$ is observed at the putative nematic
QCP \cite{Coldea20}, perhaps due to strong elasto-nematic coupling \cite{Paul17,Reiss20}.

The multi-band nature of the FeSC also opens the door for more exotic
pairing states besides $s^{+-}$. To illustrate this, consider a toy
model with one hole and two electron pockets subjected
to repulsive pairing interactions. Fig.~\ref{fig_superconductivity}(d)
shows schematically the possible pairing states obtained by tuning
the ratio between the inter-band electron-pocket/electron-pocket and electron-pocket/hole-pocket interactions, which can be different e.g. if the orbital compositions of the Fermi pockets are distinct.
When the ratio is small, an $s^{+-}$ state is obtained: the gaps
on the electron pockets are identical and have a $\pi$ phase
shift with respect to the hole-pocket gap. When the ratio is large,
the ground state is $d$-wave: the gaps on the two electron pockets
have equal magnitude but a relative $\pi$ phase, whereas the gap
on the hole pocket averages to zero. When the ratio is of order one, it is possible to realize a nematic $s+d$ superconducting state \cite{Fernandes_Chubukov}, in which the electron-pockets gaps have the same phase but distinct magnitudes. This is not to be confused with the case where nematicity onsets above $T_{{\rm c}}$
and then coexists with superconductivity, as in FeSe. Another option is a time-reversal
symmetry-breaking (TRSB) $s+id$ superconducting state \cite{Congjun09,Platt13}, in which the electron-pockets gaps have equal magnitude but their relative phase is neither $0$
(as in an $s^{+-}$ state) nor $\pi$ (as in a $d$-wave).
A different type of TRSB pairing state, called $s+is$ \cite{Stanev10}, has been proposed in
heavily K-doped BaFe$_{2}$As$_{2}$, based on muon-spin rotation
measurements \cite{Grinenko20}. 

More broadly, the variation of orbital spectral weight along the Fermi pockets [Fig. \ref{fig_phase_diagrams}(e)] endows the projected pairing interaction with an angular dependence, which can favor non-$s$-wave pairing. Microscopic calculations have in fact suggested that the $s^{+-}$ and $d$-wave interactions can be comparable in strength \cite{Mazin_review,Chubukov_review,Platt13,Hirschfeld_review}. This is
supported by the observation of certain peaks in the Raman spectrum
that have been interpreted as collective $d$-wave excitations inside
the $s^{+-}$ state \cite{Hackl13,Blumberg16}. Alternatively, such sharp peaks have also been
attributed to a collective nematic excitation \cite{Gallais16}. The non-monotonic evolution of $T_c$ with pressure in KFe$_2$As$_2$ has also been interpreted as evidence for nearly-degenerate superconducting states \cite{Tafti13}. Finally, the fact that
the small Fermi energy of some FeSC is comparable
to the gap value has motivated the search for strong-coupling superconductivity
described not by the BCS formalism, but by the Bose-Einstein condensate
(BEC) prescription of tightly-bound pre-formed Cooper pairs. Although
certain properties of FeSe and FeTe$_{1-x}$Se$_{x}$ have been described in terms of a BEC-BCS crossover \cite{Kanigel17,Shibauchi14}, direct evidence for pre-formed
pairs remains to be seen.

\section{Topological phenomena}

\begin{figure}
\begin{centering}
\includegraphics[width=0.9\columnwidth]{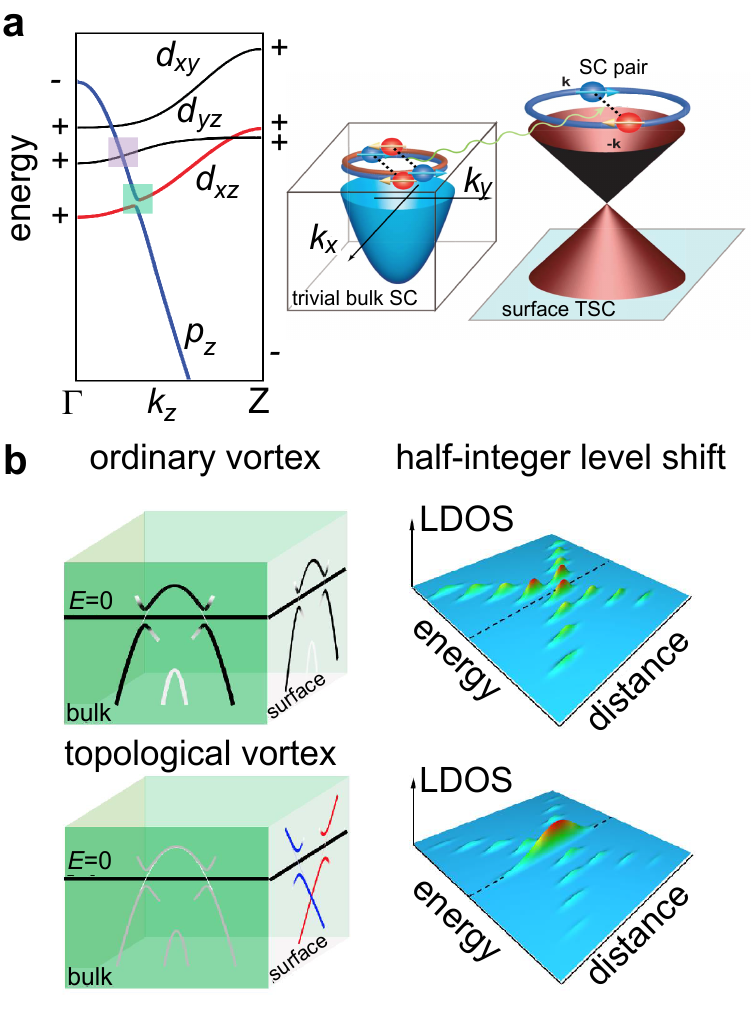}
\par\end{centering}
\caption{\textbf{Band inversion and topological phenomena.} (a) The topological
$p$-$d$ band inversion in the FeSC is illustrated in the left panel.
The downward shift of the $p_{z}$ orbital along the $\Gamma$-$Z$
direction causes different topological phenomena such as bulk Dirac
semimetal states (purple region) and helical Dirac surface states
(green region), depending on whether it crosses a $d$ orbital of
the same parity ($d_{xy}$ or $d_{yz}$) or of different parity ($d_{xz}$),
respectively. The right panel illustrates the topological superconductivity
induced on the surface Dirac states by the bulk superconducting state. Figure reproduced from \cite{Zhang18}.
(b) In the quantum limit, which is achievable in some FeSC, discrete
levels can be observed inside a vortex by probing the local density
of states (LDOS) via STM. In an ordinary vortex (upper panels), these
bound states are all at finite energies, whereas in a topological
vortex (lower panels), a sharp zero-energy mode, called a Majorana
zero mode, appears well separated from the other bound states. Figure reproduced from \cite{Kong19}.
\label{fig_topology}}
\end{figure}

One of the most recent developments in the field is the discovery
of topological properties in some FeSC (for recent reviews, see \cite{Wu_review,Kong_review,Shibauchi20,Kreisel20}). As schematically illustrated
in Fig.~\ref{fig_topology}(a), this arises from $p$-$d$ band inversions
along the $\Gamma$-$Z$ direction, involving an odd-parity anionic
$p_{z}$-band and an even-parity Fe $d$-band ($t_{2g}$), as predicted
by DFT \cite{Fang15,SCZhang16} and observed by ARPES \cite{Zhang18,Kanigel20}. Whereas the crossings with the bands
of dominant $d_{xy}$ or $d_{yz}$ character are protected, resulting
in bulk topological Dirac semimetal states (purple shaded region),
the crossing with the $d_{xz}$-dominated band is gapped, resulting
in a topological insulating state (green shaded region). The helical
surface Dirac cones emerging when the chemical potential crosses this
topological gap, as well as the bulk Dirac semimetal states, were observed by ARPES in a few FeSC, most notably FeTe$_{1-x}$Se$_{x}$ \cite{Zhang18,Zhang19}. 

Upon emergence of the $s^{+-}$-wave state in the bulk,
superconductivity can be induced on these Dirac surface states {[}right
panel of Fig.~\ref{fig_topology}(a){]}. Similarly to topological
insulator/superconductor heterostructures, the surface Dirac states
of the FeSC can also support Majorana zero modes (MZM) in the vortex
cores of the superconducting state. The important difference is that
the topological superconductivity on the FeSC surface is intrinsic
and displays higher $T_{c}$ values, while also avoiding the interfacial
complexities of the heterostructures.

Perhaps even more significantly, the FeSC usually have small Fermi
energies ($E_{{\rm F}}$) due to correlation effects, as discussed
in Section II. In the particular case of FeTe$_{1-x}$Se$_{x}$, $E_{{\rm F}}$
can be so small that it becomes comparable to the superconducting
gap $\Delta$. This is important because, inside the vortex of any
superconductor, there are discrete energy levels of $\nu\Delta^{2}/E_{{\rm F}}$.
These levels can only be distinguished in the quantum limit, where
thermal broadening is smaller than the level spacing $\Delta^{2}/E_{{\rm F}}$.
Because $\Delta$ and $E_{F}$ are of similar magnitudes, the quantum
limit is achievable in the FeSC \cite{Chen18}. In an ordinary vortex, $\nu$ is
expected to be half-integer, and the discrete levels never have zero
energy {[}upper panels of Fig.~\ref{fig_topology}(b){]}. However,
in a topological vortex, where Dirac surface states are incorporated,
$\nu$ is shifted to integer values due to the spin texture of the Dirac
fermions \cite{Kong19}. As a result, a MZM emerges as the vortex bound state with
zero energy {[}lower panels of Fig.~\ref{fig_topology}(b){]}. Experimentally,
zero-energy bound states as well as higher-energy discrete levels
have been observed in FeTe$_{1-x}$Se$_{x}$ via STM
measurements \cite{Gao18,Machida19,Kong19}, providing strong support for the existence of MZM.

Notwithstanding its simplicity and its intrinsic character, the FeSC
Majorana platform is subjected to issues such as spatial inhomogeneity
and the three-dimensional nature of the crystals. Some of these issues
may be the reason why MZM are only observed in a fraction of the vortices
realized inside the superconducting state \cite{Kong19}. Besides in the interior of vortices, signatures of Majorana fermions have
also been reported in different types of lattice defects, such as
interstitials \cite{Yin15}, line vacancies \cite{Chen20}, and crystalline domain boundaries \cite{Madhavan20}. In the latter case, a
one-dimensional dispersing Majorana mode was found. A potential link between the coherent-incoherent crossover and the onset of topological superconductivity in FeTe$_{1-x}$Se$_{x}$ was also observed \cite{Yangmu20}. 

Several theoretical proposals for realizing other exotic topological
effects based on the $p$-$d$ band inversion have been put forward \cite{Wu_review},
such as dispersing Majorana fermions \cite{Konig_Majorana} and higher-order Majorana modes
in corners and hinges of samples \cite{DasSarma19}. Topological phenomena that rely
only on the Fe $3d$ orbitals have also been proposed, such as the
realization of double Weyl points \cite{Heinsdorf21}. Theoretically, including the role
of correlations in predictions of topological effects will be an important
step forward, particularly in compounds as correlated as FeTe$_{1-x}$Se$_{x}$.
The interplay with other types of electronic order, such as nematicity
and magnetism, also remains little explored. Advances on the experimental
front will benefit from controllable tuning of MZM and from designing
a feasible pathway for braiding them \cite{Kong20}. 

\section{Outlook}

After 13 years, the FeSC continue to provide a rich and unmatched
framework to assess the interplay between correlations, unconventional
superconductivity, magnetism, nematicity, quantum criticality, and
topology. While significant advances have occurred, deep questions linger and continue to emerge   as  discussed in this review. New iron-based compounds  continue to be  regularly discovered, some with unusual structural properties promoted by the spacing layers, such as CaKFe$_4$As$_4$ with centers of inversion away from the FeAs layer, the monoclinic Ca$_{1-x}$La$_x$FeAs$_2$ \cite{Katayama13}, and the insulating ladder compound BaFe$_2$Se$_3$ \cite{Dagotto_selenides,Birgeneau20}. Pnictide compounds based not on iron, but on nickel, cobalt, and manganese have also been systematically grown and studied. These systems offer new opportunities to address some of those unanswered questions and, at the same time, venture into unexplored directions.

Many of the theoretical and experimental advances spurred by studies of the FeSC have found fertile ground in other quantum materials. The discovery of this completely new class of compounds provided a fresh testbed to compare state-of-the-art methods in correlated electronic structure calculations against a large number of experimental results. The concept of a Hund metal has also been used to explain the normal-state properties of a broad range of quantum materials, such as Sr$_2$RuO$_4$ \cite{Georges13}. Multi-orbital pairing models, such as RPA, FLEX, and (f)RG, have been extensively employed to shed new light on multi-band superconductors, such as ruthenates and nickelates. The concept of vestigial orders and the associated phenomenological models have led to important new insights into antiferromagnetic and topological superconducting materials \cite{Fernandes19}. Experimentally, symmetry-breaking strain has been recognized as a uniquely appropriate tool to probe electronic nematic order. Strain-based techniques applied to transport, thermodynamic, scattering, spectroscopic and local probe measurements are now considered mainstream, and have enabled the identification and manipulation of electronic nematicity and a variety of other electronic states in disparate materials such as cuprates and $f$-electron systems -- which have emerged as promising candidates to realize nematic quantum criticality \cite{Maharaj17}.

Conceptually, FeSC emphasize a type of correlated state -- the Hund metal -- that is different from a Mott insulator or a heavy-fermion metal. Although at low temperatures and energies it behaves as a Fermi liquid, it displays a wide temperature range where charge/orbital degrees of freedom seem itinerant but spins seem localized. On the one hand, approaches that go beyond DFT and consider the frequency dependence of the electronic interactions (such as DMFT) have been successful in describing effects related to the intermediate-energy range, such as the large fluctuating local moment and the coherent-incoherent crossover experience by the $d_{xy}$ orbital. On the other hand, phenomenological models that go beyond mean-field and perturbative approaches such as (f)RG and RPA, which focus on the long-wavelength momentum dependence of the electronic interactions, have nicely captured low-energy phenomena that emerge when the system enters its Fermi-liquid regime, such as the structure of the magnetic, superconducting, and nematic order parameters. Significant progress in this field -- and in other similar types of quantum materials -- will require novel ideas that can seamlessly combine long-wavelength theories with more accurate correlated electronic structure methods, thus smoothly interpolating between the intermediate- and low-energy behaviors. 

\begin{acknowledgements}
We thank all our co-authors and collaborators with whom we had many fruitful discussions since the discovery of the iron-based superconductors. We particularly thank H. Miao and  T. H. Lee (Figure 2), M. Christensen (Figure 4) and L.-Y. Kong (Figure 6) for their assistance in making some of the figures. R.M.F. was supported by the U.S. Department of Energy, Office of Science, Basic Energy Sciences, Materials Science and Engineering Division, under Award No. DE-SC0020045. A.I.C. acknowledges an EPSRC Career Acceleration Fellowship (EP/I004475/1) and the Oxford Centre for Applied Superconductivity (CFAS) for financial support. A.I.C is grateful to the KITP program {\it correlated20}, which was supported in part by the National Science Foundation under Grant No. NSF PHY-1748958. H.D. is supported by the National Natural Science Foundation of China (Grant Nos. 11888101, 11674371), the Strategic Priority Research Program of Chinese Academy of Sciences, China (Grant Nos. XDB28000000, XDB07000000), and the Beijing Municipal Science \& Technology Commission, China (Grant No. Z191100007219012). I.R.F. was supported by the Department of Energy, Office of Basic Energy Sciences, under contract DE-AC02-76SF00515. P.J.H. was supported by the U.S. Department of Energy, Office of Basic Sciences under Grant No. DE-FG02-05ER46236. G.K. was supported by NSF DMR-1733071.

\end{acknowledgements}

\bibliography{references}

\begin{thebibliography}{165}%
\makeatletter
\providecommand \@ifxundefined [1]{%
 \@ifx{#1\undefined}
}%
\providecommand \@ifnum [1]{%
 \ifnum #1\expandafter \@firstoftwo
 \else \expandafter \@secondoftwo
 \fi
}%
\providecommand \@ifx [1]{%
 \ifx #1\expandafter \@firstoftwo
 \else \expandafter \@secondoftwo
 \fi
}%
\providecommand \natexlab [1]{#1}%
\providecommand \enquote  [1]{``#1''}%
\providecommand \bibnamefont  [1]{#1}%
\providecommand \bibfnamefont [1]{#1}%
\providecommand \citenamefont [1]{#1}%
\providecommand \href@noop [0]{\@secondoftwo}%
\providecommand \href [0]{\begingroup \@sanitize@url \@href}%
\providecommand \@href[1]{\@@startlink{#1}\@@href}%
\providecommand \@@href[1]{\endgroup#1\@@endlink}%
\providecommand \@sanitize@url [0]{\catcode `\\12\catcode `\$12\catcode
  `\&12\catcode `\#12\catcode `\^12\catcode `\_12\catcode `\%12\relax}%
\providecommand \@@startlink[1]{}%
\providecommand \@@endlink[0]{}%
\providecommand \url  [0]{\begingroup\@sanitize@url \@url }%
\providecommand \@url [1]{\endgroup\@href {#1}{\urlprefix }}%
\providecommand \urlprefix  [0]{URL }%
\providecommand \Eprint [0]{\href }%
\providecommand \doibase [0]{https://doi.org/}%
\providecommand \selectlanguage [0]{\@gobble}%
\providecommand \bibinfo  [0]{\@secondoftwo}%
\providecommand \bibfield  [0]{\@secondoftwo}%
\providecommand \translation [1]{[#1]}%
\providecommand \BibitemOpen [0]{}%
\providecommand \bibitemStop [0]{}%
\providecommand \bibitemNoStop [0]{.\EOS\space}%
\providecommand \EOS [0]{\spacefactor3000\relax}%
\providecommand \BibitemShut  [1]{\csname bibitem#1\endcsname}%
\let\auto@bib@innerbib\@empty
\bibitem [{\citenamefont {Keimer}\ \emph {et~al.}(2015)\citenamefont {Keimer},
  \citenamefont {Kivelson}, \citenamefont {Norman}, \citenamefont {Uchida},\
  and\ \citenamefont {Zaanen}}]{Keimer15}%
  \BibitemOpen
  \bibfield  {author} {\bibinfo {author} {\bibfnamefont {B.}~\bibnamefont
  {Keimer}}, \bibinfo {author} {\bibfnamefont {S.~A.}\ \bibnamefont
  {Kivelson}}, \bibinfo {author} {\bibfnamefont {M.~R.}\ \bibnamefont
  {Norman}}, \bibinfo {author} {\bibfnamefont {S.}~\bibnamefont {Uchida}},\
  and\ \bibinfo {author} {\bibfnamefont {J.}~\bibnamefont {Zaanen}},\
  }\bibfield  {title} {\bibinfo {title} {{From quantum matter to
  high-temperature superconductivity in copper oxides}},\ }\href
  {https://doi.org/10.1038/nature14165} {\bibfield  {journal} {\bibinfo
  {journal} {Nature}\ }\textbf {\bibinfo {volume} {518}},\ \bibinfo {pages}
  {179} (\bibinfo {year} {2015})}\BibitemShut {NoStop}%
\bibitem [{\citenamefont {White}\ \emph {et~al.}(2015)\citenamefont {White},
  \citenamefont {Thompson},\ and\ \citenamefont {Maple}}]{Maple15}%
  \BibitemOpen
  \bibfield  {author} {\bibinfo {author} {\bibfnamefont {B.}~\bibnamefont
  {White}}, \bibinfo {author} {\bibfnamefont {J.}~\bibnamefont {Thompson}},\
  and\ \bibinfo {author} {\bibfnamefont {M.}~\bibnamefont {Maple}},\ }\bibfield
   {title} {\bibinfo {title} {Unconventional superconductivity in heavy-fermion
  compounds},\ }\href
  {https://doi.org/https://doi.org/10.1016/j.physc.2015.02.044} {\bibfield
  {journal} {\bibinfo  {journal} {Physica C: Superconductivity and its
  Applications}\ }\textbf {\bibinfo {volume} {514}},\ \bibinfo {pages} {246}
  (\bibinfo {year} {2015})}\BibitemShut {NoStop}%
\bibitem [{\citenamefont {Singleton}\ and\ \citenamefont
  {Mielke}(2002)}]{Singleton02}%
  \BibitemOpen
  \bibfield  {author} {\bibinfo {author} {\bibfnamefont {J.}~\bibnamefont
  {Singleton}}\ and\ \bibinfo {author} {\bibfnamefont {C.}~\bibnamefont
  {Mielke}},\ }\bibfield  {title} {\bibinfo {title} {{Quasi-two-dimensional
  organic superconductors: A review}},\ }\href
  {https://doi.org/10.1080/00107510110108681} {\bibfield  {journal} {\bibinfo
  {journal} {Contemporary Physics}\ }\textbf {\bibinfo {volume} {43}},\
  \bibinfo {pages} {63} (\bibinfo {year} {2002})}\BibitemShut {NoStop}%
\bibitem [{\citenamefont {Kamihara}\ \emph {et~al.}(2008)\citenamefont
  {Kamihara}, \citenamefont {Watanabe}, \citenamefont {Hirano},\ and\
  \citenamefont {Hosono}}]{Kamihara08}%
  \BibitemOpen
  \bibfield  {author} {\bibinfo {author} {\bibfnamefont {Y.}~\bibnamefont
  {Kamihara}}, \bibinfo {author} {\bibfnamefont {T.}~\bibnamefont {Watanabe}},
  \bibinfo {author} {\bibfnamefont {M.}~\bibnamefont {Hirano}},\ and\ \bibinfo
  {author} {\bibfnamefont {H.}~\bibnamefont {Hosono}},\ }\bibfield  {title}
  {\bibinfo {title} {{Iron-Based Layered Superconductor LaO$_{1-x}$F$_x$FeAs
  ($x = 0.05-0.12$) with $T_c = 26$ K}},\ }\href
  {https://doi.org/10.1021/ja800073m} {\bibfield  {journal} {\bibinfo
  {journal} {Journal of the American Chemical Society}\ }\textbf {\bibinfo
  {volume} {130}},\ \bibinfo {pages} {3296} (\bibinfo {year}
  {2008})}\BibitemShut {NoStop}%
\bibitem [{\citenamefont {Johnston}(2010)}]{Johnston10}%
  \BibitemOpen
  \bibfield  {author} {\bibinfo {author} {\bibfnamefont {D.~C.}\ \bibnamefont
  {Johnston}},\ }\bibfield  {title} {\bibinfo {title} {{The puzzle of high
  temperature superconductivity in layered iron pnictides and chalcogenides}},\
  }\href {https://doi.org/10.1080/00018732.2010.513480} {\bibfield  {journal}
  {\bibinfo  {journal} {Advances in Physics}\ }\textbf {\bibinfo {volume}
  {59}},\ \bibinfo {pages} {803} (\bibinfo {year} {2010})}\BibitemShut
  {NoStop}%
\bibitem [{\citenamefont {Hirschfeld}\ \emph {et~al.}(2011)\citenamefont
  {Hirschfeld}, \citenamefont {Korshunov},\ and\ \citenamefont
  {Mazin}}]{Mazin_review}%
  \BibitemOpen
  \bibfield  {author} {\bibinfo {author} {\bibfnamefont {P.~J.}\ \bibnamefont
  {Hirschfeld}}, \bibinfo {author} {\bibfnamefont {M.~M.}\ \bibnamefont
  {Korshunov}},\ and\ \bibinfo {author} {\bibfnamefont {I.~I.}\ \bibnamefont
  {Mazin}},\ }\bibfield  {title} {\bibinfo {title} {{Gap symmetry and structure
  of Fe-based superconductors}},\ }\href
  {https://doi.org/10.1088/0034-4885/74/12/124508} {\bibfield  {journal}
  {\bibinfo  {journal} {Reports on Progress in Physics}\ }\textbf {\bibinfo
  {volume} {74}},\ \bibinfo {pages} {124508} (\bibinfo {year}
  {2011})}\BibitemShut {NoStop}%
\bibitem [{\citenamefont {Chubukov}(2012)}]{Chubukov_review}%
  \BibitemOpen
  \bibfield  {author} {\bibinfo {author} {\bibfnamefont {A.~V.}\ \bibnamefont
  {Chubukov}},\ }\bibfield  {title} {\bibinfo {title} {{Pairing Mechanism in
  Fe-Based Superconductors}},\ }\href
  {https://doi.org/10.1146/annurev-conmatphys-020911-125055} {\bibfield
  {journal} {\bibinfo  {journal} {Annual Review of Condensed Matter Physics}\
  }\textbf {\bibinfo {volume} {3}},\ \bibinfo {pages} {57} (\bibinfo {year}
  {2012})}\BibitemShut {NoStop}%
\bibitem [{\citenamefont {Wang}\ and\ \citenamefont
  {Lee}(2011)}]{DHLee_review}%
  \BibitemOpen
  \bibfield  {author} {\bibinfo {author} {\bibfnamefont {F.}~\bibnamefont
  {Wang}}\ and\ \bibinfo {author} {\bibfnamefont {D.-H.}\ \bibnamefont {Lee}},\
  }\bibfield  {title} {\bibinfo {title} {{The Electron-Pairing Mechanism of
  Iron-Based Superconductors}},\ }\href
  {https://doi.org/10.1126/science.1200182} {\bibfield  {journal} {\bibinfo
  {journal} {Science}\ }\textbf {\bibinfo {volume} {332}},\ \bibinfo {pages}
  {200} (\bibinfo {year} {2011})}\BibitemShut {NoStop}%
\bibitem [{\citenamefont {Mazin}\ \emph {et~al.}(2008)\citenamefont {Mazin},
  \citenamefont {Singh}, \citenamefont {Johannes},\ and\ \citenamefont
  {Du}}]{Mazin08}%
  \BibitemOpen
  \bibfield  {author} {\bibinfo {author} {\bibfnamefont {I.~I.}\ \bibnamefont
  {Mazin}}, \bibinfo {author} {\bibfnamefont {D.~J.}\ \bibnamefont {Singh}},
  \bibinfo {author} {\bibfnamefont {M.~D.}\ \bibnamefont {Johannes}},\ and\
  \bibinfo {author} {\bibfnamefont {M.~H.}\ \bibnamefont {Du}},\ }\bibfield
  {title} {\bibinfo {title} {{Unconventional Superconductivity with a Sign
  Reversal in the Order Parameter of
  ${\mathrm{LaFeAsO}}_{1-x}{\mathrm{F}}_{x}$}},\ }\href
  {https://doi.org/10.1103/PhysRevLett.101.057003} {\bibfield  {journal}
  {\bibinfo  {journal} {Phys. Rev. Lett.}\ }\textbf {\bibinfo {volume} {101}},\
  \bibinfo {pages} {057003} (\bibinfo {year} {2008})}\BibitemShut {NoStop}%
\bibitem [{\citenamefont {Haule}\ and\ \citenamefont
  {Kotliar}(2009)}]{Haule09}%
  \BibitemOpen
  \bibfield  {author} {\bibinfo {author} {\bibfnamefont {K.}~\bibnamefont
  {Haule}}\ and\ \bibinfo {author} {\bibfnamefont {G.}~\bibnamefont
  {Kotliar}},\ }\bibfield  {title} {\bibinfo {title}
  {{Coherence{\textendash}incoherence crossover in the normal state of iron
  oxypnictides and importance of Hund{\textquotesingle}s rule coupling}},\
  }\href {https://doi.org/10.1088/1367-2630/11/2/025021} {\bibfield  {journal}
  {\bibinfo  {journal} {New Journal of Physics}\ }\textbf {\bibinfo {volume}
  {11}},\ \bibinfo {pages} {025021} (\bibinfo {year} {2009})}\BibitemShut
  {NoStop}%
\bibitem [{\citenamefont {Georges}\ \emph {et~al.}(2013)\citenamefont
  {Georges}, \citenamefont {Medici},\ and\ \citenamefont
  {Mravlje}}]{Georges13}%
  \BibitemOpen
  \bibfield  {author} {\bibinfo {author} {\bibfnamefont {A.}~\bibnamefont
  {Georges}}, \bibinfo {author} {\bibfnamefont {L.~d.}\ \bibnamefont
  {Medici}},\ and\ \bibinfo {author} {\bibfnamefont {J.}~\bibnamefont
  {Mravlje}},\ }\bibfield  {title} {\bibinfo {title} {{Strong Correlations from
  Hund's Coupling}},\ }\href
  {https://doi.org/10.1146/annurev-conmatphys-020911-125045} {\bibfield
  {journal} {\bibinfo  {journal} {Annual Review of Condensed Matter Physics}\
  }\textbf {\bibinfo {volume} {4}},\ \bibinfo {pages} {137} (\bibinfo {year}
  {2013})}\BibitemShut {NoStop}%
\bibitem [{\citenamefont {Yin}\ \emph {et~al.}(2011)\citenamefont {Yin},
  \citenamefont {Haule},\ and\ \citenamefont {Kotliar}}]{Kotliar11}%
  \BibitemOpen
  \bibfield  {author} {\bibinfo {author} {\bibfnamefont {Z.}~\bibnamefont
  {Yin}}, \bibinfo {author} {\bibfnamefont {K.}~\bibnamefont {Haule}},\ and\
  \bibinfo {author} {\bibfnamefont {G.}~\bibnamefont {Kotliar}},\ }\bibfield
  {title} {\bibinfo {title} {Kinetic frustration and the nature of the magnetic
  and paramagnetic states in iron pnictides and iron chalcogenides},\ }\href
  {https://doi.org/10.1038/nmat3120} {\bibfield  {journal} {\bibinfo  {journal}
  {Nature materials}\ }\textbf {\bibinfo {volume} {10}},\ \bibinfo {pages}
  {932} (\bibinfo {year} {2011})}\BibitemShut {NoStop}%
\bibitem [{\citenamefont {de' Medici}(2015)}]{Medici15}%
  \BibitemOpen
  \bibfield  {author} {\bibinfo {author} {\bibfnamefont {L.}~\bibnamefont {de'
  Medici}},\ }\bibinfo {title} {Weak and strong correlations in {Fe}
  superconductors},\ in\ \href {https://doi.org/10.1007/978-3-319-11254-1_11}
  {\emph {\bibinfo {booktitle} {Iron-Based Superconductivity}}},\ \bibinfo
  {editor} {edited by\ \bibinfo {editor} {\bibfnamefont {P.~D.}\ \bibnamefont
  {Johnson}}, \bibinfo {editor} {\bibfnamefont {G.}~\bibnamefont {Xu}},\ and\
  \bibinfo {editor} {\bibfnamefont {W.-G.}\ \bibnamefont {Yin}}}\ (\bibinfo
  {publisher} {Springer International Publishing},\ \bibinfo {address} {Cham},\
  \bibinfo {year} {2015})\ pp.\ \bibinfo {pages} {409--441}\BibitemShut
  {NoStop}%
\bibitem [{\citenamefont {Sprau}\ \emph {et~al.}(2017)\citenamefont {Sprau},
  \citenamefont {Kostin}, \citenamefont {Kreisel}, \citenamefont {B{\"o}hmer},
  \citenamefont {Taufour}, \citenamefont {Canfield}, \citenamefont {Mukherjee},
  \citenamefont {Hirschfeld}, \citenamefont {Andersen},\ and\ \citenamefont
  {Davis}}]{Sprau17}%
  \BibitemOpen
  \bibfield  {author} {\bibinfo {author} {\bibfnamefont {P.~O.}\ \bibnamefont
  {Sprau}}, \bibinfo {author} {\bibfnamefont {A.}~\bibnamefont {Kostin}},
  \bibinfo {author} {\bibfnamefont {A.}~\bibnamefont {Kreisel}}, \bibinfo
  {author} {\bibfnamefont {A.~E.}\ \bibnamefont {B{\"o}hmer}}, \bibinfo
  {author} {\bibfnamefont {V.}~\bibnamefont {Taufour}}, \bibinfo {author}
  {\bibfnamefont {P.~C.}\ \bibnamefont {Canfield}}, \bibinfo {author}
  {\bibfnamefont {S.}~\bibnamefont {Mukherjee}}, \bibinfo {author}
  {\bibfnamefont {P.~J.}\ \bibnamefont {Hirschfeld}}, \bibinfo {author}
  {\bibfnamefont {B.~M.}\ \bibnamefont {Andersen}},\ and\ \bibinfo {author}
  {\bibfnamefont {J.~C.~S.}\ \bibnamefont {Davis}},\ }\bibfield  {title}
  {\bibinfo {title} {{Discovery of orbital-selective Cooper pairing in FeSe}},\
  }\href {https://doi.org/10.1126/science.aal1575} {\bibfield  {journal}
  {\bibinfo  {journal} {Science}\ }\textbf {\bibinfo {volume} {357}},\ \bibinfo
  {pages} {75} (\bibinfo {year} {2017})}\BibitemShut {NoStop}%
\bibitem [{\citenamefont {Shibauchi}\ \emph {et~al.}(2014)\citenamefont
  {Shibauchi}, \citenamefont {Carrington},\ and\ \citenamefont
  {Matsuda}}]{Shibauchi14}%
  \BibitemOpen
  \bibfield  {author} {\bibinfo {author} {\bibfnamefont {T.}~\bibnamefont
  {Shibauchi}}, \bibinfo {author} {\bibfnamefont {A.}~\bibnamefont
  {Carrington}},\ and\ \bibinfo {author} {\bibfnamefont {Y.}~\bibnamefont
  {Matsuda}},\ }\bibfield  {title} {\bibinfo {title} {A quantum critical point
  lying beneath the superconducting dome in iron pnictides},\ }\href
  {https://doi.org/10.1146/annurev-conmatphys-031113-133921} {\bibfield
  {journal} {\bibinfo  {journal} {Annual Review of Condensed Matter Physics}\
  }\textbf {\bibinfo {volume} {5}},\ \bibinfo {pages} {113} (\bibinfo {year}
  {2014})}\BibitemShut {NoStop}%
\bibitem [{\citenamefont {Coldea}\ and\ \citenamefont
  {Watson}(2018)}]{Coldea18}%
  \BibitemOpen
  \bibfield  {author} {\bibinfo {author} {\bibfnamefont {A.~I.}\ \bibnamefont
  {Coldea}}\ and\ \bibinfo {author} {\bibfnamefont {M.~D.}\ \bibnamefont
  {Watson}},\ }\bibfield  {title} {\bibinfo {title} {{The Key Ingredients of
  the Electronic Structure of FeSe}},\ }\href
  {https://doi.org/10.1146/annurev-conmatphys-033117-054137} {\bibfield
  {journal} {\bibinfo  {journal} {Annual Review of Condensed Matter Physics}\
  }\textbf {\bibinfo {volume} {9}},\ \bibinfo {pages} {125} (\bibinfo {year}
  {2018})}\BibitemShut {NoStop}%
\bibitem [{\citenamefont {Kreisel}\ \emph {et~al.}(2020)\citenamefont
  {Kreisel}, \citenamefont {Hirschfeld},\ and\ \citenamefont
  {Andersen}}]{Kreisel20}%
  \BibitemOpen
  \bibfield  {author} {\bibinfo {author} {\bibfnamefont {A.}~\bibnamefont
  {Kreisel}}, \bibinfo {author} {\bibfnamefont {P.~J.}\ \bibnamefont
  {Hirschfeld}},\ and\ \bibinfo {author} {\bibfnamefont {B.~M.}\ \bibnamefont
  {Andersen}},\ }\bibfield  {title} {\bibinfo {title} {{On the Remarkable
  Superconductivity of FeSe and Its Close Cousins}},\ }\bibfield  {journal}
  {\bibinfo  {journal} {Symmetry}\ }\textbf {\bibinfo {volume} {12}},\ \href
  {https://doi.org/10.3390/sym12091402} {10.3390/sym12091402} (\bibinfo {year}
  {2020})\BibitemShut {NoStop}%
\bibitem [{\citenamefont {Dai}(2015)}]{Dai15}%
  \BibitemOpen
  \bibfield  {author} {\bibinfo {author} {\bibfnamefont {P.}~\bibnamefont
  {Dai}},\ }\bibfield  {title} {\bibinfo {title} {Antiferromagnetic order and
  spin dynamics in iron-based superconductors},\ }\href
  {https://doi.org/10.1103/RevModPhys.87.855} {\bibfield  {journal} {\bibinfo
  {journal} {Rev. Mod. Phys.}\ }\textbf {\bibinfo {volume} {87}},\ \bibinfo
  {pages} {855} (\bibinfo {year} {2015})}\BibitemShut {NoStop}%
\bibitem [{\citenamefont {Lumsden}\ and\ \citenamefont
  {Christianson}(2010)}]{Lumsden10}%
  \BibitemOpen
  \bibfield  {author} {\bibinfo {author} {\bibfnamefont {M.~D.}\ \bibnamefont
  {Lumsden}}\ and\ \bibinfo {author} {\bibfnamefont {A.~D.}\ \bibnamefont
  {Christianson}},\ }\bibfield  {title} {\bibinfo {title} {{Magnetism in
  Fe-based superconductors}},\ }\href
  {https://doi.org/10.1088/0953-8984/22/20/203203} {\bibfield  {journal}
  {\bibinfo  {journal} {Journal of Physics: Condensed Matter}\ }\textbf
  {\bibinfo {volume} {22}},\ \bibinfo {pages} {203203} (\bibinfo {year}
  {2010})}\BibitemShut {NoStop}%
\bibitem [{\citenamefont {Inosov}(2016)}]{Inosov2016}%
  \BibitemOpen
  \bibfield  {author} {\bibinfo {author} {\bibfnamefont {D.~S.}\ \bibnamefont
  {Inosov}},\ }\bibfield  {title} {\bibinfo {title} {Spin fluctuations in iron
  pnictides and chalcogenides: From antiferromagnetism to superconductivity},\
  }\href {https://doi.org/https://doi.org/10.1016/j.crhy.2015.03.001}
  {\bibfield  {journal} {\bibinfo  {journal} {Comptes Rendus Physique}\
  }\textbf {\bibinfo {volume} {17}},\ \bibinfo {pages} {60} (\bibinfo {year}
  {2016})}\BibitemShut {NoStop}%
\bibitem [{\citenamefont {Scalapino}(2012)}]{Scalapino_review}%
  \BibitemOpen
  \bibfield  {author} {\bibinfo {author} {\bibfnamefont {D.~J.}\ \bibnamefont
  {Scalapino}},\ }\bibfield  {title} {\bibinfo {title} {A common thread: The
  pairing interaction for unconventional superconductors},\ }\href
  {https://doi.org/10.1103/RevModPhys.84.1383} {\bibfield  {journal} {\bibinfo
  {journal} {Rev. Mod. Phys.}\ }\textbf {\bibinfo {volume} {84}},\ \bibinfo
  {pages} {1383} (\bibinfo {year} {2012})}\BibitemShut {NoStop}%
\bibitem [{\citenamefont {Fernandes}\ \emph {et~al.}(2014)\citenamefont
  {Fernandes}, \citenamefont {Chubukov},\ and\ \citenamefont
  {Schmalian}}]{Fernandes14}%
  \BibitemOpen
  \bibfield  {author} {\bibinfo {author} {\bibfnamefont {R.~M.}\ \bibnamefont
  {Fernandes}}, \bibinfo {author} {\bibfnamefont {A.~V.}\ \bibnamefont
  {Chubukov}},\ and\ \bibinfo {author} {\bibfnamefont {J.}~\bibnamefont
  {Schmalian}},\ }\bibfield  {title} {\bibinfo {title} {What drives nematic
  order in iron-based superconductors?},\ }\href
  {https://doi.org/10.1038/nphys2877} {\bibfield  {journal} {\bibinfo
  {journal} {Nature Physics}\ }\textbf {\bibinfo {volume} {10}},\ \bibinfo
  {pages} {97} (\bibinfo {year} {2014})}\BibitemShut {NoStop}%
\bibitem [{\citenamefont {Fradkin}\ \emph {et~al.}(2010)\citenamefont
  {Fradkin}, \citenamefont {Kivelson}, \citenamefont {Lawler}, \citenamefont
  {Eisenstein},\ and\ \citenamefont {Mackenzie}}]{Fradkin10}%
  \BibitemOpen
  \bibfield  {author} {\bibinfo {author} {\bibfnamefont {E.}~\bibnamefont
  {Fradkin}}, \bibinfo {author} {\bibfnamefont {S.~A.}\ \bibnamefont
  {Kivelson}}, \bibinfo {author} {\bibfnamefont {M.~J.}\ \bibnamefont
  {Lawler}}, \bibinfo {author} {\bibfnamefont {J.~P.}\ \bibnamefont
  {Eisenstein}},\ and\ \bibinfo {author} {\bibfnamefont {A.~P.}\ \bibnamefont
  {Mackenzie}},\ }\bibfield  {title} {\bibinfo {title} {{Nematic Fermi Fluids
  in Condensed Matter Physics}},\ }\href
  {https://doi.org/10.1146/annurev-conmatphys-070909-103925} {\bibfield
  {journal} {\bibinfo  {journal} {Annual Review of Condensed Matter Physics}\
  }\textbf {\bibinfo {volume} {1}},\ \bibinfo {pages} {153} (\bibinfo {year}
  {2010})}\BibitemShut {NoStop}%
\bibitem [{\citenamefont {Chu}\ \emph {et~al.}(2012)\citenamefont {Chu},
  \citenamefont {Kuo}, \citenamefont {Analytis},\ and\ \citenamefont
  {Fisher}}]{Chu12}%
  \BibitemOpen
  \bibfield  {author} {\bibinfo {author} {\bibfnamefont {J.-H.}\ \bibnamefont
  {Chu}}, \bibinfo {author} {\bibfnamefont {H.-H.}\ \bibnamefont {Kuo}},
  \bibinfo {author} {\bibfnamefont {J.~G.}\ \bibnamefont {Analytis}},\ and\
  \bibinfo {author} {\bibfnamefont {I.~R.}\ \bibnamefont {Fisher}},\ }\bibfield
   {title} {\bibinfo {title} {{Divergent nematic susceptibility in an iron
  arsenide superconductor}},\ }\href {https://doi.org/10.1126/science.1221713}
  {\bibfield  {journal} {\bibinfo  {journal} {Science}\ }\textbf {\bibinfo
  {volume} {337}},\ \bibinfo {pages} {710} (\bibinfo {year}
  {2012})}\BibitemShut {NoStop}%
\bibitem [{\citenamefont {Bohmer}\ and\ \citenamefont
  {Meingast}(2016)}]{Bohmer16}%
  \BibitemOpen
  \bibfield  {author} {\bibinfo {author} {\bibfnamefont {A.~E.}\ \bibnamefont
  {Bohmer}}\ and\ \bibinfo {author} {\bibfnamefont {C.}~\bibnamefont
  {Meingast}},\ }\bibfield  {title} {\bibinfo {title} {Electronic nematic
  susceptibility of iron-based superconductors},\ }\href
  {https://doi.org/10.1016/j.crhy.2015.07.001} {\bibfield  {journal} {\bibinfo
  {journal} {Comptes Rendus Physique}\ }\textbf {\bibinfo {volume} {17}},\
  \bibinfo {pages} {90} (\bibinfo {year} {2016})}\BibitemShut {NoStop}%
\bibitem [{\citenamefont {Gallais}\ and\ \citenamefont
  {Paul}(2016)}]{Gallais_review}%
  \BibitemOpen
  \bibfield  {author} {\bibinfo {author} {\bibfnamefont {Y.}~\bibnamefont
  {Gallais}}\ and\ \bibinfo {author} {\bibfnamefont {I.}~\bibnamefont {Paul}},\
  }\bibfield  {title} {\bibinfo {title} {Charge nematicity and electronic raman
  scattering in iron-based superconductors},\ }\href
  {https://doi.org/http://dx.doi.org/10.1016/j.crhy.2015.10.001} {\bibfield
  {journal} {\bibinfo  {journal} {Comptes Rendus Physique}\ }\textbf {\bibinfo
  {volume} {17}},\ \bibinfo {pages} {113} (\bibinfo {year} {2016})}\BibitemShut
  {NoStop}%
\bibitem [{\citenamefont {Zhang}\ \emph {et~al.}(2018)\citenamefont {Zhang},
  \citenamefont {Yaji}, \citenamefont {Hashimoto}, \citenamefont {Ota},
  \citenamefont {Kondo}, \citenamefont {Okazaki}, \citenamefont {Wang},
  \citenamefont {Wen}, \citenamefont {Gu}, \citenamefont {Ding},\ and\
  \citenamefont {Shin}}]{Zhang18}%
  \BibitemOpen
  \bibfield  {author} {\bibinfo {author} {\bibfnamefont {P.}~\bibnamefont
  {Zhang}}, \bibinfo {author} {\bibfnamefont {K.}~\bibnamefont {Yaji}},
  \bibinfo {author} {\bibfnamefont {T.}~\bibnamefont {Hashimoto}}, \bibinfo
  {author} {\bibfnamefont {Y.}~\bibnamefont {Ota}}, \bibinfo {author}
  {\bibfnamefont {T.}~\bibnamefont {Kondo}}, \bibinfo {author} {\bibfnamefont
  {K.}~\bibnamefont {Okazaki}}, \bibinfo {author} {\bibfnamefont
  {Z.}~\bibnamefont {Wang}}, \bibinfo {author} {\bibfnamefont {J.}~\bibnamefont
  {Wen}}, \bibinfo {author} {\bibfnamefont {G.~D.}\ \bibnamefont {Gu}},
  \bibinfo {author} {\bibfnamefont {H.}~\bibnamefont {Ding}},\ and\ \bibinfo
  {author} {\bibfnamefont {S.}~\bibnamefont {Shin}},\ }\bibfield  {title}
  {\bibinfo {title} {Observation of topological superconductivity on the
  surface of an iron-based superconductor},\ }\href
  {https://doi.org/10.1126/science.aan4596} {\bibfield  {journal} {\bibinfo
  {journal} {Science}\ }\textbf {\bibinfo {volume} {360}},\ \bibinfo {pages}
  {182} (\bibinfo {year} {2018})}\BibitemShut {NoStop}%
\bibitem [{\citenamefont {Singh}\ and\ \citenamefont {Du}(2008)}]{Singh08}%
  \BibitemOpen
  \bibfield  {author} {\bibinfo {author} {\bibfnamefont {D.~J.}\ \bibnamefont
  {Singh}}\ and\ \bibinfo {author} {\bibfnamefont {M.-H.}\ \bibnamefont {Du}},\
  }\bibfield  {title} {\bibinfo {title} {{Density Functional Study of
  ${\mathrm{LaFeAsO}}_{1\ensuremath{-}x}{\mathrm{F}}_{x}$: A Low Carrier
  Density Superconductor Near Itinerant Magnetism}},\ }\href
  {https://doi.org/10.1103/PhysRevLett.100.237003} {\bibfield  {journal}
  {\bibinfo  {journal} {Phys. Rev. Lett.}\ }\textbf {\bibinfo {volume} {100}},\
  \bibinfo {pages} {237003} (\bibinfo {year} {2008})}\BibitemShut {NoStop}%
\bibitem [{\citenamefont {Cvetkovic}\ and\ \citenamefont
  {Vafek}(2013)}]{Vafek13}%
  \BibitemOpen
  \bibfield  {author} {\bibinfo {author} {\bibfnamefont {V.}~\bibnamefont
  {Cvetkovic}}\ and\ \bibinfo {author} {\bibfnamefont {O.}~\bibnamefont
  {Vafek}},\ }\bibfield  {title} {\bibinfo {title} {{Space group symmetry,
  spin-orbit coupling, and the low-energy effective Hamiltonian for iron-based
  superconductors}},\ }\href {https://doi.org/10.1103/PhysRevB.88.134510}
  {\bibfield  {journal} {\bibinfo  {journal} {Phys. Rev. B}\ }\textbf {\bibinfo
  {volume} {88}},\ \bibinfo {pages} {134510} (\bibinfo {year}
  {2013})}\BibitemShut {NoStop}%
\bibitem [{\citenamefont {Eschrig}\ and\ \citenamefont
  {Koepernik}(2009)}]{Eschrig09}%
  \BibitemOpen
  \bibfield  {author} {\bibinfo {author} {\bibfnamefont {H.}~\bibnamefont
  {Eschrig}}\ and\ \bibinfo {author} {\bibfnamefont {K.}~\bibnamefont
  {Koepernik}},\ }\bibfield  {title} {\bibinfo {title} {Tight-binding models
  for the iron-based superconductors},\ }\href
  {https://doi.org/10.1103/PhysRevB.80.104503} {\bibfield  {journal} {\bibinfo
  {journal} {Phys. Rev. B}\ }\textbf {\bibinfo {volume} {80}},\ \bibinfo
  {pages} {104503} (\bibinfo {year} {2009})}\BibitemShut {NoStop}%
\bibitem [{\citenamefont {Borisenko}\ \emph {et~al.}(2016)\citenamefont
  {Borisenko}, \citenamefont {Evtushinsky}, \citenamefont {Liu}, \citenamefont
  {Morozov}, \citenamefont {Kappenberger}, \citenamefont {Wurmehl},
  \citenamefont {B{\"u}chner}, \citenamefont {Yaresko}, \citenamefont {Kim},
  \citenamefont {Hoesch} \emph {et~al.}}]{Borisenko16}%
  \BibitemOpen
  \bibfield  {author} {\bibinfo {author} {\bibfnamefont {S.}~\bibnamefont
  {Borisenko}}, \bibinfo {author} {\bibfnamefont {D.}~\bibnamefont
  {Evtushinsky}}, \bibinfo {author} {\bibfnamefont {Z.-H.}\ \bibnamefont
  {Liu}}, \bibinfo {author} {\bibfnamefont {I.}~\bibnamefont {Morozov}},
  \bibinfo {author} {\bibfnamefont {R.}~\bibnamefont {Kappenberger}}, \bibinfo
  {author} {\bibfnamefont {S.}~\bibnamefont {Wurmehl}}, \bibinfo {author}
  {\bibfnamefont {B.}~\bibnamefont {B{\"u}chner}}, \bibinfo {author}
  {\bibfnamefont {A.}~\bibnamefont {Yaresko}}, \bibinfo {author} {\bibfnamefont
  {T.}~\bibnamefont {Kim}}, \bibinfo {author} {\bibfnamefont {M.}~\bibnamefont
  {Hoesch}}, \emph {et~al.},\ }\bibfield  {title} {\bibinfo {title} {{Direct
  observation of spin-orbit coupling in iron-based superconductors}},\ }\href
  {https://doi.org/10.1038/nphys3594} {\bibfield  {journal} {\bibinfo
  {journal} {Nature Physics}\ }\textbf {\bibinfo {volume} {12}},\ \bibinfo
  {pages} {311} (\bibinfo {year} {2016})}\BibitemShut {NoStop}%
\bibitem [{\citenamefont {Wang}\ \emph
  {et~al.}(2015{\natexlab{a}})\citenamefont {Wang}, \citenamefont {Zhang},
  \citenamefont {Xu}, \citenamefont {Zeng}, \citenamefont {Miao}, \citenamefont
  {Xu}, \citenamefont {Qian}, \citenamefont {Weng}, \citenamefont {Richard},
  \citenamefont {Fedorov}, \citenamefont {Ding}, \citenamefont {Dai},\ and\
  \citenamefont {Fang}}]{Fang15}%
  \BibitemOpen
  \bibfield  {author} {\bibinfo {author} {\bibfnamefont {Z.}~\bibnamefont
  {Wang}}, \bibinfo {author} {\bibfnamefont {P.}~\bibnamefont {Zhang}},
  \bibinfo {author} {\bibfnamefont {G.}~\bibnamefont {Xu}}, \bibinfo {author}
  {\bibfnamefont {L.~K.}\ \bibnamefont {Zeng}}, \bibinfo {author}
  {\bibfnamefont {H.}~\bibnamefont {Miao}}, \bibinfo {author} {\bibfnamefont
  {X.}~\bibnamefont {Xu}}, \bibinfo {author} {\bibfnamefont {T.}~\bibnamefont
  {Qian}}, \bibinfo {author} {\bibfnamefont {H.}~\bibnamefont {Weng}}, \bibinfo
  {author} {\bibfnamefont {P.}~\bibnamefont {Richard}}, \bibinfo {author}
  {\bibfnamefont {A.~V.}\ \bibnamefont {Fedorov}}, \bibinfo {author}
  {\bibfnamefont {H.}~\bibnamefont {Ding}}, \bibinfo {author} {\bibfnamefont
  {X.}~\bibnamefont {Dai}},\ and\ \bibinfo {author} {\bibfnamefont
  {Z.}~\bibnamefont {Fang}},\ }\bibfield  {title} {\bibinfo {title}
  {{Topological nature of the ${\mathrm{FeSe}}_{0.5}{\mathrm{Te}}_{0.5}$
  superconductor}},\ }\href {https://doi.org/10.1103/PhysRevB.92.115119}
  {\bibfield  {journal} {\bibinfo  {journal} {Phys. Rev. B}\ }\textbf {\bibinfo
  {volume} {92}},\ \bibinfo {pages} {115119} (\bibinfo {year}
  {2015}{\natexlab{a}})}\BibitemShut {NoStop}%
\bibitem [{\citenamefont {Xu}\ \emph {et~al.}(2016)\citenamefont {Xu},
  \citenamefont {Lian}, \citenamefont {Tang}, \citenamefont {Qi},\ and\
  \citenamefont {Zhang}}]{SCZhang16}%
  \BibitemOpen
  \bibfield  {author} {\bibinfo {author} {\bibfnamefont {G.}~\bibnamefont
  {Xu}}, \bibinfo {author} {\bibfnamefont {B.}~\bibnamefont {Lian}}, \bibinfo
  {author} {\bibfnamefont {P.}~\bibnamefont {Tang}}, \bibinfo {author}
  {\bibfnamefont {X.-L.}\ \bibnamefont {Qi}},\ and\ \bibinfo {author}
  {\bibfnamefont {S.-C.}\ \bibnamefont {Zhang}},\ }\bibfield  {title} {\bibinfo
  {title} {{Topological Superconductivity on the Surface of Fe-Based
  Superconductors}},\ }\href {https://doi.org/10.1103/PhysRevLett.117.047001}
  {\bibfield  {journal} {\bibinfo  {journal} {Phys. Rev. Lett.}\ }\textbf
  {\bibinfo {volume} {117}},\ \bibinfo {pages} {047001} (\bibinfo {year}
  {2016})}\BibitemShut {NoStop}%
\bibitem [{\citenamefont {Yang}\ \emph {et~al.}(2009)\citenamefont {Yang},
  \citenamefont {Sorini}, \citenamefont {Chen}, \citenamefont {Moritz},
  \citenamefont {Lee}, \citenamefont {Vernay}, \citenamefont {Olalde-Velasco},
  \citenamefont {Denlinger}, \citenamefont {Delley}, \citenamefont {Chu},
  \citenamefont {Analytis}, \citenamefont {Fisher}, \citenamefont {Ren},
  \citenamefont {Yang}, \citenamefont {Lu}, \citenamefont {Zhao}, \citenamefont
  {van~den Brink}, \citenamefont {Hussain}, \citenamefont {Shen},\ and\
  \citenamefont {Devereaux}}]{XR1}%
  \BibitemOpen
  \bibfield  {author} {\bibinfo {author} {\bibfnamefont {W.~L.}\ \bibnamefont
  {Yang}}, \bibinfo {author} {\bibfnamefont {A.~P.}\ \bibnamefont {Sorini}},
  \bibinfo {author} {\bibfnamefont {C.-C.}\ \bibnamefont {Chen}}, \bibinfo
  {author} {\bibfnamefont {B.}~\bibnamefont {Moritz}}, \bibinfo {author}
  {\bibfnamefont {W.-S.}\ \bibnamefont {Lee}}, \bibinfo {author} {\bibfnamefont
  {F.}~\bibnamefont {Vernay}}, \bibinfo {author} {\bibfnamefont
  {P.}~\bibnamefont {Olalde-Velasco}}, \bibinfo {author} {\bibfnamefont
  {J.~D.}\ \bibnamefont {Denlinger}}, \bibinfo {author} {\bibfnamefont
  {B.}~\bibnamefont {Delley}}, \bibinfo {author} {\bibfnamefont {J.-H.}\
  \bibnamefont {Chu}}, \bibinfo {author} {\bibfnamefont {J.~G.}\ \bibnamefont
  {Analytis}}, \bibinfo {author} {\bibfnamefont {I.~R.}\ \bibnamefont
  {Fisher}}, \bibinfo {author} {\bibfnamefont {Z.~A.}\ \bibnamefont {Ren}},
  \bibinfo {author} {\bibfnamefont {J.}~\bibnamefont {Yang}}, \bibinfo {author}
  {\bibfnamefont {W.}~\bibnamefont {Lu}}, \bibinfo {author} {\bibfnamefont
  {Z.~X.}\ \bibnamefont {Zhao}}, \bibinfo {author} {\bibfnamefont
  {J.}~\bibnamefont {van~den Brink}}, \bibinfo {author} {\bibfnamefont
  {Z.}~\bibnamefont {Hussain}}, \bibinfo {author} {\bibfnamefont {Z.-X.}\
  \bibnamefont {Shen}},\ and\ \bibinfo {author} {\bibfnamefont {T.~P.}\
  \bibnamefont {Devereaux}},\ }\bibfield  {title} {\bibinfo {title} {Evidence
  for weak electronic correlations in iron pnictides},\ }\href
  {https://doi.org/10.1103/PhysRevB.80.014508} {\bibfield  {journal} {\bibinfo
  {journal} {Phys. Rev. B}\ }\textbf {\bibinfo {volume} {80}},\ \bibinfo
  {pages} {014508} (\bibinfo {year} {2009})}\BibitemShut {NoStop}%
\bibitem [{\citenamefont {Richard}\ \emph {et~al.}(2015)\citenamefont
  {Richard}, \citenamefont {Qian},\ and\ \citenamefont {Ding}}]{Richard15}%
  \BibitemOpen
  \bibfield  {author} {\bibinfo {author} {\bibfnamefont {P.}~\bibnamefont
  {Richard}}, \bibinfo {author} {\bibfnamefont {T.}~\bibnamefont {Qian}},\ and\
  \bibinfo {author} {\bibfnamefont {H.}~\bibnamefont {Ding}},\ }\bibfield
  {title} {\bibinfo {title} {{ARPES measurements of the superconducting gap of
  Fe-based superconductors and their implications to the pairing mechanism}},\
  }\href {https://doi.org/10.1088/0953-8984/27/29/293203} {\bibfield  {journal}
  {\bibinfo  {journal} {Journal of Physics: Condensed Matter}\ }\textbf
  {\bibinfo {volume} {27}},\ \bibinfo {pages} {293203} (\bibinfo {year}
  {2015})}\BibitemShut {NoStop}%
\bibitem [{\citenamefont {Yi}\ \emph {et~al.}(2017)\citenamefont {Yi},
  \citenamefont {Zhang}, \citenamefont {Shen},\ and\ \citenamefont
  {Lu}}]{Yi_review}%
  \BibitemOpen
  \bibfield  {author} {\bibinfo {author} {\bibfnamefont {M.}~\bibnamefont
  {Yi}}, \bibinfo {author} {\bibfnamefont {Y.}~\bibnamefont {Zhang}}, \bibinfo
  {author} {\bibfnamefont {Z.-X.}\ \bibnamefont {Shen}},\ and\ \bibinfo
  {author} {\bibfnamefont {D.}~\bibnamefont {Lu}},\ }\bibfield  {title}
  {\bibinfo {title} {Role of the orbital degree of freedom in iron-based
  superconductors},\ }\href {https://doi.org/10.1038/s41535-017-0059-y}
  {\bibfield  {journal} {\bibinfo  {journal} {npj Quantum Materials}\ }\textbf
  {\bibinfo {volume} {2}},\ \bibinfo {pages} {57} (\bibinfo {year}
  {2017})}\BibitemShut {NoStop}%
\bibitem [{\citenamefont {Carrington}(2011)}]{Carrington11}%
  \BibitemOpen
  \bibfield  {author} {\bibinfo {author} {\bibfnamefont {A.}~\bibnamefont
  {Carrington}},\ }\bibfield  {title} {\bibinfo {title} {{Quantum oscillation
  studies of the Fermi surface of iron-pnictide superconductors}},\ }\href
  {https://doi.org/10.1088/0034-4885/74/12/124507} {\bibfield  {journal}
  {\bibinfo  {journal} {Reports on Progress in Physics}\ }\textbf {\bibinfo
  {volume} {74}},\ \bibinfo {pages} {124507} (\bibinfo {year}
  {2011})}\BibitemShut {NoStop}%
\bibitem [{\citenamefont {Qazilbash}\ \emph {et~al.}(2009)\citenamefont
  {Qazilbash}, \citenamefont {Hamlin}, \citenamefont {Baumbach}, \citenamefont
  {Zhang}, \citenamefont {Singh}, \citenamefont {Maple},\ and\ \citenamefont
  {Basov}}]{Qazilbash09}%
  \BibitemOpen
  \bibfield  {author} {\bibinfo {author} {\bibfnamefont {M.}~\bibnamefont
  {Qazilbash}}, \bibinfo {author} {\bibfnamefont {J.}~\bibnamefont {Hamlin}},
  \bibinfo {author} {\bibfnamefont {R.}~\bibnamefont {Baumbach}}, \bibinfo
  {author} {\bibfnamefont {L.}~\bibnamefont {Zhang}}, \bibinfo {author}
  {\bibfnamefont {D.~J.}\ \bibnamefont {Singh}}, \bibinfo {author}
  {\bibfnamefont {M.}~\bibnamefont {Maple}},\ and\ \bibinfo {author}
  {\bibfnamefont {D.}~\bibnamefont {Basov}},\ }\bibfield  {title} {\bibinfo
  {title} {Electronic correlations in the iron pnictides},\ }\href
  {https://doi.org/10.1038/nphys1343} {\bibfield  {journal} {\bibinfo
  {journal} {Nature Physics}\ }\textbf {\bibinfo {volume} {5}},\ \bibinfo
  {pages} {647} (\bibinfo {year} {2009})}\BibitemShut {NoStop}%
\bibitem [{\citenamefont {Haule}\ \emph {et~al.}(2008)\citenamefont {Haule},
  \citenamefont {Shim},\ and\ \citenamefont {Kotliar}}]{Haule08_PRL}%
  \BibitemOpen
  \bibfield  {author} {\bibinfo {author} {\bibfnamefont {K.}~\bibnamefont
  {Haule}}, \bibinfo {author} {\bibfnamefont {J.~H.}\ \bibnamefont {Shim}},\
  and\ \bibinfo {author} {\bibfnamefont {G.}~\bibnamefont {Kotliar}},\
  }\bibfield  {title} {\bibinfo {title} {{Correlated Electronic Structure of
  ${\mathrm{LaO}}_{1\ensuremath{-}x}{\mathrm{F}}_{x}\mathrm{FeAs}$}},\ }\href
  {https://doi.org/10.1103/PhysRevLett.100.226402} {\bibfield  {journal}
  {\bibinfo  {journal} {Phys. Rev. Lett.}\ }\textbf {\bibinfo {volume} {100}},\
  \bibinfo {pages} {226402} (\bibinfo {year} {2008})}\BibitemShut {NoStop}%
\bibitem [{\citenamefont {Skornyakov}\ \emph {et~al.}(2009)\citenamefont
  {Skornyakov}, \citenamefont {Efremov}, \citenamefont {Skorikov},
  \citenamefont {Korotin}, \citenamefont {Izyumov}, \citenamefont {Anisimov},
  \citenamefont {Kozhevnikov},\ and\ \citenamefont {Vollhardt}}]{Vollhardt09}%
  \BibitemOpen
  \bibfield  {author} {\bibinfo {author} {\bibfnamefont {S.~L.}\ \bibnamefont
  {Skornyakov}}, \bibinfo {author} {\bibfnamefont {A.~V.}\ \bibnamefont
  {Efremov}}, \bibinfo {author} {\bibfnamefont {N.~A.}\ \bibnamefont
  {Skorikov}}, \bibinfo {author} {\bibfnamefont {M.~A.}\ \bibnamefont
  {Korotin}}, \bibinfo {author} {\bibfnamefont {Y.~A.}\ \bibnamefont
  {Izyumov}}, \bibinfo {author} {\bibfnamefont {V.~I.}\ \bibnamefont
  {Anisimov}}, \bibinfo {author} {\bibfnamefont {A.~V.}\ \bibnamefont
  {Kozhevnikov}},\ and\ \bibinfo {author} {\bibfnamefont {D.}~\bibnamefont
  {Vollhardt}},\ }\bibfield  {title} {\bibinfo {title} {{Classification of the
  electronic correlation strength in the iron pnictides: The case of the parent
  compound ${\text{BaFe}}_{2}{\text{As}}_{2}$}},\ }\href
  {https://doi.org/10.1103/PhysRevB.80.092501} {\bibfield  {journal} {\bibinfo
  {journal} {Phys. Rev. B}\ }\textbf {\bibinfo {volume} {80}},\ \bibinfo
  {pages} {092501} (\bibinfo {year} {2009})}\BibitemShut {NoStop}%
\bibitem [{\citenamefont {Werner}\ \emph {et~al.}(2012)\citenamefont {Werner},
  \citenamefont {Casula}, \citenamefont {Miyake}, \citenamefont {Aryasetiawan},
  \citenamefont {Millis},\ and\ \citenamefont {Biermann}}]{Werner12}%
  \BibitemOpen
  \bibfield  {author} {\bibinfo {author} {\bibfnamefont {P.}~\bibnamefont
  {Werner}}, \bibinfo {author} {\bibfnamefont {M.}~\bibnamefont {Casula}},
  \bibinfo {author} {\bibfnamefont {T.}~\bibnamefont {Miyake}}, \bibinfo
  {author} {\bibfnamefont {F.}~\bibnamefont {Aryasetiawan}}, \bibinfo {author}
  {\bibfnamefont {A.~J.}\ \bibnamefont {Millis}},\ and\ \bibinfo {author}
  {\bibfnamefont {S.}~\bibnamefont {Biermann}},\ }\bibfield  {title} {\bibinfo
  {title} {{Satellites and large doping and temperature dependence of
  electronic properties in hole-doped BaFe$_2$As$_2$}},\ }\href
  {https://doi.org/10.1038/nphys2250} {\bibfield  {journal} {\bibinfo
  {journal} {Nature Physics}\ }\textbf {\bibinfo {volume} {8}},\ \bibinfo
  {pages} {331} (\bibinfo {year} {2012})}\BibitemShut {NoStop}%
\bibitem [{\citenamefont {Ferber}\ \emph {et~al.}(2012)\citenamefont {Ferber},
  \citenamefont {Foyevtsova}, \citenamefont {Valent\'{\i}},\ and\ \citenamefont
  {Jeschke}}]{Valenti12}%
  \BibitemOpen
  \bibfield  {author} {\bibinfo {author} {\bibfnamefont {J.}~\bibnamefont
  {Ferber}}, \bibinfo {author} {\bibfnamefont {K.}~\bibnamefont {Foyevtsova}},
  \bibinfo {author} {\bibfnamefont {R.}~\bibnamefont {Valent\'{\i}}},\ and\
  \bibinfo {author} {\bibfnamefont {H.~O.}\ \bibnamefont {Jeschke}},\
  }\bibfield  {title} {\bibinfo {title} {{LDA $+$ DMFT study of the effects of
  correlation in LiFeAs}},\ }\href {https://doi.org/10.1103/PhysRevB.85.094505}
  {\bibfield  {journal} {\bibinfo  {journal} {Phys. Rev. B}\ }\textbf {\bibinfo
  {volume} {85}},\ \bibinfo {pages} {094505} (\bibinfo {year}
  {2012})}\BibitemShut {NoStop}%
\bibitem [{\citenamefont {{van Roekeghem}}\ \emph {et~al.}(2016)\citenamefont
  {{van Roekeghem}}, \citenamefont {Richard}, \citenamefont {Ding},\ and\
  \citenamefont {Biermann}}]{Biermann16}%
  \BibitemOpen
  \bibfield  {author} {\bibinfo {author} {\bibfnamefont {A.}~\bibnamefont {{van
  Roekeghem}}}, \bibinfo {author} {\bibfnamefont {P.}~\bibnamefont {Richard}},
  \bibinfo {author} {\bibfnamefont {H.}~\bibnamefont {Ding}},\ and\ \bibinfo
  {author} {\bibfnamefont {S.}~\bibnamefont {Biermann}},\ }\bibfield  {title}
  {\bibinfo {title} {{Spectral properties of transition metal pnictides and
  chalcogenides: Angle-resolved photoemission spectroscopy and dynamical
  mean-field theory}},\ }\href
  {https://doi.org/https://doi.org/10.1016/j.crhy.2015.11.003} {\bibfield
  {journal} {\bibinfo  {journal} {Comptes Rendus Physique}\ }\textbf {\bibinfo
  {volume} {17}},\ \bibinfo {pages} {140} (\bibinfo {year} {2016})}\BibitemShut
  {NoStop}%
\bibitem [{\citenamefont {Coldea}\ \emph {et~al.}(2008)\citenamefont {Coldea},
  \citenamefont {Fletcher}, \citenamefont {Carrington}, \citenamefont
  {Analytis}, \citenamefont {Bangura}, \citenamefont {Chu}, \citenamefont
  {Erickson}, \citenamefont {Fisher}, \citenamefont {Hussey},\ and\
  \citenamefont {McDonald}}]{Coldea08}%
  \BibitemOpen
  \bibfield  {author} {\bibinfo {author} {\bibfnamefont {A.~I.}\ \bibnamefont
  {Coldea}}, \bibinfo {author} {\bibfnamefont {J.~D.}\ \bibnamefont
  {Fletcher}}, \bibinfo {author} {\bibfnamefont {A.}~\bibnamefont
  {Carrington}}, \bibinfo {author} {\bibfnamefont {J.~G.}\ \bibnamefont
  {Analytis}}, \bibinfo {author} {\bibfnamefont {A.~F.}\ \bibnamefont
  {Bangura}}, \bibinfo {author} {\bibfnamefont {J.-H.}\ \bibnamefont {Chu}},
  \bibinfo {author} {\bibfnamefont {A.~S.}\ \bibnamefont {Erickson}}, \bibinfo
  {author} {\bibfnamefont {I.~R.}\ \bibnamefont {Fisher}}, \bibinfo {author}
  {\bibfnamefont {N.~E.}\ \bibnamefont {Hussey}},\ and\ \bibinfo {author}
  {\bibfnamefont {R.~D.}\ \bibnamefont {McDonald}},\ }\bibfield  {title}
  {\bibinfo {title} {{Fermi Surface of Superconducting LaFePO Determined from
  Quantum Oscillations}},\ }\href
  {https://doi.org/10.1103/PhysRevLett.101.216402} {\bibfield  {journal}
  {\bibinfo  {journal} {Phys. Rev. Lett.}\ }\textbf {\bibinfo {volume} {101}},\
  \bibinfo {pages} {216402} (\bibinfo {year} {2008})}\BibitemShut {NoStop}%
\bibitem [{\citenamefont {Kordyuk}\ \emph {et~al.}(2013)\citenamefont
  {Kordyuk}, \citenamefont {Zabolotnyy}, \citenamefont {Evtushinsky},
  \citenamefont {Yaresko}, \citenamefont {B{\"u}chner},\ and\ \citenamefont
  {Borisenko}}]{Kordyuk13}%
  \BibitemOpen
  \bibfield  {author} {\bibinfo {author} {\bibfnamefont {A.}~\bibnamefont
  {Kordyuk}}, \bibinfo {author} {\bibfnamefont {V.}~\bibnamefont {Zabolotnyy}},
  \bibinfo {author} {\bibfnamefont {D.}~\bibnamefont {Evtushinsky}}, \bibinfo
  {author} {\bibfnamefont {A.}~\bibnamefont {Yaresko}}, \bibinfo {author}
  {\bibfnamefont {B.}~\bibnamefont {B{\"u}chner}},\ and\ \bibinfo {author}
  {\bibfnamefont {S.}~\bibnamefont {Borisenko}},\ }\bibfield  {title} {\bibinfo
  {title} {{Electronic band structure of ferro-pnictide superconductors from
  ARPES experiment}},\ }\href {https://doi.org/10.1007/s10948-013-2210-8}
  {\bibfield  {journal} {\bibinfo  {journal} {Journal of Superconductivity and
  Novel Magnetism}\ }\textbf {\bibinfo {volume} {26}},\ \bibinfo {pages} {2837}
  (\bibinfo {year} {2013})}\BibitemShut {NoStop}%
\bibitem [{\citenamefont {Dhaka}\ \emph {et~al.}(2013)\citenamefont {Dhaka},
  \citenamefont {Hahn}, \citenamefont {Razzoli}, \citenamefont {Jiang},
  \citenamefont {Shi}, \citenamefont {Harmon}, \citenamefont {Thaler},
  \citenamefont {Bud'ko}, \citenamefont {Canfield},\ and\ \citenamefont
  {Kaminski}}]{Kaminski13}%
  \BibitemOpen
  \bibfield  {author} {\bibinfo {author} {\bibfnamefont {R.~S.}\ \bibnamefont
  {Dhaka}}, \bibinfo {author} {\bibfnamefont {S.~E.}\ \bibnamefont {Hahn}},
  \bibinfo {author} {\bibfnamefont {E.}~\bibnamefont {Razzoli}}, \bibinfo
  {author} {\bibfnamefont {R.}~\bibnamefont {Jiang}}, \bibinfo {author}
  {\bibfnamefont {M.}~\bibnamefont {Shi}}, \bibinfo {author} {\bibfnamefont
  {B.~N.}\ \bibnamefont {Harmon}}, \bibinfo {author} {\bibfnamefont
  {A.}~\bibnamefont {Thaler}}, \bibinfo {author} {\bibfnamefont {S.~L.}\
  \bibnamefont {Bud'ko}}, \bibinfo {author} {\bibfnamefont {P.~C.}\
  \bibnamefont {Canfield}},\ and\ \bibinfo {author} {\bibfnamefont
  {A.}~\bibnamefont {Kaminski}},\ }\bibfield  {title} {\bibinfo {title}
  {{Unusual Temperature Dependence of Band Dispersion in
  $\mathrm{Ba}({\mathrm{Fe}}_{1\ensuremath{-}x}{\mathrm{Ru}}_{x}{)}_{2}{\mathrm{As}}_{2}$
  and its Consequences for Antiferromagnetic Ordering}},\ }\href
  {https://doi.org/10.1103/PhysRevLett.110.067002} {\bibfield  {journal}
  {\bibinfo  {journal} {Phys. Rev. Lett.}\ }\textbf {\bibinfo {volume} {110}},\
  \bibinfo {pages} {067002} (\bibinfo {year} {2013})}\BibitemShut {NoStop}%
\bibitem [{\citenamefont {Brouet}\ \emph {et~al.}(2013)\citenamefont {Brouet},
  \citenamefont {Lin}, \citenamefont {Texier}, \citenamefont {Bobroff},
  \citenamefont {Taleb-Ibrahimi}, \citenamefont {Le~F\`evre}, \citenamefont
  {Bertran}, \citenamefont {Casula}, \citenamefont {Werner}, \citenamefont
  {Biermann}, \citenamefont {Rullier-Albenque}, \citenamefont {Forget},\ and\
  \citenamefont {Colson}}]{Brouet13}%
  \BibitemOpen
  \bibfield  {author} {\bibinfo {author} {\bibfnamefont {V.}~\bibnamefont
  {Brouet}}, \bibinfo {author} {\bibfnamefont {P.-H.}\ \bibnamefont {Lin}},
  \bibinfo {author} {\bibfnamefont {Y.}~\bibnamefont {Texier}}, \bibinfo
  {author} {\bibfnamefont {J.}~\bibnamefont {Bobroff}}, \bibinfo {author}
  {\bibfnamefont {A.}~\bibnamefont {Taleb-Ibrahimi}}, \bibinfo {author}
  {\bibfnamefont {P.}~\bibnamefont {Le~F\`evre}}, \bibinfo {author}
  {\bibfnamefont {F.}~\bibnamefont {Bertran}}, \bibinfo {author} {\bibfnamefont
  {M.}~\bibnamefont {Casula}}, \bibinfo {author} {\bibfnamefont
  {P.}~\bibnamefont {Werner}}, \bibinfo {author} {\bibfnamefont
  {S.}~\bibnamefont {Biermann}}, \bibinfo {author} {\bibfnamefont
  {F.}~\bibnamefont {Rullier-Albenque}}, \bibinfo {author} {\bibfnamefont
  {A.}~\bibnamefont {Forget}},\ and\ \bibinfo {author} {\bibfnamefont
  {D.}~\bibnamefont {Colson}},\ }\bibfield  {title} {\bibinfo {title} {{Large
  Temperature Dependence of the Number of Carriers in Co-Doped
  ${\mathrm{BaFe}}_{2}{\mathrm{As}}_{2}$}},\ }\href
  {https://doi.org/10.1103/PhysRevLett.110.167002} {\bibfield  {journal}
  {\bibinfo  {journal} {Phys. Rev. Lett.}\ }\textbf {\bibinfo {volume} {110}},\
  \bibinfo {pages} {167002} (\bibinfo {year} {2013})}\BibitemShut {NoStop}%
\bibitem [{\citenamefont {Ortenzi}\ \emph {et~al.}(2009)\citenamefont
  {Ortenzi}, \citenamefont {Cappelluti}, \citenamefont {Benfatto},\ and\
  \citenamefont {Pietronero}}]{Ortenzi09}%
  \BibitemOpen
  \bibfield  {author} {\bibinfo {author} {\bibfnamefont {L.}~\bibnamefont
  {Ortenzi}}, \bibinfo {author} {\bibfnamefont {E.}~\bibnamefont {Cappelluti}},
  \bibinfo {author} {\bibfnamefont {L.}~\bibnamefont {Benfatto}},\ and\
  \bibinfo {author} {\bibfnamefont {L.}~\bibnamefont {Pietronero}},\ }\bibfield
   {title} {\bibinfo {title} {{Fermi-Surface Shrinking and Interband Coupling
  in Iron-Based Pnictides}},\ }\href
  {https://doi.org/10.1103/PhysRevLett.103.046404} {\bibfield  {journal}
  {\bibinfo  {journal} {Phys. Rev. Lett.}\ }\textbf {\bibinfo {volume} {103}},\
  \bibinfo {pages} {046404} (\bibinfo {year} {2009})}\BibitemShut {NoStop}%
\bibitem [{\citenamefont {Fernandes}\ and\ \citenamefont
  {Chubukov}(2016)}]{Fernandes_Chubukov}%
  \BibitemOpen
  \bibfield  {author} {\bibinfo {author} {\bibfnamefont {R.~M.}\ \bibnamefont
  {Fernandes}}\ and\ \bibinfo {author} {\bibfnamefont {A.~V.}\ \bibnamefont
  {Chubukov}},\ }\bibfield  {title} {\bibinfo {title} {Low-energy microscopic
  models for iron-based superconductors: a review},\ }\href
  {https://doi.org/10.1088/1361-6633/80/1/014503} {\bibfield  {journal}
  {\bibinfo  {journal} {Reports on Progress in Physics}\ }\textbf {\bibinfo
  {volume} {80}},\ \bibinfo {pages} {014503} (\bibinfo {year}
  {2016})}\BibitemShut {NoStop}%
\bibitem [{\citenamefont {Zantout}\ \emph {et~al.}(2019)\citenamefont
  {Zantout}, \citenamefont {Backes},\ and\ \citenamefont
  {Valent\'{\i}}}]{Zantout19}%
  \BibitemOpen
  \bibfield  {author} {\bibinfo {author} {\bibfnamefont {K.}~\bibnamefont
  {Zantout}}, \bibinfo {author} {\bibfnamefont {S.}~\bibnamefont {Backes}},\
  and\ \bibinfo {author} {\bibfnamefont {R.}~\bibnamefont {Valent\'{\i}}},\
  }\bibfield  {title} {\bibinfo {title} {{Effect of Nonlocal Correlations on
  the Electronic Structure of LiFeAs}},\ }\href
  {https://doi.org/10.1103/PhysRevLett.123.256401} {\bibfield  {journal}
  {\bibinfo  {journal} {Phys. Rev. Lett.}\ }\textbf {\bibinfo {volume} {123}},\
  \bibinfo {pages} {256401} (\bibinfo {year} {2019})}\BibitemShut {NoStop}%
\bibitem [{\citenamefont {Bhattacharyya}\ \emph {et~al.}(2020)\citenamefont
  {Bhattacharyya}, \citenamefont {Bj\"ornson}, \citenamefont {Zantout},
  \citenamefont {Steffensen}, \citenamefont {Fanfarillo}, \citenamefont
  {Kreisel}, \citenamefont {Valent\'{\i}}, \citenamefont {Andersen},\ and\
  \citenamefont {Hirschfeld}}]{Bhattacharyya20}%
  \BibitemOpen
  \bibfield  {author} {\bibinfo {author} {\bibfnamefont {S.}~\bibnamefont
  {Bhattacharyya}}, \bibinfo {author} {\bibfnamefont {K.}~\bibnamefont
  {Bj\"ornson}}, \bibinfo {author} {\bibfnamefont {K.}~\bibnamefont {Zantout}},
  \bibinfo {author} {\bibfnamefont {D.}~\bibnamefont {Steffensen}}, \bibinfo
  {author} {\bibfnamefont {L.}~\bibnamefont {Fanfarillo}}, \bibinfo {author}
  {\bibfnamefont {A.}~\bibnamefont {Kreisel}}, \bibinfo {author} {\bibfnamefont
  {R.}~\bibnamefont {Valent\'{\i}}}, \bibinfo {author} {\bibfnamefont {B.~M.}\
  \bibnamefont {Andersen}},\ and\ \bibinfo {author} {\bibfnamefont {P.~J.}\
  \bibnamefont {Hirschfeld}},\ }\bibfield  {title} {\bibinfo {title} {Nonlocal
  correlations in iron pnictides and chalcogenides},\ }\href
  {https://doi.org/10.1103/PhysRevB.102.035109} {\bibfield  {journal} {\bibinfo
   {journal} {Phys. Rev. B}\ }\textbf {\bibinfo {volume} {102}},\ \bibinfo
  {pages} {035109} (\bibinfo {year} {2020})}\BibitemShut {NoStop}%
\bibitem [{\citenamefont {Kim}\ \emph {et~al.}(2020)\citenamefont {Kim},
  \citenamefont {Miao}, \citenamefont {Choi}, \citenamefont {Zingl},
  \citenamefont {Georges},\ and\ \citenamefont {Kotliar}}]{Kim20}%
  \BibitemOpen
  \bibfield  {author} {\bibinfo {author} {\bibfnamefont {M.}~\bibnamefont
  {Kim}}, \bibinfo {author} {\bibfnamefont {H.}~\bibnamefont {Miao}}, \bibinfo
  {author} {\bibfnamefont {S.}~\bibnamefont {Choi}}, \bibinfo {author}
  {\bibfnamefont {M.}~\bibnamefont {Zingl}}, \bibinfo {author} {\bibfnamefont
  {A.}~\bibnamefont {Georges}},\ and\ \bibinfo {author} {\bibfnamefont
  {G.}~\bibnamefont {Kotliar}},\ }\bibfield  {title} {\bibinfo {title} {{On the
  Spatial Locality of Electronic Correlations in LiFeAs}},\ }\href@noop {}
  {\bibfield  {journal} {\bibinfo  {journal} {arXiv:2009.10577}\ } (\bibinfo
  {year} {2020})}\BibitemShut {NoStop}%
\bibitem [{\citenamefont {van~der Marel}\ and\ \citenamefont
  {Sawatzky}(1988)}]{Sawatzky88}%
  \BibitemOpen
  \bibfield  {author} {\bibinfo {author} {\bibfnamefont {D.}~\bibnamefont
  {van~der Marel}}\ and\ \bibinfo {author} {\bibfnamefont {G.~A.}\ \bibnamefont
  {Sawatzky}},\ }\bibfield  {title} {\bibinfo {title} {{Electron-electron
  interaction and localization in d and f transition metals}},\ }\href
  {https://doi.org/10.1103/PhysRevB.37.10674} {\bibfield  {journal} {\bibinfo
  {journal} {Phys. Rev. B}\ }\textbf {\bibinfo {volume} {37}},\ \bibinfo
  {pages} {10674} (\bibinfo {year} {1988})}\BibitemShut {NoStop}%
\bibitem [{\citenamefont {Hardy}\ \emph {et~al.}(2013)\citenamefont {Hardy},
  \citenamefont {B\"ohmer}, \citenamefont {Aoki}, \citenamefont {Burger},
  \citenamefont {Wolf}, \citenamefont {Schweiss}, \citenamefont {Heid},
  \citenamefont {Adelmann}, \citenamefont {Yao}, \citenamefont {Kotliar},
  \citenamefont {Schmalian},\ and\ \citenamefont
  {Meingast}}]{Schmalian_Kotliar}%
  \BibitemOpen
  \bibfield  {author} {\bibinfo {author} {\bibfnamefont {F.}~\bibnamefont
  {Hardy}}, \bibinfo {author} {\bibfnamefont {A.~E.}\ \bibnamefont {B\"ohmer}},
  \bibinfo {author} {\bibfnamefont {D.}~\bibnamefont {Aoki}}, \bibinfo {author}
  {\bibfnamefont {P.}~\bibnamefont {Burger}}, \bibinfo {author} {\bibfnamefont
  {T.}~\bibnamefont {Wolf}}, \bibinfo {author} {\bibfnamefont {P.}~\bibnamefont
  {Schweiss}}, \bibinfo {author} {\bibfnamefont {R.}~\bibnamefont {Heid}},
  \bibinfo {author} {\bibfnamefont {P.}~\bibnamefont {Adelmann}}, \bibinfo
  {author} {\bibfnamefont {Y.~X.}\ \bibnamefont {Yao}}, \bibinfo {author}
  {\bibfnamefont {G.}~\bibnamefont {Kotliar}}, \bibinfo {author} {\bibfnamefont
  {J.}~\bibnamefont {Schmalian}},\ and\ \bibinfo {author} {\bibfnamefont
  {C.}~\bibnamefont {Meingast}},\ }\bibfield  {title} {\bibinfo {title}
  {{Evidence of Strong Correlations and Coherence-Incoherence Crossover in the
  Iron Pnictide Superconductor ${\mathrm{KFe}}_{2}{\mathrm{As}}_{2}$}},\ }\href
  {https://doi.org/10.1103/PhysRevLett.111.027002} {\bibfield  {journal}
  {\bibinfo  {journal} {Phys. Rev. Lett.}\ }\textbf {\bibinfo {volume} {111}},\
  \bibinfo {pages} {027002} (\bibinfo {year} {2013})}\BibitemShut {NoStop}%
\bibitem [{\citenamefont {Yin}\ \emph {et~al.}(2012)\citenamefont {Yin},
  \citenamefont {Haule},\ and\ \citenamefont {Kotliar}}]{Kotliar12}%
  \BibitemOpen
  \bibfield  {author} {\bibinfo {author} {\bibfnamefont {Z.~P.}\ \bibnamefont
  {Yin}}, \bibinfo {author} {\bibfnamefont {K.}~\bibnamefont {Haule}},\ and\
  \bibinfo {author} {\bibfnamefont {G.}~\bibnamefont {Kotliar}},\ }\bibfield
  {title} {\bibinfo {title} {{Fractional power-law behavior and its origin in
  iron-chalcogenide and ruthenate superconductors: Insights from
  first-principles calculations}},\ }\href
  {https://doi.org/10.1103/PhysRevB.86.195141} {\bibfield  {journal} {\bibinfo
  {journal} {Phys. Rev. B}\ }\textbf {\bibinfo {volume} {86}},\ \bibinfo
  {pages} {195141} (\bibinfo {year} {2012})}\BibitemShut {NoStop}%
\bibitem [{\citenamefont {Yu}\ and\ \citenamefont {Si}(2013)}]{Si13}%
  \BibitemOpen
  \bibfield  {author} {\bibinfo {author} {\bibfnamefont {R.}~\bibnamefont
  {Yu}}\ and\ \bibinfo {author} {\bibfnamefont {Q.}~\bibnamefont {Si}},\
  }\bibfield  {title} {\bibinfo {title} {{Orbital-Selective Mott Phase in
  Multiorbital Models for Alkaline Iron Selenides
  ${\mathrm{K}}_{1\ensuremath{-}x}{\mathrm{Fe}}_{2\ensuremath{-}y}{\mathrm{Se}}_{2}$}},\
  }\href {https://doi.org/10.1103/PhysRevLett.110.146402} {\bibfield  {journal}
  {\bibinfo  {journal} {Phys. Rev. Lett.}\ }\textbf {\bibinfo {volume} {110}},\
  \bibinfo {pages} {146402} (\bibinfo {year} {2013})}\BibitemShut {NoStop}%
\bibitem [{\citenamefont {de' Medici}\ \emph {et~al.}(2014)\citenamefont {de'
  Medici}, \citenamefont {Giovannetti},\ and\ \citenamefont
  {Capone}}]{Medici14}%
  \BibitemOpen
  \bibfield  {author} {\bibinfo {author} {\bibfnamefont {L.}~\bibnamefont {de'
  Medici}}, \bibinfo {author} {\bibfnamefont {G.}~\bibnamefont {Giovannetti}},\
  and\ \bibinfo {author} {\bibfnamefont {M.}~\bibnamefont {Capone}},\
  }\bibfield  {title} {\bibinfo {title} {{Selective Mott Physics as a Key to
  Iron Superconductors}},\ }\href
  {https://doi.org/10.1103/PhysRevLett.112.177001} {\bibfield  {journal}
  {\bibinfo  {journal} {Phys. Rev. Lett.}\ }\textbf {\bibinfo {volume} {112}},\
  \bibinfo {pages} {177001} (\bibinfo {year} {2014})}\BibitemShut {NoStop}%
\bibitem [{\citenamefont {Si}\ \emph {et~al.}(2016)\citenamefont {Si},
  \citenamefont {Yu},\ and\ \citenamefont {Abrahams}}]{Si16}%
  \BibitemOpen
  \bibfield  {author} {\bibinfo {author} {\bibfnamefont {Q.}~\bibnamefont
  {Si}}, \bibinfo {author} {\bibfnamefont {R.}~\bibnamefont {Yu}},\ and\
  \bibinfo {author} {\bibfnamefont {E.}~\bibnamefont {Abrahams}},\ }\bibfield
  {title} {\bibinfo {title} {High-temperature superconductivity in iron
  pnictides and chalcogenides},\ }\href
  {https://doi.org/10.1038/natrevmats.2016.17} {\bibfield  {journal} {\bibinfo
  {journal} {Nature Reviews Materials}\ }\textbf {\bibinfo {volume} {1}},\
  \bibinfo {pages} {16017} (\bibinfo {year} {2016})}\BibitemShut {NoStop}%
\bibitem [{\citenamefont {Bascones}\ \emph {et~al.}(2016)\citenamefont
  {Bascones}, \citenamefont {Valenzuela},\ and\ \citenamefont
  {Calder\'on}}]{Bascones_review}%
  \BibitemOpen
  \bibfield  {author} {\bibinfo {author} {\bibfnamefont {E.}~\bibnamefont
  {Bascones}}, \bibinfo {author} {\bibfnamefont {B.}~\bibnamefont
  {Valenzuela}},\ and\ \bibinfo {author} {\bibfnamefont {M.~J.}\ \bibnamefont
  {Calder\'on}},\ }\bibfield  {title} {\bibinfo {title} {{Magnetic interactions
  in iron superconductors: A review}},\ }\href
  {https://doi.org/10.1016/j.crhy.2015.05.004} {\bibfield  {journal} {\bibinfo
  {journal} {Comptes Rendus Physique}\ }\textbf {\bibinfo {volume} {17}},\
  \bibinfo {pages} {36} (\bibinfo {year} {2016})}\BibitemShut {NoStop}%
\bibitem [{\citenamefont {Rhodes}\ \emph {et~al.}(2018)\citenamefont {Rhodes},
  \citenamefont {Watson}, \citenamefont {Haghighirad}, \citenamefont
  {Evtushinsky}, \citenamefont {Eschrig},\ and\ \citenamefont
  {Kim}}]{Rhodes18}%
  \BibitemOpen
  \bibfield  {author} {\bibinfo {author} {\bibfnamefont {L.~C.}\ \bibnamefont
  {Rhodes}}, \bibinfo {author} {\bibfnamefont {M.~D.}\ \bibnamefont {Watson}},
  \bibinfo {author} {\bibfnamefont {A.~A.}\ \bibnamefont {Haghighirad}},
  \bibinfo {author} {\bibfnamefont {D.~V.}\ \bibnamefont {Evtushinsky}},
  \bibinfo {author} {\bibfnamefont {M.}~\bibnamefont {Eschrig}},\ and\ \bibinfo
  {author} {\bibfnamefont {T.~K.}\ \bibnamefont {Kim}},\ }\bibfield  {title}
  {\bibinfo {title} {{Scaling of the superconducting gap with orbital character
  in FeSe}},\ }\href {https://doi.org/10.1103/PhysRevB.98.180503} {\bibfield
  {journal} {\bibinfo  {journal} {Phys. Rev. B}\ }\textbf {\bibinfo {volume}
  {98}},\ \bibinfo {pages} {180503} (\bibinfo {year} {2018})}\BibitemShut
  {NoStop}%
\bibitem [{\citenamefont {Liu}\ \emph {et~al.}(2018)\citenamefont {Liu},
  \citenamefont {Li}, \citenamefont {Huang}, \citenamefont {Lei}, \citenamefont
  {Wang}, \citenamefont {Wu}, \citenamefont {Shen}, \citenamefont {Gao},
  \citenamefont {Zhang}, \citenamefont {Liu}, \citenamefont {Hu}, \citenamefont
  {Xu}, \citenamefont {Liang}, \citenamefont {Liu}, \citenamefont {Ai},
  \citenamefont {Zhao}, \citenamefont {He}, \citenamefont {Yu}, \citenamefont
  {Liu}, \citenamefont {Mao}, \citenamefont {Dong}, \citenamefont {Jia},
  \citenamefont {Zhang}, \citenamefont {Zhang}, \citenamefont {Yang},
  \citenamefont {Wang}, \citenamefont {Peng}, \citenamefont {Shi},
  \citenamefont {Hu}, \citenamefont {Xiang}, \citenamefont {Chen},
  \citenamefont {Xu}, \citenamefont {Chen},\ and\ \citenamefont
  {Zhou}}]{Zhou18}%
  \BibitemOpen
  \bibfield  {author} {\bibinfo {author} {\bibfnamefont {D.}~\bibnamefont
  {Liu}}, \bibinfo {author} {\bibfnamefont {C.}~\bibnamefont {Li}}, \bibinfo
  {author} {\bibfnamefont {J.}~\bibnamefont {Huang}}, \bibinfo {author}
  {\bibfnamefont {B.}~\bibnamefont {Lei}}, \bibinfo {author} {\bibfnamefont
  {L.}~\bibnamefont {Wang}}, \bibinfo {author} {\bibfnamefont {X.}~\bibnamefont
  {Wu}}, \bibinfo {author} {\bibfnamefont {B.}~\bibnamefont {Shen}}, \bibinfo
  {author} {\bibfnamefont {Q.}~\bibnamefont {Gao}}, \bibinfo {author}
  {\bibfnamefont {Y.}~\bibnamefont {Zhang}}, \bibinfo {author} {\bibfnamefont
  {X.}~\bibnamefont {Liu}}, \bibinfo {author} {\bibfnamefont {Y.}~\bibnamefont
  {Hu}}, \bibinfo {author} {\bibfnamefont {Y.}~\bibnamefont {Xu}}, \bibinfo
  {author} {\bibfnamefont {A.}~\bibnamefont {Liang}}, \bibinfo {author}
  {\bibfnamefont {J.}~\bibnamefont {Liu}}, \bibinfo {author} {\bibfnamefont
  {P.}~\bibnamefont {Ai}}, \bibinfo {author} {\bibfnamefont {L.}~\bibnamefont
  {Zhao}}, \bibinfo {author} {\bibfnamefont {S.}~\bibnamefont {He}}, \bibinfo
  {author} {\bibfnamefont {L.}~\bibnamefont {Yu}}, \bibinfo {author}
  {\bibfnamefont {G.}~\bibnamefont {Liu}}, \bibinfo {author} {\bibfnamefont
  {Y.}~\bibnamefont {Mao}}, \bibinfo {author} {\bibfnamefont {X.}~\bibnamefont
  {Dong}}, \bibinfo {author} {\bibfnamefont {X.}~\bibnamefont {Jia}}, \bibinfo
  {author} {\bibfnamefont {F.}~\bibnamefont {Zhang}}, \bibinfo {author}
  {\bibfnamefont {S.}~\bibnamefont {Zhang}}, \bibinfo {author} {\bibfnamefont
  {F.}~\bibnamefont {Yang}}, \bibinfo {author} {\bibfnamefont {Z.}~\bibnamefont
  {Wang}}, \bibinfo {author} {\bibfnamefont {Q.}~\bibnamefont {Peng}}, \bibinfo
  {author} {\bibfnamefont {Y.}~\bibnamefont {Shi}}, \bibinfo {author}
  {\bibfnamefont {J.}~\bibnamefont {Hu}}, \bibinfo {author} {\bibfnamefont
  {T.}~\bibnamefont {Xiang}}, \bibinfo {author} {\bibfnamefont
  {X.}~\bibnamefont {Chen}}, \bibinfo {author} {\bibfnamefont {Z.}~\bibnamefont
  {Xu}}, \bibinfo {author} {\bibfnamefont {C.}~\bibnamefont {Chen}},\ and\
  \bibinfo {author} {\bibfnamefont {X.~J.}\ \bibnamefont {Zhou}},\ }\bibfield
  {title} {\bibinfo {title} {{Orbital Origin of Extremely Anisotropic
  Superconducting Gap in Nematic Phase of FeSe Superconductor}},\ }\href
  {https://doi.org/10.1103/PhysRevX.8.031033} {\bibfield  {journal} {\bibinfo
  {journal} {Phys. Rev. X}\ }\textbf {\bibinfo {volume} {8}},\ \bibinfo {pages}
  {031033} (\bibinfo {year} {2018})}\BibitemShut {NoStop}%
\bibitem [{\citenamefont {Yin}\ \emph {et~al.}(2014)\citenamefont {Yin},
  \citenamefont {Haule},\ and\ \citenamefont {Kotliar}}]{Kotliar_antiphase}%
  \BibitemOpen
  \bibfield  {author} {\bibinfo {author} {\bibfnamefont {Z.}~\bibnamefont
  {Yin}}, \bibinfo {author} {\bibfnamefont {K.}~\bibnamefont {Haule}},\ and\
  \bibinfo {author} {\bibfnamefont {G.}~\bibnamefont {Kotliar}},\ }\bibfield
  {title} {\bibinfo {title} {Spin dynamics and orbital-antiphase pairing
  symmetry in iron-based superconductors},\ }\href
  {https://doi.org/10.1038/nphys3116} {\bibfield  {journal} {\bibinfo
  {journal} {Nature Physics}\ }\textbf {\bibinfo {volume} {10}},\ \bibinfo
  {pages} {845} (\bibinfo {year} {2014})}\BibitemShut {NoStop}%
\bibitem [{\citenamefont {Wang}\ \emph {et~al.}(2013)\citenamefont {Wang},
  \citenamefont {Zhang}, \citenamefont {Lu}, \citenamefont {Tan}, \citenamefont
  {Luo}, \citenamefont {Song}, \citenamefont {Wang}, \citenamefont {Zhang},
  \citenamefont {Goremychkin}, \citenamefont {Perring}, \citenamefont {Maier},
  \citenamefont {Yin}, \citenamefont {Haule}, \citenamefont {Kotliar},\ and\
  \citenamefont {Dai}}]{Wang13}%
  \BibitemOpen
  \bibfield  {author} {\bibinfo {author} {\bibfnamefont {M.}~\bibnamefont
  {Wang}}, \bibinfo {author} {\bibfnamefont {C.}~\bibnamefont {Zhang}},
  \bibinfo {author} {\bibfnamefont {X.}~\bibnamefont {Lu}}, \bibinfo {author}
  {\bibfnamefont {G.}~\bibnamefont {Tan}}, \bibinfo {author} {\bibfnamefont
  {H.}~\bibnamefont {Luo}}, \bibinfo {author} {\bibfnamefont {Y.}~\bibnamefont
  {Song}}, \bibinfo {author} {\bibfnamefont {M.}~\bibnamefont {Wang}}, \bibinfo
  {author} {\bibfnamefont {X.}~\bibnamefont {Zhang}}, \bibinfo {author}
  {\bibfnamefont {E.}~\bibnamefont {Goremychkin}}, \bibinfo {author}
  {\bibfnamefont {T.}~\bibnamefont {Perring}}, \bibinfo {author} {\bibfnamefont
  {T.}~\bibnamefont {Maier}}, \bibinfo {author} {\bibfnamefont
  {Z.}~\bibnamefont {Yin}}, \bibinfo {author} {\bibfnamefont {K.}~\bibnamefont
  {Haule}}, \bibinfo {author} {\bibfnamefont {G.}~\bibnamefont {Kotliar}},\
  and\ \bibinfo {author} {\bibfnamefont {P.}~\bibnamefont {Dai}},\ }\bibfield
  {title} {\bibinfo {title} {Doping dependence of spin excitations and its
  correlations with high-temperature superconductivity in iron pnictides},\
  }\href {https://doi.org/10.1038/ncomms3874} {\bibfield  {journal} {\bibinfo
  {journal} {Nature Communications}\ }\textbf {\bibinfo {volume} {4}},\
  \bibinfo {pages} {2874} (\bibinfo {year} {2013})}\BibitemShut {NoStop}%
\bibitem [{\citenamefont {Stadler}\ \emph {et~al.}(2015)\citenamefont
  {Stadler}, \citenamefont {Yin}, \citenamefont {von Delft}, \citenamefont
  {Kotliar},\ and\ \citenamefont
  {Weichselbaum}}]{von_delft_PhysRevLett.115.136401}%
  \BibitemOpen
  \bibfield  {author} {\bibinfo {author} {\bibfnamefont {K.~M.}\ \bibnamefont
  {Stadler}}, \bibinfo {author} {\bibfnamefont {Z.~P.}\ \bibnamefont {Yin}},
  \bibinfo {author} {\bibfnamefont {J.}~\bibnamefont {von Delft}}, \bibinfo
  {author} {\bibfnamefont {G.}~\bibnamefont {Kotliar}},\ and\ \bibinfo {author}
  {\bibfnamefont {A.}~\bibnamefont {Weichselbaum}},\ }\bibfield  {title}
  {\bibinfo {title} {Dynamical mean-field theory plus numerical
  renormalization-group study of spin-orbital separation in a three-band hund
  metal},\ }\href {https://doi.org/10.1103/PhysRevLett.115.136401} {\bibfield
  {journal} {\bibinfo  {journal} {Phys. Rev. Lett.}\ }\textbf {\bibinfo
  {volume} {115}},\ \bibinfo {pages} {136401} (\bibinfo {year}
  {2015})}\BibitemShut {NoStop}%
\bibitem [{\citenamefont {Meier}\ \emph {et~al.}(2018)\citenamefont {Meier},
  \citenamefont {Ding}, \citenamefont {Kreyssig}, \citenamefont {Bud’ko},
  \citenamefont {Sapkota}, \citenamefont {Kothapalli}, \citenamefont {Borisov},
  \citenamefont {Valent{\'\i}}, \citenamefont {Batista}, \citenamefont {Orth},
  \citenamefont {Fernandes}, \citenamefont {Goldman}, \citenamefont {Furukawa},
  \citenamefont {B\"ohmer},\ and\ \citenamefont {Canfield}}]{Meier18}%
  \BibitemOpen
  \bibfield  {author} {\bibinfo {author} {\bibfnamefont {W.~R.}\ \bibnamefont
  {Meier}}, \bibinfo {author} {\bibfnamefont {Q.-P.}\ \bibnamefont {Ding}},
  \bibinfo {author} {\bibfnamefont {A.}~\bibnamefont {Kreyssig}}, \bibinfo
  {author} {\bibfnamefont {S.~L.}\ \bibnamefont {Bud’ko}}, \bibinfo {author}
  {\bibfnamefont {A.}~\bibnamefont {Sapkota}}, \bibinfo {author} {\bibfnamefont
  {K.}~\bibnamefont {Kothapalli}}, \bibinfo {author} {\bibfnamefont
  {V.}~\bibnamefont {Borisov}}, \bibinfo {author} {\bibfnamefont
  {R.}~\bibnamefont {Valent{\'\i}}}, \bibinfo {author} {\bibfnamefont {C.~D.}\
  \bibnamefont {Batista}}, \bibinfo {author} {\bibfnamefont {P.~P.}\
  \bibnamefont {Orth}}, \bibinfo {author} {\bibfnamefont {R.~M.}\ \bibnamefont
  {Fernandes}}, \bibinfo {author} {\bibfnamefont {A.~I.}\ \bibnamefont
  {Goldman}}, \bibinfo {author} {\bibfnamefont {Y.}~\bibnamefont {Furukawa}},
  \bibinfo {author} {\bibfnamefont {A.~E.}\ \bibnamefont {B\"ohmer}},\ and\
  \bibinfo {author} {\bibfnamefont {P.~C.}\ \bibnamefont {Canfield}},\
  }\bibfield  {title} {\bibinfo {title} {Hedgehog spin-vortex crystal
  stabilized in a hole-doped iron-based superconductor},\ }\href
  {https://doi.org/10.1038/s41535-017-0076-x} {\bibfield  {journal} {\bibinfo
  {journal} {npj Quantum Materials}\ }\textbf {\bibinfo {volume} {3}},\
  \bibinfo {pages} {5} (\bibinfo {year} {2018})}\BibitemShut {NoStop}%
\bibitem [{\citenamefont {Allred}\ \emph {et~al.}(2016)\citenamefont {Allred},
  \citenamefont {Taddei}, \citenamefont {Bugaris}, \citenamefont {Krogstad},
  \citenamefont {Lapidus}, \citenamefont {Chung}, \citenamefont {Claus},
  \citenamefont {Kanatzidis}, \citenamefont {Brown}, \citenamefont {Kang},
  \citenamefont {Fernandes}, \citenamefont {Eremin}, \citenamefont
  {Rosenkranz}, \citenamefont {Chmaissem},\ and\ \citenamefont
  {Osborn}}]{Allred16}%
  \BibitemOpen
  \bibfield  {author} {\bibinfo {author} {\bibfnamefont {J.~M.}\ \bibnamefont
  {Allred}}, \bibinfo {author} {\bibfnamefont {K.~M.}\ \bibnamefont {Taddei}},
  \bibinfo {author} {\bibfnamefont {D.~E.}\ \bibnamefont {Bugaris}}, \bibinfo
  {author} {\bibfnamefont {M.~J.}\ \bibnamefont {Krogstad}}, \bibinfo {author}
  {\bibfnamefont {S.~H.}\ \bibnamefont {Lapidus}}, \bibinfo {author}
  {\bibfnamefont {D.~Y.}\ \bibnamefont {Chung}}, \bibinfo {author}
  {\bibfnamefont {H.}~\bibnamefont {Claus}}, \bibinfo {author} {\bibfnamefont
  {M.~G.}\ \bibnamefont {Kanatzidis}}, \bibinfo {author} {\bibfnamefont
  {D.~E.}\ \bibnamefont {Brown}}, \bibinfo {author} {\bibfnamefont
  {J.}~\bibnamefont {Kang}}, \bibinfo {author} {\bibfnamefont {R.~M.}\
  \bibnamefont {Fernandes}}, \bibinfo {author} {\bibfnamefont {I.}~\bibnamefont
  {Eremin}}, \bibinfo {author} {\bibfnamefont {S.}~\bibnamefont {Rosenkranz}},
  \bibinfo {author} {\bibfnamefont {O.}~\bibnamefont {Chmaissem}},\ and\
  \bibinfo {author} {\bibfnamefont {R.}~\bibnamefont {Osborn}},\ }\bibfield
  {title} {\bibinfo {title} {{Double-Q spin-density wave in iron arsenide
  superconductors}},\ }\href {https://doi.org/10.1038/nphys3629} {\bibfield
  {journal} {\bibinfo  {journal} {Nature Physics}\ }\textbf {\bibinfo {volume}
  {12}},\ \bibinfo {pages} {493} (\bibinfo {year} {2016})}\BibitemShut
  {NoStop}%
\bibitem [{\citenamefont {Christensen}\ \emph {et~al.}(2015)\citenamefont
  {Christensen}, \citenamefont {Kang}, \citenamefont {Andersen}, \citenamefont
  {Eremin},\ and\ \citenamefont {Fernandes}}]{Christensen15}%
  \BibitemOpen
  \bibfield  {author} {\bibinfo {author} {\bibfnamefont {M.~H.}\ \bibnamefont
  {Christensen}}, \bibinfo {author} {\bibfnamefont {J.}~\bibnamefont {Kang}},
  \bibinfo {author} {\bibfnamefont {B.~M.}\ \bibnamefont {Andersen}}, \bibinfo
  {author} {\bibfnamefont {I.}~\bibnamefont {Eremin}},\ and\ \bibinfo {author}
  {\bibfnamefont {R.~M.}\ \bibnamefont {Fernandes}},\ }\bibfield  {title}
  {\bibinfo {title} {{Spin reorientation driven by the interplay between
  spin-orbit coupling and Hund's rule coupling in iron pnictides}},\ }\href
  {https://doi.org/10.1103/PhysRevB.92.214509} {\bibfield  {journal} {\bibinfo
  {journal} {Phys. Rev. B}\ }\textbf {\bibinfo {volume} {92}},\ \bibinfo
  {pages} {214509} (\bibinfo {year} {2015})}\BibitemShut {NoStop}%
\bibitem [{\citenamefont {Qureshi}\ \emph {et~al.}(2012)\citenamefont
  {Qureshi}, \citenamefont {Steffens}, \citenamefont {Drees}, \citenamefont
  {Komarek}, \citenamefont {Lamago}, \citenamefont {Sidis}, \citenamefont
  {Harnagea}, \citenamefont {Grafe}, \citenamefont {Wurmehl}, \citenamefont
  {B\"uchner},\ and\ \citenamefont {Braden}}]{Braden12}%
  \BibitemOpen
  \bibfield  {author} {\bibinfo {author} {\bibfnamefont {N.}~\bibnamefont
  {Qureshi}}, \bibinfo {author} {\bibfnamefont {P.}~\bibnamefont {Steffens}},
  \bibinfo {author} {\bibfnamefont {Y.}~\bibnamefont {Drees}}, \bibinfo
  {author} {\bibfnamefont {A.~C.}\ \bibnamefont {Komarek}}, \bibinfo {author}
  {\bibfnamefont {D.}~\bibnamefont {Lamago}}, \bibinfo {author} {\bibfnamefont
  {Y.}~\bibnamefont {Sidis}}, \bibinfo {author} {\bibfnamefont
  {L.}~\bibnamefont {Harnagea}}, \bibinfo {author} {\bibfnamefont {H.-J.}\
  \bibnamefont {Grafe}}, \bibinfo {author} {\bibfnamefont {S.}~\bibnamefont
  {Wurmehl}}, \bibinfo {author} {\bibfnamefont {B.}~\bibnamefont {B\"uchner}},\
  and\ \bibinfo {author} {\bibfnamefont {M.}~\bibnamefont {Braden}},\
  }\bibfield  {title} {\bibinfo {title} {{Inelastic Neutron-Scattering
  Measurements of Incommensurate Magnetic Excitations on Superconducting LiFeAs
  Single Crystals}},\ }\href {https://doi.org/10.1103/PhysRevLett.108.117001}
  {\bibfield  {journal} {\bibinfo  {journal} {Phys. Rev. Lett.}\ }\textbf
  {\bibinfo {volume} {108}},\ \bibinfo {pages} {117001} (\bibinfo {year}
  {2012})}\BibitemShut {NoStop}%
\bibitem [{\citenamefont {Wang}\ \emph {et~al.}(2016)\citenamefont {Wang},
  \citenamefont {Shen}, \citenamefont {Pan}, \citenamefont {Zhang},
  \citenamefont {Ikeuchi}, \citenamefont {Iida}, \citenamefont {Christianson},
  \citenamefont {Walker}, \citenamefont {Adroja}, \citenamefont {Abdel-Hafiez},
  \citenamefont {Chen}, \citenamefont {Chareev}, \citenamefont {Vasiliev},\
  and\ \citenamefont {Zhao}}]{Wang2016}%
  \BibitemOpen
  \bibfield  {author} {\bibinfo {author} {\bibfnamefont {Q.}~\bibnamefont
  {Wang}}, \bibinfo {author} {\bibfnamefont {Y.}~\bibnamefont {Shen}}, \bibinfo
  {author} {\bibfnamefont {B.}~\bibnamefont {Pan}}, \bibinfo {author}
  {\bibfnamefont {X.}~\bibnamefont {Zhang}}, \bibinfo {author} {\bibfnamefont
  {K.}~\bibnamefont {Ikeuchi}}, \bibinfo {author} {\bibfnamefont
  {K.}~\bibnamefont {Iida}}, \bibinfo {author} {\bibfnamefont {A.}~\bibnamefont
  {Christianson}}, \bibinfo {author} {\bibfnamefont {H.}~\bibnamefont
  {Walker}}, \bibinfo {author} {\bibfnamefont {D.}~\bibnamefont {Adroja}},
  \bibinfo {author} {\bibfnamefont {M.}~\bibnamefont {Abdel-Hafiez}}, \bibinfo
  {author} {\bibfnamefont {X.}~\bibnamefont {Chen}}, \bibinfo {author}
  {\bibfnamefont {D.}~\bibnamefont {Chareev}}, \bibinfo {author} {\bibfnamefont
  {A.}~\bibnamefont {Vasiliev}},\ and\ \bibinfo {author} {\bibfnamefont
  {J.}~\bibnamefont {Zhao}},\ }\bibfield  {title} {\bibinfo {title} {{Magnetic
  ground state of FeSe}},\ }\href {https://doi.org/10.1038/ncomms12182}
  {\bibfield  {journal} {\bibinfo  {journal} {Nature Communications}\ }\textbf
  {\bibinfo {volume} {7}},\ \bibinfo {pages} {12182} (\bibinfo {year}
  {2016})}\BibitemShut {NoStop}%
\bibitem [{\citenamefont {Lumsden}\ \emph {et~al.}(2010)\citenamefont
  {Lumsden}, \citenamefont {Christianson}, \citenamefont {Goremychkin},
  \citenamefont {Nagler}, \citenamefont {Mook}, \citenamefont {Stone},
  \citenamefont {Abernathy}, \citenamefont {Guidi}, \citenamefont {MacDougall},
  \citenamefont {de~la Cruz}, \citenamefont {Sefat}, \citenamefont {McGuire},
  \citenamefont {Sales},\ and\ \citenamefont {Mandrus}}]{Mandrus10}%
  \BibitemOpen
  \bibfield  {author} {\bibinfo {author} {\bibfnamefont {M.~D.}\ \bibnamefont
  {Lumsden}}, \bibinfo {author} {\bibfnamefont {A.~D.}\ \bibnamefont
  {Christianson}}, \bibinfo {author} {\bibfnamefont {E.}~\bibnamefont
  {Goremychkin}}, \bibinfo {author} {\bibfnamefont {S.~E.}\ \bibnamefont
  {Nagler}}, \bibinfo {author} {\bibfnamefont {H.}~\bibnamefont {Mook}},
  \bibinfo {author} {\bibfnamefont {M.~B.}\ \bibnamefont {Stone}}, \bibinfo
  {author} {\bibfnamefont {D.~L.}\ \bibnamefont {Abernathy}}, \bibinfo {author}
  {\bibfnamefont {T.}~\bibnamefont {Guidi}}, \bibinfo {author} {\bibfnamefont
  {G.~J.}\ \bibnamefont {MacDougall}}, \bibinfo {author} {\bibfnamefont
  {C.}~\bibnamefont {de~la Cruz}}, \bibinfo {author} {\bibfnamefont
  {A.}~\bibnamefont {Sefat}}, \bibinfo {author} {\bibfnamefont
  {M.}~\bibnamefont {McGuire}}, \bibinfo {author} {\bibfnamefont
  {B.}~\bibnamefont {Sales}},\ and\ \bibinfo {author} {\bibfnamefont
  {D.}~\bibnamefont {Mandrus}},\ }\bibfield  {title} {\bibinfo {title}
  {{Evolution of spin excitations into the superconducting state in
  FeTe$_{1-x}$Se$_x$}},\ }\href {https://doi.org/10.1038/nphys1512} {\bibfield
  {journal} {\bibinfo  {journal} {Nature Physics}\ }\textbf {\bibinfo {volume}
  {6}},\ \bibinfo {pages} {182} (\bibinfo {year} {2010})}\BibitemShut {NoStop}%
\bibitem [{\citenamefont {Liu}\ \emph {et~al.}(2010)\citenamefont {Liu},
  \citenamefont {Hu}, \citenamefont {Qian}, \citenamefont {Fobes},
  \citenamefont {Mao}, \citenamefont {Bao}, \citenamefont {Reehuis},
  \citenamefont {Kimber}, \citenamefont {Proke{\v{s}}}, \citenamefont {Matas},
  \citenamefont {Argyriou}, \citenamefont {Hiess}, \citenamefont {Rotaru},
  \citenamefont {Pham}, \citenamefont {Spinu}, \citenamefont {Qiu},
  \citenamefont {Thampy}, \citenamefont {Savici}, \citenamefont {Rodriguez},\
  and\ \citenamefont {Broholm}}]{Broholm10}%
  \BibitemOpen
  \bibfield  {author} {\bibinfo {author} {\bibfnamefont {T.}~\bibnamefont
  {Liu}}, \bibinfo {author} {\bibfnamefont {J.}~\bibnamefont {Hu}}, \bibinfo
  {author} {\bibfnamefont {B.}~\bibnamefont {Qian}}, \bibinfo {author}
  {\bibfnamefont {D.}~\bibnamefont {Fobes}}, \bibinfo {author} {\bibfnamefont
  {Z.~Q.}\ \bibnamefont {Mao}}, \bibinfo {author} {\bibfnamefont
  {W.}~\bibnamefont {Bao}}, \bibinfo {author} {\bibfnamefont {M.}~\bibnamefont
  {Reehuis}}, \bibinfo {author} {\bibfnamefont {S.}~\bibnamefont {Kimber}},
  \bibinfo {author} {\bibfnamefont {K.}~\bibnamefont {Proke{\v{s}}}}, \bibinfo
  {author} {\bibfnamefont {S.}~\bibnamefont {Matas}}, \bibinfo {author}
  {\bibfnamefont {D.}~\bibnamefont {Argyriou}}, \bibinfo {author}
  {\bibfnamefont {A.}~\bibnamefont {Hiess}}, \bibinfo {author} {\bibfnamefont
  {A.}~\bibnamefont {Rotaru}}, \bibinfo {author} {\bibfnamefont
  {H.}~\bibnamefont {Pham}}, \bibinfo {author} {\bibfnamefont {L.}~\bibnamefont
  {Spinu}}, \bibinfo {author} {\bibfnamefont {Y.}~\bibnamefont {Qiu}}, \bibinfo
  {author} {\bibfnamefont {V.}~\bibnamefont {Thampy}}, \bibinfo {author}
  {\bibfnamefont {A.}~\bibnamefont {Savici}}, \bibinfo {author} {\bibfnamefont
  {J.}~\bibnamefont {Rodriguez}},\ and\ \bibinfo {author} {\bibfnamefont
  {C.}~\bibnamefont {Broholm}},\ }\bibfield  {title} {\bibinfo {title} {{From
  ($\pi$, 0) magnetic order to superconductivity with ($\pi$, $\pi$) magnetic
  resonance in Fe$_{1.02}$Te$_{1-x}$Se$_x$}},\ }\href
  {https://doi.org/10.1038/nmat2800} {\bibfield  {journal} {\bibinfo  {journal}
  {Nature Materials}\ }\textbf {\bibinfo {volume} {9}},\ \bibinfo {pages} {718}
  (\bibinfo {year} {2010})}\BibitemShut {NoStop}%
\bibitem [{\citenamefont {Gastiasoro}\ and\ \citenamefont
  {Andersen}(2014)}]{Gastiasoro14}%
  \BibitemOpen
  \bibfield  {author} {\bibinfo {author} {\bibfnamefont {M.~N.}\ \bibnamefont
  {Gastiasoro}}\ and\ \bibinfo {author} {\bibfnamefont {B.~M.}\ \bibnamefont
  {Andersen}},\ }\bibfield  {title} {\bibinfo {title} {Enhancement of magnetic
  stripe order in iron-pnictide superconductors from the interaction between
  conduction electrons and magnetic impurities},\ }\href
  {https://doi.org/10.1103/PhysRevLett.113.067002} {\bibfield  {journal}
  {\bibinfo  {journal} {Phys. Rev. Lett.}\ }\textbf {\bibinfo {volume} {113}},\
  \bibinfo {pages} {067002} (\bibinfo {year} {2014})}\BibitemShut {NoStop}%
\bibitem [{\citenamefont {Pratt}\ \emph {et~al.}(2011)\citenamefont {Pratt},
  \citenamefont {Kim}, \citenamefont {Kreyssig}, \citenamefont {Lee},
  \citenamefont {Tucker}, \citenamefont {Thaler}, \citenamefont {Tian},
  \citenamefont {Zarestky}, \citenamefont {Bud'ko}, \citenamefont {Canfield},
  \citenamefont {Harmon}, \citenamefont {Goldman},\ and\ \citenamefont
  {McQueeney}}]{Pratt11}%
  \BibitemOpen
  \bibfield  {author} {\bibinfo {author} {\bibfnamefont {D.~K.}\ \bibnamefont
  {Pratt}}, \bibinfo {author} {\bibfnamefont {M.~G.}\ \bibnamefont {Kim}},
  \bibinfo {author} {\bibfnamefont {A.}~\bibnamefont {Kreyssig}}, \bibinfo
  {author} {\bibfnamefont {Y.~B.}\ \bibnamefont {Lee}}, \bibinfo {author}
  {\bibfnamefont {G.~S.}\ \bibnamefont {Tucker}}, \bibinfo {author}
  {\bibfnamefont {A.}~\bibnamefont {Thaler}}, \bibinfo {author} {\bibfnamefont
  {W.}~\bibnamefont {Tian}}, \bibinfo {author} {\bibfnamefont {J.~L.}\
  \bibnamefont {Zarestky}}, \bibinfo {author} {\bibfnamefont {S.~L.}\
  \bibnamefont {Bud'ko}}, \bibinfo {author} {\bibfnamefont {P.~C.}\
  \bibnamefont {Canfield}}, \bibinfo {author} {\bibfnamefont {B.~N.}\
  \bibnamefont {Harmon}}, \bibinfo {author} {\bibfnamefont {A.~I.}\
  \bibnamefont {Goldman}},\ and\ \bibinfo {author} {\bibfnamefont {R.~J.}\
  \bibnamefont {McQueeney}},\ }\bibfield  {title} {\bibinfo {title}
  {{Incommensurate Spin-Density Wave Order in Electron-Doped
  ${\mathrm{BaFe}}_{2}{\mathrm{As}}_{2}$ Superconductors}},\ }\href
  {https://doi.org/10.1103/PhysRevLett.106.257001} {\bibfield  {journal}
  {\bibinfo  {journal} {Phys. Rev. Lett.}\ }\textbf {\bibinfo {volume} {106}},\
  \bibinfo {pages} {257001} (\bibinfo {year} {2011})}\BibitemShut {NoStop}%
\bibitem [{\citenamefont {Sheveleva}\ \emph {et~al.}(2020)\citenamefont
  {Sheveleva}, \citenamefont {Xu}, \citenamefont {Marsik}, \citenamefont
  {Lyzwa}, \citenamefont {Mallett}, \citenamefont {Willa}, \citenamefont
  {Meingast}, \citenamefont {Wolf}, \citenamefont {Shevtsova}, \citenamefont
  {Pashkevich},\ and\ \citenamefont {Bernhard}}]{Sheveleva20}%
  \BibitemOpen
  \bibfield  {author} {\bibinfo {author} {\bibfnamefont {E.}~\bibnamefont
  {Sheveleva}}, \bibinfo {author} {\bibfnamefont {B.}~\bibnamefont {Xu}},
  \bibinfo {author} {\bibfnamefont {P.}~\bibnamefont {Marsik}}, \bibinfo
  {author} {\bibfnamefont {F.}~\bibnamefont {Lyzwa}}, \bibinfo {author}
  {\bibfnamefont {B.~P.~P.}\ \bibnamefont {Mallett}}, \bibinfo {author}
  {\bibfnamefont {K.}~\bibnamefont {Willa}}, \bibinfo {author} {\bibfnamefont
  {C.}~\bibnamefont {Meingast}}, \bibinfo {author} {\bibfnamefont
  {T.}~\bibnamefont {Wolf}}, \bibinfo {author} {\bibfnamefont {T.}~\bibnamefont
  {Shevtsova}}, \bibinfo {author} {\bibfnamefont {Y.~G.}\ \bibnamefont
  {Pashkevich}},\ and\ \bibinfo {author} {\bibfnamefont {C.}~\bibnamefont
  {Bernhard}},\ }\bibfield  {title} {\bibinfo {title} {{Muon spin rotation and
  infrared spectroscopy study of
  ${\mathrm{Ba}}_{1\ensuremath{-}x}{\mathrm{Na}}_{x}{\mathrm{Fe}}_{2}{\mathrm{As}}_{2}$}},\
  }\href {https://doi.org/10.1103/PhysRevB.101.224515} {\bibfield  {journal}
  {\bibinfo  {journal} {Phys. Rev. B}\ }\textbf {\bibinfo {volume} {101}},\
  \bibinfo {pages} {224515} (\bibinfo {year} {2020})}\BibitemShut {NoStop}%
\bibitem [{\citenamefont {Lorenzana}\ \emph {et~al.}(2008)\citenamefont
  {Lorenzana}, \citenamefont {Seibold}, \citenamefont {Ortix},\ and\
  \citenamefont {Grilli}}]{Lorenzana08}%
  \BibitemOpen
  \bibfield  {author} {\bibinfo {author} {\bibfnamefont {J.}~\bibnamefont
  {Lorenzana}}, \bibinfo {author} {\bibfnamefont {G.}~\bibnamefont {Seibold}},
  \bibinfo {author} {\bibfnamefont {C.}~\bibnamefont {Ortix}},\ and\ \bibinfo
  {author} {\bibfnamefont {M.}~\bibnamefont {Grilli}},\ }\bibfield  {title}
  {\bibinfo {title} {{Competing Orders in FeAs Layers}},\ }\href
  {https://doi.org/10.1103/PhysRevLett.101.186402} {\bibfield  {journal}
  {\bibinfo  {journal} {Phys. Rev. Lett.}\ }\textbf {\bibinfo {volume} {101}},\
  \bibinfo {pages} {186402} (\bibinfo {year} {2008})}\BibitemShut {NoStop}%
\bibitem [{\citenamefont {Fernandes}\ \emph {et~al.}(2016)\citenamefont
  {Fernandes}, \citenamefont {Kivelson},\ and\ \citenamefont
  {Berg}}]{RMF_Berg}%
  \BibitemOpen
  \bibfield  {author} {\bibinfo {author} {\bibfnamefont {R.~M.}\ \bibnamefont
  {Fernandes}}, \bibinfo {author} {\bibfnamefont {S.~A.}\ \bibnamefont
  {Kivelson}},\ and\ \bibinfo {author} {\bibfnamefont {E.}~\bibnamefont
  {Berg}},\ }\bibfield  {title} {\bibinfo {title} {{Vestigial chiral and charge
  orders from bidirectional spin-density waves: Application to the iron-based
  superconductors}},\ }\href {https://doi.org/10.1103/PhysRevB.93.014511}
  {\bibfield  {journal} {\bibinfo  {journal} {Phys. Rev. B}\ }\textbf {\bibinfo
  {volume} {93}},\ \bibinfo {pages} {014511} (\bibinfo {year}
  {2016})}\BibitemShut {NoStop}%
\bibitem [{\citenamefont {Seo}\ \emph {et~al.}(2008)\citenamefont {Seo},
  \citenamefont {Bernevig},\ and\ \citenamefont {Hu}}]{JPHu08}%
  \BibitemOpen
  \bibfield  {author} {\bibinfo {author} {\bibfnamefont {K.}~\bibnamefont
  {Seo}}, \bibinfo {author} {\bibfnamefont {B.~A.}\ \bibnamefont {Bernevig}},\
  and\ \bibinfo {author} {\bibfnamefont {J.}~\bibnamefont {Hu}},\ }\bibfield
  {title} {\bibinfo {title} {Pairing symmetry in a two-orbital exchange
  coupling model of oxypnictides},\ }\href
  {https://doi.org/10.1103/PhysRevLett.101.206404} {\bibfield  {journal}
  {\bibinfo  {journal} {Phys. Rev. Lett.}\ }\textbf {\bibinfo {volume} {101}},\
  \bibinfo {pages} {206404} (\bibinfo {year} {2008})}\BibitemShut {NoStop}%
\bibitem [{\citenamefont {Dai}\ \emph {et~al.}(2012)\citenamefont {Dai},
  \citenamefont {Hu},\ and\ \citenamefont {Dagotto}}]{Dagotto_review}%
  \BibitemOpen
  \bibfield  {author} {\bibinfo {author} {\bibfnamefont {P.}~\bibnamefont
  {Dai}}, \bibinfo {author} {\bibfnamefont {J.}~\bibnamefont {Hu}},\ and\
  \bibinfo {author} {\bibfnamefont {E.}~\bibnamefont {Dagotto}},\ }\bibfield
  {title} {\bibinfo {title} {Magnetism and its microscopic origin in iron-based
  high-temperature superconductors},\ }\href
  {https://doi.org/10.1038/nphys2438} {\bibfield  {journal} {\bibinfo
  {journal} {Nature Physics}\ }\textbf {\bibinfo {volume} {8}},\ \bibinfo
  {pages} {709} (\bibinfo {year} {2012})}\BibitemShut {NoStop}%
\bibitem [{\citenamefont {Eremin}\ and\ \citenamefont
  {Chubukov}(2010)}]{Eremin11}%
  \BibitemOpen
  \bibfield  {author} {\bibinfo {author} {\bibfnamefont {I.}~\bibnamefont
  {Eremin}}\ and\ \bibinfo {author} {\bibfnamefont {A.~V.}\ \bibnamefont
  {Chubukov}},\ }\bibfield  {title} {\bibinfo {title} {Magnetic degeneracy and
  hidden metallicity of the spin-density-wave state in ferropnictides},\ }\href
  {https://doi.org/10.1103/PhysRevB.81.024511} {\bibfield  {journal} {\bibinfo
  {journal} {Phys. Rev. B}\ }\textbf {\bibinfo {volume} {81}},\ \bibinfo
  {pages} {024511} (\bibinfo {year} {2010})}\BibitemShut {NoStop}%
\bibitem [{\citenamefont {Vorontsov}\ \emph {et~al.}(2010)\citenamefont
  {Vorontsov}, \citenamefont {Vavilov},\ and\ \citenamefont
  {Chubukov}}]{Vorontsov10}%
  \BibitemOpen
  \bibfield  {author} {\bibinfo {author} {\bibfnamefont {A.~B.}\ \bibnamefont
  {Vorontsov}}, \bibinfo {author} {\bibfnamefont {M.~G.}\ \bibnamefont
  {Vavilov}},\ and\ \bibinfo {author} {\bibfnamefont {A.~V.}\ \bibnamefont
  {Chubukov}},\ }\bibfield  {title} {\bibinfo {title} {Superconductivity and
  spin-density waves in multiband metals},\ }\href
  {https://doi.org/10.1103/PhysRevB.81.174538} {\bibfield  {journal} {\bibinfo
  {journal} {Phys. Rev. B}\ }\textbf {\bibinfo {volume} {81}},\ \bibinfo
  {pages} {174538} (\bibinfo {year} {2010})}\BibitemShut {NoStop}%
\bibitem [{\citenamefont {Yildirim}(2008)}]{Yildirim08}%
  \BibitemOpen
  \bibfield  {author} {\bibinfo {author} {\bibfnamefont {T.}~\bibnamefont
  {Yildirim}},\ }\bibfield  {title} {\bibinfo {title} {{Origin of the 150-K
  Anomaly in LaFeAsO: Competing Antiferromagnetic Interactions, Frustration,
  and a Structural Phase Transition}},\ }\href
  {https://doi.org/10.1103/PhysRevLett.101.057010} {\bibfield  {journal}
  {\bibinfo  {journal} {Phys. Rev. Lett.}\ }\textbf {\bibinfo {volume} {101}},\
  \bibinfo {pages} {057010} (\bibinfo {year} {2008})}\BibitemShut {NoStop}%
\bibitem [{\citenamefont {Glasbrenner}\ \emph {et~al.}(2015)\citenamefont
  {Glasbrenner}, \citenamefont {Mazin}, \citenamefont {Jeschke}, \citenamefont
  {Hirschfeld}, \citenamefont {Fernandes},\ and\ \citenamefont
  {Valent{\'\i}}}]{Glasbrenner15}%
  \BibitemOpen
  \bibfield  {author} {\bibinfo {author} {\bibfnamefont {J.}~\bibnamefont
  {Glasbrenner}}, \bibinfo {author} {\bibfnamefont {I.}~\bibnamefont {Mazin}},
  \bibinfo {author} {\bibfnamefont {H.~O.}\ \bibnamefont {Jeschke}}, \bibinfo
  {author} {\bibfnamefont {P.}~\bibnamefont {Hirschfeld}}, \bibinfo {author}
  {\bibfnamefont {R.}~\bibnamefont {Fernandes}},\ and\ \bibinfo {author}
  {\bibfnamefont {R.}~\bibnamefont {Valent{\'\i}}},\ }\bibfield  {title}
  {\bibinfo {title} {{Effect of magnetic frustration on nematicity and
  superconductivity in iron chalcogenides}},\ }\href
  {https://doi.org/10.1038/nphys3434} {\bibfield  {journal} {\bibinfo
  {journal} {Nature Physics}\ }\textbf {\bibinfo {volume} {11}},\ \bibinfo
  {pages} {953} (\bibinfo {year} {2015})}\BibitemShut {NoStop}%
\bibitem [{\citenamefont {Hirayama}\ \emph {et~al.}(2015)\citenamefont
  {Hirayama}, \citenamefont {Misawa}, \citenamefont {Miyake},\ and\
  \citenamefont {Imada}}]{Hirayama15}%
  \BibitemOpen
  \bibfield  {author} {\bibinfo {author} {\bibfnamefont {M.}~\bibnamefont
  {Hirayama}}, \bibinfo {author} {\bibfnamefont {T.}~\bibnamefont {Misawa}},
  \bibinfo {author} {\bibfnamefont {T.}~\bibnamefont {Miyake}},\ and\ \bibinfo
  {author} {\bibfnamefont {M.}~\bibnamefont {Imada}},\ }\bibfield  {title}
  {\bibinfo {title} {Ab initio studies of magnetism in the iron chalcogenides
  fete and fese},\ }\href {https://doi.org/10.7566/JPSJ.84.093703} {\bibfield
  {journal} {\bibinfo  {journal} {Journal of the Physical Society of Japan}\
  }\textbf {\bibinfo {volume} {84}},\ \bibinfo {pages} {093703} (\bibinfo
  {year} {2015})}\BibitemShut {NoStop}%
\bibitem [{\citenamefont {Abrahams}\ and\ \citenamefont
  {Si}(2011)}]{Abrahams11}%
  \BibitemOpen
  \bibfield  {author} {\bibinfo {author} {\bibfnamefont {E.}~\bibnamefont
  {Abrahams}}\ and\ \bibinfo {author} {\bibfnamefont {Q.}~\bibnamefont {Si}},\
  }\bibfield  {title} {\bibinfo {title} {Quantum criticality in the iron
  pnictides and chalcogenides},\ }\href
  {https://doi.org/10.1088/0953-8984/23/22/223201} {\bibfield  {journal}
  {\bibinfo  {journal} {Journal of Physics: Condensed Matter}\ }\textbf
  {\bibinfo {volume} {23}},\ \bibinfo {pages} {223201} (\bibinfo {year}
  {2011})}\BibitemShut {NoStop}%
\bibitem [{\citenamefont {Hayes}\ \emph {et~al.}(2016)\citenamefont {Hayes},
  \citenamefont {McDonald}, \citenamefont {Breznay}, \citenamefont {Helm},
  \citenamefont {Moll}, \citenamefont {Wartenbe}, \citenamefont {Shekhter},\
  and\ \citenamefont {Analytis}}]{Hayes16}%
  \BibitemOpen
  \bibfield  {author} {\bibinfo {author} {\bibfnamefont {I.~M.}\ \bibnamefont
  {Hayes}}, \bibinfo {author} {\bibfnamefont {R.~D.}\ \bibnamefont {McDonald}},
  \bibinfo {author} {\bibfnamefont {N.~P.}\ \bibnamefont {Breznay}}, \bibinfo
  {author} {\bibfnamefont {T.}~\bibnamefont {Helm}}, \bibinfo {author}
  {\bibfnamefont {P.~J.}\ \bibnamefont {Moll}}, \bibinfo {author}
  {\bibfnamefont {M.}~\bibnamefont {Wartenbe}}, \bibinfo {author}
  {\bibfnamefont {A.}~\bibnamefont {Shekhter}},\ and\ \bibinfo {author}
  {\bibfnamefont {J.~G.}\ \bibnamefont {Analytis}},\ }\bibfield  {title}
  {\bibinfo {title} {{Scaling between magnetic field and temperature in the
  high-temperature superconductor BaFe$_2$ (As$_{1-x}$P$_x$)$_2$}},\ }\href
  {https://doi.org/10.1038/nphys3773} {\bibfield  {journal} {\bibinfo
  {journal} {Nature Physics}\ }\textbf {\bibinfo {volume} {12}},\ \bibinfo
  {pages} {916} (\bibinfo {year} {2016})}\BibitemShut {NoStop}%
\bibitem [{\citenamefont {Chowdhury}\ \emph {et~al.}(2013)\citenamefont
  {Chowdhury}, \citenamefont {Swingle}, \citenamefont {Berg},\ and\
  \citenamefont {Sachdev}}]{Chowdhury13}%
  \BibitemOpen
  \bibfield  {author} {\bibinfo {author} {\bibfnamefont {D.}~\bibnamefont
  {Chowdhury}}, \bibinfo {author} {\bibfnamefont {B.}~\bibnamefont {Swingle}},
  \bibinfo {author} {\bibfnamefont {E.}~\bibnamefont {Berg}},\ and\ \bibinfo
  {author} {\bibfnamefont {S.}~\bibnamefont {Sachdev}},\ }\bibfield  {title}
  {\bibinfo {title} {{Singularity of the London Penetration Depth at Quantum
  Critical Points in Superconductors}},\ }\href
  {https://doi.org/10.1103/PhysRevLett.111.157004} {\bibfield  {journal}
  {\bibinfo  {journal} {Phys. Rev. Lett.}\ }\textbf {\bibinfo {volume} {111}},\
  \bibinfo {pages} {157004} (\bibinfo {year} {2013})}\BibitemShut {NoStop}%
\bibitem [{\citenamefont {Levchenko}\ \emph {et~al.}(2013)\citenamefont
  {Levchenko}, \citenamefont {Vavilov}, \citenamefont {Khodas},\ and\
  \citenamefont {Chubukov}}]{Levchenko13}%
  \BibitemOpen
  \bibfield  {author} {\bibinfo {author} {\bibfnamefont {A.}~\bibnamefont
  {Levchenko}}, \bibinfo {author} {\bibfnamefont {M.~G.}\ \bibnamefont
  {Vavilov}}, \bibinfo {author} {\bibfnamefont {M.}~\bibnamefont {Khodas}},\
  and\ \bibinfo {author} {\bibfnamefont {A.~V.}\ \bibnamefont {Chubukov}},\
  }\bibfield  {title} {\bibinfo {title} {{Enhancement of the London Penetration
  Depth in Pnictides at the Onset of Spin-Density-Wave Order under
  Superconducting Dome}},\ }\href
  {https://doi.org/10.1103/PhysRevLett.110.177003} {\bibfield  {journal}
  {\bibinfo  {journal} {Phys. Rev. Lett.}\ }\textbf {\bibinfo {volume} {110}},\
  \bibinfo {pages} {177003} (\bibinfo {year} {2013})}\BibitemShut {NoStop}%
\bibitem [{\citenamefont {Kasahara}\ \emph {et~al.}(2010)\citenamefont
  {Kasahara}, \citenamefont {Shibauchi}, \citenamefont {Hashimoto},
  \citenamefont {Ikada}, \citenamefont {Tonegawa}, \citenamefont {Okazaki},
  \citenamefont {Shishido}, \citenamefont {Ikeda}, \citenamefont {Takeya},
  \citenamefont {Hirata}, \citenamefont {Terashima},\ and\ \citenamefont
  {Matsuda}}]{Kasahara10}%
  \BibitemOpen
  \bibfield  {author} {\bibinfo {author} {\bibfnamefont {S.}~\bibnamefont
  {Kasahara}}, \bibinfo {author} {\bibfnamefont {T.}~\bibnamefont {Shibauchi}},
  \bibinfo {author} {\bibfnamefont {K.}~\bibnamefont {Hashimoto}}, \bibinfo
  {author} {\bibfnamefont {K.}~\bibnamefont {Ikada}}, \bibinfo {author}
  {\bibfnamefont {S.}~\bibnamefont {Tonegawa}}, \bibinfo {author}
  {\bibfnamefont {R.}~\bibnamefont {Okazaki}}, \bibinfo {author} {\bibfnamefont
  {H.}~\bibnamefont {Shishido}}, \bibinfo {author} {\bibfnamefont
  {H.}~\bibnamefont {Ikeda}}, \bibinfo {author} {\bibfnamefont
  {H.}~\bibnamefont {Takeya}}, \bibinfo {author} {\bibfnamefont
  {K.}~\bibnamefont {Hirata}}, \bibinfo {author} {\bibfnamefont
  {T.}~\bibnamefont {Terashima}},\ and\ \bibinfo {author} {\bibfnamefont
  {Y.}~\bibnamefont {Matsuda}},\ }\bibfield  {title} {\bibinfo {title}
  {{Evolution from non-Fermi- to Fermi-liquid transport via isovalent doping in
  ${\text{BaFe}}_{2}{({\text{As}}_{1\ensuremath{-}x}{\text{P}}_{x})}_{2}$
  superconductors}},\ }\href {https://doi.org/10.1103/PhysRevB.81.184519}
  {\bibfield  {journal} {\bibinfo  {journal} {Phys. Rev. B}\ }\textbf {\bibinfo
  {volume} {81}},\ \bibinfo {pages} {184519} (\bibinfo {year}
  {2010})}\BibitemShut {NoStop}%
\bibitem [{\citenamefont {Lu}\ \emph {et~al.}(2014)\citenamefont {Lu},
  \citenamefont {Park}, \citenamefont {Zhang}, \citenamefont {Luo},
  \citenamefont {Nevidomskyy}, \citenamefont {Si},\ and\ \citenamefont
  {Dai}}]{PCDai14}%
  \BibitemOpen
  \bibfield  {author} {\bibinfo {author} {\bibfnamefont {X.}~\bibnamefont
  {Lu}}, \bibinfo {author} {\bibfnamefont {J.}~\bibnamefont {Park}}, \bibinfo
  {author} {\bibfnamefont {R.}~\bibnamefont {Zhang}}, \bibinfo {author}
  {\bibfnamefont {H.}~\bibnamefont {Luo}}, \bibinfo {author} {\bibfnamefont
  {A.~H.}\ \bibnamefont {Nevidomskyy}}, \bibinfo {author} {\bibfnamefont
  {Q.}~\bibnamefont {Si}},\ and\ \bibinfo {author} {\bibfnamefont
  {P.}~\bibnamefont {Dai}},\ }\bibfield  {title} {\bibinfo {title} {{Nematic
  spin correlations in the tetragonal state of uniaxial-strained
  BaFe$_{2-x}$Ni$_x$As$_2$}},\ }\href {https://doi.org/10.1126/science.1251853}
  {\bibfield  {journal} {\bibinfo  {journal} {Science}\ }\textbf {\bibinfo
  {volume} {345}},\ \bibinfo {pages} {657} (\bibinfo {year}
  {2014})}\BibitemShut {NoStop}%
\bibitem [{\citenamefont {Chu}\ \emph {et~al.}(2010)\citenamefont {Chu},
  \citenamefont {Analytis}, \citenamefont {De~Greve}, \citenamefont {McMahon},
  \citenamefont {Islam}, \citenamefont {Yamamoto},\ and\ \citenamefont
  {Fisher}}]{Chu10}%
  \BibitemOpen
  \bibfield  {author} {\bibinfo {author} {\bibfnamefont {J.-H.}\ \bibnamefont
  {Chu}}, \bibinfo {author} {\bibfnamefont {J.~G.}\ \bibnamefont {Analytis}},
  \bibinfo {author} {\bibfnamefont {K.}~\bibnamefont {De~Greve}}, \bibinfo
  {author} {\bibfnamefont {P.~L.}\ \bibnamefont {McMahon}}, \bibinfo {author}
  {\bibfnamefont {Z.}~\bibnamefont {Islam}}, \bibinfo {author} {\bibfnamefont
  {Y.}~\bibnamefont {Yamamoto}},\ and\ \bibinfo {author} {\bibfnamefont
  {I.~R.}\ \bibnamefont {Fisher}},\ }\bibfield  {title} {\bibinfo {title}
  {{In-Plane Resistivity Anisotropy in an Underdoped Iron Arsenide
  Superconductor}},\ }\href {https://doi.org/10.1126/science.1190482}
  {\bibfield  {journal} {\bibinfo  {journal} {Science}\ }\textbf {\bibinfo
  {volume} {329}},\ \bibinfo {pages} {824} (\bibinfo {year}
  {2010})}\BibitemShut {NoStop}%
\bibitem [{\citenamefont {Tanatar}\ \emph {et~al.}(2010)\citenamefont
  {Tanatar}, \citenamefont {Blomberg}, \citenamefont {Kreyssig}, \citenamefont
  {Kim}, \citenamefont {Ni}, \citenamefont {Thaler}, \citenamefont {Bud'ko},
  \citenamefont {Canfield}, \citenamefont {Goldman}, \citenamefont {Mazin},\
  and\ \citenamefont {Prozorov}}]{Tanatar10}%
  \BibitemOpen
  \bibfield  {author} {\bibinfo {author} {\bibfnamefont {M.~A.}\ \bibnamefont
  {Tanatar}}, \bibinfo {author} {\bibfnamefont {E.~C.}\ \bibnamefont
  {Blomberg}}, \bibinfo {author} {\bibfnamefont {A.}~\bibnamefont {Kreyssig}},
  \bibinfo {author} {\bibfnamefont {M.~G.}\ \bibnamefont {Kim}}, \bibinfo
  {author} {\bibfnamefont {N.}~\bibnamefont {Ni}}, \bibinfo {author}
  {\bibfnamefont {A.}~\bibnamefont {Thaler}}, \bibinfo {author} {\bibfnamefont
  {S.~L.}\ \bibnamefont {Bud'ko}}, \bibinfo {author} {\bibfnamefont {P.~C.}\
  \bibnamefont {Canfield}}, \bibinfo {author} {\bibfnamefont {A.~I.}\
  \bibnamefont {Goldman}}, \bibinfo {author} {\bibfnamefont {I.~I.}\
  \bibnamefont {Mazin}},\ and\ \bibinfo {author} {\bibfnamefont
  {R.}~\bibnamefont {Prozorov}},\ }\bibfield  {title} {\bibinfo {title}
  {{Uniaxial-strain mechanical detwinning of ${\text{CaFe}}_{2}{\text{As}}_{2}$
  and ${\text{BaFe}}_{2}{\text{As}}_{2}$ crystals: Optical and transport
  study}},\ }\href {https://doi.org/10.1103/PhysRevB.81.184508} {\bibfield
  {journal} {\bibinfo  {journal} {Phys. Rev. B}\ }\textbf {\bibinfo {volume}
  {81}},\ \bibinfo {pages} {184508} (\bibinfo {year} {2010})}\BibitemShut
  {NoStop}%
\bibitem [{\citenamefont {Mirri}\ \emph {et~al.}(2015)\citenamefont {Mirri},
  \citenamefont {Dusza}, \citenamefont {Bastelberger}, \citenamefont
  {Chinotti}, \citenamefont {Degiorgi}, \citenamefont {Chu}, \citenamefont
  {Kuo},\ and\ \citenamefont {Fisher}}]{Degiorgi15}%
  \BibitemOpen
  \bibfield  {author} {\bibinfo {author} {\bibfnamefont {C.}~\bibnamefont
  {Mirri}}, \bibinfo {author} {\bibfnamefont {A.}~\bibnamefont {Dusza}},
  \bibinfo {author} {\bibfnamefont {S.}~\bibnamefont {Bastelberger}}, \bibinfo
  {author} {\bibfnamefont {M.}~\bibnamefont {Chinotti}}, \bibinfo {author}
  {\bibfnamefont {L.}~\bibnamefont {Degiorgi}}, \bibinfo {author}
  {\bibfnamefont {J.-H.}\ \bibnamefont {Chu}}, \bibinfo {author} {\bibfnamefont
  {H.-H.}\ \bibnamefont {Kuo}},\ and\ \bibinfo {author} {\bibfnamefont {I.~R.}\
  \bibnamefont {Fisher}},\ }\bibfield  {title} {\bibinfo {title} {{Origin of
  the Resistive Anisotropy in the Electronic Nematic Phase of
  ${\mathrm{BaFe}}_{2}{\mathrm{As}}_{2}$ Revealed by Optical Spectroscopy}},\
  }\href {https://doi.org/10.1103/PhysRevLett.115.107001} {\bibfield  {journal}
  {\bibinfo  {journal} {Phys. Rev. Lett.}\ }\textbf {\bibinfo {volume} {115}},\
  \bibinfo {pages} {107001} (\bibinfo {year} {2015})}\BibitemShut {NoStop}%
\bibitem [{\citenamefont {Chuang}\ \emph {et~al.}(2010)\citenamefont {Chuang},
  \citenamefont {Allan}, \citenamefont {Lee}, \citenamefont {Xie},
  \citenamefont {Ni}, \citenamefont {Bud'ko}, \citenamefont {Boebinger},
  \citenamefont {Canfield},\ and\ \citenamefont {Davis}}]{JCDavis10}%
  \BibitemOpen
  \bibfield  {author} {\bibinfo {author} {\bibfnamefont {T.-M.}\ \bibnamefont
  {Chuang}}, \bibinfo {author} {\bibfnamefont {M.~P.}\ \bibnamefont {Allan}},
  \bibinfo {author} {\bibfnamefont {J.}~\bibnamefont {Lee}}, \bibinfo {author}
  {\bibfnamefont {Y.}~\bibnamefont {Xie}}, \bibinfo {author} {\bibfnamefont
  {N.}~\bibnamefont {Ni}}, \bibinfo {author} {\bibfnamefont {S.~L.}\
  \bibnamefont {Bud'ko}}, \bibinfo {author} {\bibfnamefont {G.~S.}\
  \bibnamefont {Boebinger}}, \bibinfo {author} {\bibfnamefont {P.~C.}\
  \bibnamefont {Canfield}},\ and\ \bibinfo {author} {\bibfnamefont {J.~C.}\
  \bibnamefont {Davis}},\ }\bibfield  {title} {\bibinfo {title} {{Nematic
  Electronic Structure in the {\textquotedblleft}Parent{\textquotedblright}
  State of the Iron-Based Superconductor Ca(Fe$_{1-x}$Co$_x$)$_2$As$_2$}},\
  }\href {https://doi.org/10.1126/science.1181083} {\bibfield  {journal}
  {\bibinfo  {journal} {Science}\ }\textbf {\bibinfo {volume} {327}},\ \bibinfo
  {pages} {181} (\bibinfo {year} {2010})}\BibitemShut {NoStop}%
\bibitem [{\citenamefont {Rosenthal}\ \emph {et~al.}(2014)\citenamefont
  {Rosenthal}, \citenamefont {Andrade}, \citenamefont {Arguello}, \citenamefont
  {Fernandes}, \citenamefont {Xing}, \citenamefont {Wang}, \citenamefont {Jin},
  \citenamefont {Millis},\ and\ \citenamefont {Pasupathy}}]{Rosenthal14}%
  \BibitemOpen
  \bibfield  {author} {\bibinfo {author} {\bibfnamefont {E.~P.}\ \bibnamefont
  {Rosenthal}}, \bibinfo {author} {\bibfnamefont {E.~F.}\ \bibnamefont
  {Andrade}}, \bibinfo {author} {\bibfnamefont {C.~J.}\ \bibnamefont
  {Arguello}}, \bibinfo {author} {\bibfnamefont {R.~M.}\ \bibnamefont
  {Fernandes}}, \bibinfo {author} {\bibfnamefont {L.~Y.}\ \bibnamefont {Xing}},
  \bibinfo {author} {\bibfnamefont {X.}~\bibnamefont {Wang}}, \bibinfo {author}
  {\bibfnamefont {C.}~\bibnamefont {Jin}}, \bibinfo {author} {\bibfnamefont
  {A.~J.}\ \bibnamefont {Millis}},\ and\ \bibinfo {author} {\bibfnamefont
  {A.~N.}\ \bibnamefont {Pasupathy}},\ }\bibfield  {title} {\bibinfo {title}
  {{Visualization of electron nematicity and unidirectional antiferroic
  fluctuations at high temperatures in NaFeAs}},\ }\href
  {https://doi.org/10.1038/nphys2870} {\bibfield  {journal} {\bibinfo
  {journal} {Nature Physics}\ }\textbf {\bibinfo {volume} {10}},\ \bibinfo
  {pages} {225} (\bibinfo {year} {2014})}\BibitemShut {NoStop}%
\bibitem [{\citenamefont {Liang}\ \emph {et~al.}(2013)\citenamefont {Liang},
  \citenamefont {Moreo},\ and\ \citenamefont {Dagotto}}]{Dagotto13}%
  \BibitemOpen
  \bibfield  {author} {\bibinfo {author} {\bibfnamefont {S.}~\bibnamefont
  {Liang}}, \bibinfo {author} {\bibfnamefont {A.}~\bibnamefont {Moreo}},\ and\
  \bibinfo {author} {\bibfnamefont {E.}~\bibnamefont {Dagotto}},\ }\bibfield
  {title} {\bibinfo {title} {{Nematic State of Pnictides Stabilized by
  Interplay between Spin, Orbital, and Lattice Degrees of Freedom}},\ }\href
  {https://doi.org/10.1103/PhysRevLett.111.047004} {\bibfield  {journal}
  {\bibinfo  {journal} {Phys. Rev. Lett.}\ }\textbf {\bibinfo {volume} {111}},\
  \bibinfo {pages} {047004} (\bibinfo {year} {2013})}\BibitemShut {NoStop}%
\bibitem [{\citenamefont {Fanfarillo}\ \emph {et~al.}(2015)\citenamefont
  {Fanfarillo}, \citenamefont {Cortijo},\ and\ \citenamefont
  {Valenzuela}}]{Fanfarillo15}%
  \BibitemOpen
  \bibfield  {author} {\bibinfo {author} {\bibfnamefont {L.}~\bibnamefont
  {Fanfarillo}}, \bibinfo {author} {\bibfnamefont {A.}~\bibnamefont
  {Cortijo}},\ and\ \bibinfo {author} {\bibfnamefont {B.}~\bibnamefont
  {Valenzuela}},\ }\bibfield  {title} {\bibinfo {title} {Spin-orbital interplay
  and topology in the nematic phase of iron pnictides},\ }\href
  {https://doi.org/10.1103/PhysRevB.91.214515} {\bibfield  {journal} {\bibinfo
  {journal} {Phys. Rev. B}\ }\textbf {\bibinfo {volume} {91}},\ \bibinfo
  {pages} {214515} (\bibinfo {year} {2015})}\BibitemShut {NoStop}%
\bibitem [{\citenamefont {Lee}\ \emph {et~al.}(2009{\natexlab{a}})\citenamefont
  {Lee}, \citenamefont {Yin},\ and\ \citenamefont {Ku}}]{Ku09}%
  \BibitemOpen
  \bibfield  {author} {\bibinfo {author} {\bibfnamefont {C.-C.}\ \bibnamefont
  {Lee}}, \bibinfo {author} {\bibfnamefont {W.-G.}\ \bibnamefont {Yin}},\ and\
  \bibinfo {author} {\bibfnamefont {W.}~\bibnamefont {Ku}},\ }\bibfield
  {title} {\bibinfo {title} {{Ferro-Orbital Order and Strong Magnetic
  Anisotropy in the Parent Compounds of Iron-Pnictide Superconductors}},\
  }\href {https://doi.org/10.1103/PhysRevLett.103.267001} {\bibfield  {journal}
  {\bibinfo  {journal} {Phys. Rev. Lett.}\ }\textbf {\bibinfo {volume} {103}},\
  \bibinfo {pages} {267001} (\bibinfo {year} {2009}{\natexlab{a}})}\BibitemShut
  {NoStop}%
\bibitem [{\citenamefont {Lv}\ \emph {et~al.}(2010)\citenamefont {Lv},
  \citenamefont {Kr\"uger},\ and\ \citenamefont {Phillips}}]{Phillips10}%
  \BibitemOpen
  \bibfield  {author} {\bibinfo {author} {\bibfnamefont {W.}~\bibnamefont
  {Lv}}, \bibinfo {author} {\bibfnamefont {F.}~\bibnamefont {Kr\"uger}},\ and\
  \bibinfo {author} {\bibfnamefont {P.}~\bibnamefont {Phillips}},\ }\bibfield
  {title} {\bibinfo {title} {{Orbital ordering and unfrustrated
  $(\ensuremath{\pi},0)$ magnetism from degenerate double exchange in the iron
  pnictides}},\ }\href {https://doi.org/10.1103/PhysRevB.82.045125} {\bibfield
  {journal} {\bibinfo  {journal} {Phys. Rev. B}\ }\textbf {\bibinfo {volume}
  {82}},\ \bibinfo {pages} {045125} (\bibinfo {year} {2010})}\BibitemShut
  {NoStop}%
\bibitem [{\citenamefont {Yamakawa}\ \emph {et~al.}(2016)\citenamefont
  {Yamakawa}, \citenamefont {Onari},\ and\ \citenamefont
  {Kontani}}]{Kontani16}%
  \BibitemOpen
  \bibfield  {author} {\bibinfo {author} {\bibfnamefont {Y.}~\bibnamefont
  {Yamakawa}}, \bibinfo {author} {\bibfnamefont {S.}~\bibnamefont {Onari}},\
  and\ \bibinfo {author} {\bibfnamefont {H.}~\bibnamefont {Kontani}},\
  }\bibfield  {title} {\bibinfo {title} {{Nematicity and Magnetism in FeSe and
  Other Families of Fe-Based Superconductors}},\ }\href
  {https://doi.org/10.1103/PhysRevX.6.021032} {\bibfield  {journal} {\bibinfo
  {journal} {Phys. Rev. X}\ }\textbf {\bibinfo {volume} {6}},\ \bibinfo {pages}
  {021032} (\bibinfo {year} {2016})}\BibitemShut {NoStop}%
\bibitem [{\citenamefont {Fang}\ \emph {et~al.}(2008)\citenamefont {Fang},
  \citenamefont {Yao}, \citenamefont {Tsai}, \citenamefont {Hu},\ and\
  \citenamefont {Kivelson}}]{Kivelson08}%
  \BibitemOpen
  \bibfield  {author} {\bibinfo {author} {\bibfnamefont {C.}~\bibnamefont
  {Fang}}, \bibinfo {author} {\bibfnamefont {H.}~\bibnamefont {Yao}}, \bibinfo
  {author} {\bibfnamefont {W.-F.}\ \bibnamefont {Tsai}}, \bibinfo {author}
  {\bibfnamefont {J.}~\bibnamefont {Hu}},\ and\ \bibinfo {author}
  {\bibfnamefont {S.~A.}\ \bibnamefont {Kivelson}},\ }\bibfield  {title}
  {\bibinfo {title} {{Theory of electron nematic order in LaFeAsO}},\ }\href
  {https://doi.org/10.1103/PhysRevB.77.224509} {\bibfield  {journal} {\bibinfo
  {journal} {Phys. Rev. B}\ }\textbf {\bibinfo {volume} {77}},\ \bibinfo
  {pages} {224509} (\bibinfo {year} {2008})}\BibitemShut {NoStop}%
\bibitem [{\citenamefont {Xu}\ \emph {et~al.}(2008)\citenamefont {Xu},
  \citenamefont {M\"uller},\ and\ \citenamefont {Sachdev}}]{Sachdev08}%
  \BibitemOpen
  \bibfield  {author} {\bibinfo {author} {\bibfnamefont {C.}~\bibnamefont
  {Xu}}, \bibinfo {author} {\bibfnamefont {M.}~\bibnamefont {M\"uller}},\ and\
  \bibinfo {author} {\bibfnamefont {S.}~\bibnamefont {Sachdev}},\ }\bibfield
  {title} {\bibinfo {title} {Ising and spin orders in the iron-based
  superconductors},\ }\href {https://doi.org/10.1103/PhysRevB.78.020501}
  {\bibfield  {journal} {\bibinfo  {journal} {Phys. Rev. B}\ }\textbf {\bibinfo
  {volume} {78}},\ \bibinfo {pages} {020501} (\bibinfo {year}
  {2008})}\BibitemShut {NoStop}%
\bibitem [{\citenamefont {Fernandes}\ \emph {et~al.}(2019)\citenamefont
  {Fernandes}, \citenamefont {Orth},\ and\ \citenamefont
  {Schmalian}}]{Fernandes19}%
  \BibitemOpen
  \bibfield  {author} {\bibinfo {author} {\bibfnamefont {R.~M.}\ \bibnamefont
  {Fernandes}}, \bibinfo {author} {\bibfnamefont {P.~P.}\ \bibnamefont
  {Orth}},\ and\ \bibinfo {author} {\bibfnamefont {J.}~\bibnamefont
  {Schmalian}},\ }\bibfield  {title} {\bibinfo {title} {{Intertwined Vestigial
  Order in Quantum Materials: Nematicity and Beyond}},\ }\href
  {https://doi.org/10.1146/annurev-conmatphys-031218-013200} {\bibfield
  {journal} {\bibinfo  {journal} {Annual Review of Condensed Matter Physics}\
  }\textbf {\bibinfo {volume} {10}},\ \bibinfo {pages} {133} (\bibinfo {year}
  {2019})}\BibitemShut {NoStop}%
\bibitem [{\citenamefont {Wang}\ \emph
  {et~al.}(2015{\natexlab{b}})\citenamefont {Wang}, \citenamefont {Kivelson},\
  and\ \citenamefont {Lee}}]{Wang15}%
  \BibitemOpen
  \bibfield  {author} {\bibinfo {author} {\bibfnamefont {F.}~\bibnamefont
  {Wang}}, \bibinfo {author} {\bibfnamefont {S.~A.}\ \bibnamefont {Kivelson}},\
  and\ \bibinfo {author} {\bibfnamefont {D.-H.}\ \bibnamefont {Lee}},\
  }\bibfield  {title} {\bibinfo {title} {Nematicity and quantum paramagnetism
  in fese},\ }\href {https://doi.org/10.1038/nphys3456} {\bibfield  {journal}
  {\bibinfo  {journal} {Nature Physics}\ }\textbf {\bibinfo {volume} {11}},\
  \bibinfo {pages} {959} (\bibinfo {year} {2015}{\natexlab{b}})}\BibitemShut
  {NoStop}%
\bibitem [{\citenamefont {Fernandes}\ \emph {et~al.}(2012)\citenamefont
  {Fernandes}, \citenamefont {Chubukov}, \citenamefont {Knolle}, \citenamefont
  {Eremin},\ and\ \citenamefont {Schmalian}}]{Fernandes12}%
  \BibitemOpen
  \bibfield  {author} {\bibinfo {author} {\bibfnamefont {R.~M.}\ \bibnamefont
  {Fernandes}}, \bibinfo {author} {\bibfnamefont {A.~V.}\ \bibnamefont
  {Chubukov}}, \bibinfo {author} {\bibfnamefont {J.}~\bibnamefont {Knolle}},
  \bibinfo {author} {\bibfnamefont {I.}~\bibnamefont {Eremin}},\ and\ \bibinfo
  {author} {\bibfnamefont {J.}~\bibnamefont {Schmalian}},\ }\bibfield  {title}
  {\bibinfo {title} {Preemptive nematic order, pseudogap, and orbital order in
  the iron pnictides},\ }\href {https://doi.org/10.1103/PhysRevB.85.024534}
  {\bibfield  {journal} {\bibinfo  {journal} {Phys. Rev. B}\ }\textbf {\bibinfo
  {volume} {85}},\ \bibinfo {pages} {024534} (\bibinfo {year}
  {2012})}\BibitemShut {NoStop}%
\bibitem [{\citenamefont {Gati}\ \emph {et~al.}(2019)\citenamefont {Gati},
  \citenamefont {Xiang}, \citenamefont {Bud'ko},\ and\ \citenamefont
  {Canfield}}]{Gati19}%
  \BibitemOpen
  \bibfield  {author} {\bibinfo {author} {\bibfnamefont {E.}~\bibnamefont
  {Gati}}, \bibinfo {author} {\bibfnamefont {L.}~\bibnamefont {Xiang}},
  \bibinfo {author} {\bibfnamefont {S.~L.}\ \bibnamefont {Bud'ko}},\ and\
  \bibinfo {author} {\bibfnamefont {P.~C.}\ \bibnamefont {Canfield}},\
  }\bibfield  {title} {\bibinfo {title} {{Role of the Fermi surface for the
  pressure-tuned nematic transition in the
  ${\mathrm{BaFe}}_{2}{\mathrm{As}}_{2}$ family}},\ }\href
  {https://doi.org/10.1103/PhysRevB.100.064512} {\bibfield  {journal} {\bibinfo
   {journal} {Phys. Rev. B}\ }\textbf {\bibinfo {volume} {100}},\ \bibinfo
  {pages} {064512} (\bibinfo {year} {2019})}\BibitemShut {NoStop}%
\bibitem [{\citenamefont {Fernandes}\ \emph {et~al.}(2013)\citenamefont
  {Fernandes}, \citenamefont {B\"ohmer}, \citenamefont {Meingast},\ and\
  \citenamefont {Schmalian}}]{Fernandes13}%
  \BibitemOpen
  \bibfield  {author} {\bibinfo {author} {\bibfnamefont {R.~M.}\ \bibnamefont
  {Fernandes}}, \bibinfo {author} {\bibfnamefont {A.~E.}\ \bibnamefont
  {B\"ohmer}}, \bibinfo {author} {\bibfnamefont {C.}~\bibnamefont {Meingast}},\
  and\ \bibinfo {author} {\bibfnamefont {J.}~\bibnamefont {Schmalian}},\
  }\bibfield  {title} {\bibinfo {title} {{Scaling between Magnetic and Lattice
  Fluctuations in Iron Pnictide Superconductors}},\ }\href
  {https://doi.org/10.1103/PhysRevLett.111.137001} {\bibfield  {journal}
  {\bibinfo  {journal} {Phys. Rev. Lett.}\ }\textbf {\bibinfo {volume} {111}},\
  \bibinfo {pages} {137001} (\bibinfo {year} {2013})}\BibitemShut {NoStop}%
\bibitem [{\citenamefont {Baek}\ \emph {et~al.}(2015)\citenamefont {Baek},
  \citenamefont {Efremov}, \citenamefont {Ok}, \citenamefont {Kim},
  \citenamefont {van~den Brink},\ and\ \citenamefont {B{\"u}chner}}]{Baek15}%
  \BibitemOpen
  \bibfield  {author} {\bibinfo {author} {\bibfnamefont {S.}~\bibnamefont
  {Baek}}, \bibinfo {author} {\bibfnamefont {D.}~\bibnamefont {Efremov}},
  \bibinfo {author} {\bibfnamefont {J.}~\bibnamefont {Ok}}, \bibinfo {author}
  {\bibfnamefont {J.}~\bibnamefont {Kim}}, \bibinfo {author} {\bibfnamefont
  {J.}~\bibnamefont {van~den Brink}},\ and\ \bibinfo {author} {\bibfnamefont
  {B.}~\bibnamefont {B{\"u}chner}},\ }\bibfield  {title} {\bibinfo {title}
  {{Orbital-driven nematicity in FeSe}},\ }\href
  {https://doi.org/10.1038/nmat4138} {\bibfield  {journal} {\bibinfo  {journal}
  {Nature Materials}\ }\textbf {\bibinfo {volume} {14}},\ \bibinfo {pages}
  {210} (\bibinfo {year} {2015})}\BibitemShut {NoStop}%
\bibitem [{\citenamefont {B\"ohmer}\ \emph {et~al.}(2019)\citenamefont
  {B\"ohmer}, \citenamefont {Kothapalli}, \citenamefont {Jayasekara},
  \citenamefont {Wilde}, \citenamefont {Li}, \citenamefont {Sapkota},
  \citenamefont {Ueland}, \citenamefont {Das}, \citenamefont {Xiao},
  \citenamefont {Bi}, \citenamefont {Zhao}, \citenamefont {Alp}, \citenamefont
  {Bud'ko}, \citenamefont {Canfield}, \citenamefont {Goldman},\ and\
  \citenamefont {Kreyssig}}]{Bohmer19}%
  \BibitemOpen
  \bibfield  {author} {\bibinfo {author} {\bibfnamefont {A.~E.}\ \bibnamefont
  {B\"ohmer}}, \bibinfo {author} {\bibfnamefont {K.}~\bibnamefont
  {Kothapalli}}, \bibinfo {author} {\bibfnamefont {W.~T.}\ \bibnamefont
  {Jayasekara}}, \bibinfo {author} {\bibfnamefont {J.~M.}\ \bibnamefont
  {Wilde}}, \bibinfo {author} {\bibfnamefont {B.}~\bibnamefont {Li}}, \bibinfo
  {author} {\bibfnamefont {A.}~\bibnamefont {Sapkota}}, \bibinfo {author}
  {\bibfnamefont {B.~G.}\ \bibnamefont {Ueland}}, \bibinfo {author}
  {\bibfnamefont {P.}~\bibnamefont {Das}}, \bibinfo {author} {\bibfnamefont
  {Y.}~\bibnamefont {Xiao}}, \bibinfo {author} {\bibfnamefont {W.}~\bibnamefont
  {Bi}}, \bibinfo {author} {\bibfnamefont {J.}~\bibnamefont {Zhao}}, \bibinfo
  {author} {\bibfnamefont {E.~E.}\ \bibnamefont {Alp}}, \bibinfo {author}
  {\bibfnamefont {S.~L.}\ \bibnamefont {Bud'ko}}, \bibinfo {author}
  {\bibfnamefont {P.~C.}\ \bibnamefont {Canfield}}, \bibinfo {author}
  {\bibfnamefont {A.~I.}\ \bibnamefont {Goldman}},\ and\ \bibinfo {author}
  {\bibfnamefont {A.}~\bibnamefont {Kreyssig}},\ }\bibfield  {title} {\bibinfo
  {title} {{Distinct pressure evolution of coupled nematic and magnetic orders
  in FeSe}},\ }\href {https://doi.org/10.1103/PhysRevB.100.064515} {\bibfield
  {journal} {\bibinfo  {journal} {Phys. Rev. B}\ }\textbf {\bibinfo {volume}
  {100}},\ \bibinfo {pages} {064515} (\bibinfo {year} {2019})}\BibitemShut
  {NoStop}%
\bibitem [{\citenamefont {Zhang}\ \emph {et~al.}(2015)\citenamefont {Zhang},
  \citenamefont {Qian}, \citenamefont {Richard}, \citenamefont {Wang},
  \citenamefont {Miao}, \citenamefont {Lv}, \citenamefont {Fu}, \citenamefont
  {Wolf}, \citenamefont {Meingast}, \citenamefont {Wu}, \citenamefont {Wang},
  \citenamefont {Hu},\ and\ \citenamefont {Ding}}]{Ding15}%
  \BibitemOpen
  \bibfield  {author} {\bibinfo {author} {\bibfnamefont {P.}~\bibnamefont
  {Zhang}}, \bibinfo {author} {\bibfnamefont {T.}~\bibnamefont {Qian}},
  \bibinfo {author} {\bibfnamefont {P.}~\bibnamefont {Richard}}, \bibinfo
  {author} {\bibfnamefont {X.~P.}\ \bibnamefont {Wang}}, \bibinfo {author}
  {\bibfnamefont {H.}~\bibnamefont {Miao}}, \bibinfo {author} {\bibfnamefont
  {B.~Q.}\ \bibnamefont {Lv}}, \bibinfo {author} {\bibfnamefont {B.~B.}\
  \bibnamefont {Fu}}, \bibinfo {author} {\bibfnamefont {T.}~\bibnamefont
  {Wolf}}, \bibinfo {author} {\bibfnamefont {C.}~\bibnamefont {Meingast}},
  \bibinfo {author} {\bibfnamefont {X.~X.}\ \bibnamefont {Wu}}, \bibinfo
  {author} {\bibfnamefont {Z.~Q.}\ \bibnamefont {Wang}}, \bibinfo {author}
  {\bibfnamefont {J.~P.}\ \bibnamefont {Hu}},\ and\ \bibinfo {author}
  {\bibfnamefont {H.}~\bibnamefont {Ding}},\ }\bibfield  {title} {\bibinfo
  {title} {{Observation of two distinct ${d}_{xz}$/${d}_{yz}$ band splittings
  in FeSe}},\ }\href {https://doi.org/10.1103/PhysRevB.91.214503} {\bibfield
  {journal} {\bibinfo  {journal} {Phys. Rev. B}\ }\textbf {\bibinfo {volume}
  {91}},\ \bibinfo {pages} {214503} (\bibinfo {year} {2015})}\BibitemShut
  {NoStop}%
\bibitem [{\citenamefont {Pfau}\ \emph {et~al.}(2019)\citenamefont {Pfau},
  \citenamefont {Chen}, \citenamefont {Yi}, \citenamefont {Hashimoto},
  \citenamefont {Rotundu}, \citenamefont {Palmstrom}, \citenamefont {Chen},
  \citenamefont {Dai}, \citenamefont {Straquadine}, \citenamefont {Hristov},
  \citenamefont {Birgeneau}, \citenamefont {Fisher}, \citenamefont {Lu},\ and\
  \citenamefont {Shen}}]{Pfau19}%
  \BibitemOpen
  \bibfield  {author} {\bibinfo {author} {\bibfnamefont {H.}~\bibnamefont
  {Pfau}}, \bibinfo {author} {\bibfnamefont {S.~D.}\ \bibnamefont {Chen}},
  \bibinfo {author} {\bibfnamefont {M.}~\bibnamefont {Yi}}, \bibinfo {author}
  {\bibfnamefont {M.}~\bibnamefont {Hashimoto}}, \bibinfo {author}
  {\bibfnamefont {C.~R.}\ \bibnamefont {Rotundu}}, \bibinfo {author}
  {\bibfnamefont {J.~C.}\ \bibnamefont {Palmstrom}}, \bibinfo {author}
  {\bibfnamefont {T.}~\bibnamefont {Chen}}, \bibinfo {author} {\bibfnamefont
  {P.-C.}\ \bibnamefont {Dai}}, \bibinfo {author} {\bibfnamefont
  {J.}~\bibnamefont {Straquadine}}, \bibinfo {author} {\bibfnamefont
  {A.}~\bibnamefont {Hristov}}, \bibinfo {author} {\bibfnamefont {R.~J.}\
  \bibnamefont {Birgeneau}}, \bibinfo {author} {\bibfnamefont {I.~R.}\
  \bibnamefont {Fisher}}, \bibinfo {author} {\bibfnamefont {D.}~\bibnamefont
  {Lu}},\ and\ \bibinfo {author} {\bibfnamefont {Z.-X.}\ \bibnamefont {Shen}},\
  }\bibfield  {title} {\bibinfo {title} {{Momentum Dependence of the Nematic
  Order Parameter in Iron-Based Superconductors}},\ }\href
  {https://doi.org/10.1103/PhysRevLett.123.066402} {\bibfield  {journal}
  {\bibinfo  {journal} {Phys. Rev. Lett.}\ }\textbf {\bibinfo {volume} {123}},\
  \bibinfo {pages} {066402} (\bibinfo {year} {2019})}\BibitemShut {NoStop}%
\bibitem [{\citenamefont {Lederer}\ \emph {et~al.}(2017)\citenamefont
  {Lederer}, \citenamefont {Schattner}, \citenamefont {Berg},\ and\
  \citenamefont {Kivelson}}]{Lederer17}%
  \BibitemOpen
  \bibfield  {author} {\bibinfo {author} {\bibfnamefont {S.}~\bibnamefont
  {Lederer}}, \bibinfo {author} {\bibfnamefont {Y.}~\bibnamefont {Schattner}},
  \bibinfo {author} {\bibfnamefont {E.}~\bibnamefont {Berg}},\ and\ \bibinfo
  {author} {\bibfnamefont {S.~A.}\ \bibnamefont {Kivelson}},\ }\bibfield
  {title} {\bibinfo {title} {{Superconductivity and non-Fermi liquid behavior
  near a nematic quantum critical point}},\ }\href
  {https://doi.org/10.1073/pnas.1620651114} {\bibfield  {journal} {\bibinfo
  {journal} {Proceedings of the National Academy of Sciences}\ }\textbf
  {\bibinfo {volume} {114}},\ \bibinfo {pages} {4905} (\bibinfo {year}
  {2017})}\BibitemShut {NoStop}%
\bibitem [{\citenamefont {Klein}\ and\ \citenamefont
  {Chubukov}(2018)}]{Klein18}%
  \BibitemOpen
  \bibfield  {author} {\bibinfo {author} {\bibfnamefont {A.}~\bibnamefont
  {Klein}}\ and\ \bibinfo {author} {\bibfnamefont {A.~V.}\ \bibnamefont
  {Chubukov}},\ }\bibfield  {title} {\bibinfo {title} {{Superconductivity near
  a nematic quantum critical point: Interplay between hot and lukewarm
  regions}},\ }\href {https://doi.org/10.1103/PhysRevB.98.220501} {\bibfield
  {journal} {\bibinfo  {journal} {Phys. Rev. B}\ }\textbf {\bibinfo {volume}
  {98}},\ \bibinfo {pages} {220501} (\bibinfo {year} {2018})}\BibitemShut
  {NoStop}%
\bibitem [{\citenamefont {Worasaran}\ \emph {et~al.}(2020)\citenamefont
  {Worasaran}, \citenamefont {Ikeda}, \citenamefont {Palmstrom}, \citenamefont
  {Straquadine}, \citenamefont {Kivelson},\ and\ \citenamefont
  {Fisher}}]{Worasaran20}%
  \BibitemOpen
  \bibfield  {author} {\bibinfo {author} {\bibfnamefont {T.}~\bibnamefont
  {Worasaran}}, \bibinfo {author} {\bibfnamefont {M.~S.}\ \bibnamefont
  {Ikeda}}, \bibinfo {author} {\bibfnamefont {J.~C.}\ \bibnamefont
  {Palmstrom}}, \bibinfo {author} {\bibfnamefont {J.~A.}\ \bibnamefont
  {Straquadine}}, \bibinfo {author} {\bibfnamefont {S.~A.}\ \bibnamefont
  {Kivelson}},\ and\ \bibinfo {author} {\bibfnamefont {I.~R.}\ \bibnamefont
  {Fisher}},\ }\bibfield  {title} {\bibinfo {title} {{Nematic quantum
  criticality in an Fe-based superconductor revealed by strain-tuning}},\
  }\href@noop {} {\bibfield  {journal} {\bibinfo  {journal} {arXiv:2003.12202}\
  } (\bibinfo {year} {2020})}\BibitemShut {NoStop}%
\bibitem [{\citenamefont {Shibauchi}\ \emph {et~al.}(2020)\citenamefont
  {Shibauchi}, \citenamefont {Hanaguri},\ and\ \citenamefont
  {Matsuda}}]{Shibauchi20}%
  \BibitemOpen
  \bibfield  {author} {\bibinfo {author} {\bibfnamefont {T.}~\bibnamefont
  {Shibauchi}}, \bibinfo {author} {\bibfnamefont {T.}~\bibnamefont
  {Hanaguri}},\ and\ \bibinfo {author} {\bibfnamefont {Y.}~\bibnamefont
  {Matsuda}},\ }\bibfield  {title} {\bibinfo {title} {{Exotic Superconducting
  States in FeSe-based Materials}},\ }\href
  {https://doi.org/10.7566/JPSJ.89.102002} {\bibfield  {journal} {\bibinfo
  {journal} {Journal of the Physical Society of Japan}\ }\textbf {\bibinfo
  {volume} {89}},\ \bibinfo {pages} {102002} (\bibinfo {year}
  {2020})}\BibitemShut {NoStop}%
\bibitem [{\citenamefont {Coldea}(2020)}]{Coldea20}%
  \BibitemOpen
  \bibfield  {author} {\bibinfo {author} {\bibfnamefont {A.~I.}\ \bibnamefont
  {Coldea}},\ }\bibfield  {title} {\bibinfo {title} {{Electronic nematic states
  tuned by isoelectronic substitution in bulk FeSe$_{1-x}$S$_x$}},\ }\href@noop
  {} {\bibfield  {journal} {\bibinfo  {journal} {arXiv:2009.05523}\ } (\bibinfo
  {year} {2020})}\BibitemShut {NoStop}%
\bibitem [{\citenamefont {Okazaki}\ \emph {et~al.}(2012)\citenamefont
  {Okazaki}, \citenamefont {Ota}, \citenamefont {Kotani}, \citenamefont
  {Malaeb}, \citenamefont {Ishida}, \citenamefont {Shimojima}, \citenamefont
  {Kiss}, \citenamefont {Watanabe}, \citenamefont {Chen}, \citenamefont
  {Kihou}, \citenamefont {Lee}, \citenamefont {Iyo}, \citenamefont {Eisaki},
  \citenamefont {Saito}, \citenamefont {Fukazawa}, \citenamefont {Kohori},
  \citenamefont {Hashimoto}, \citenamefont {Shibauchi}, \citenamefont
  {Matsuda}, \citenamefont {Ikeda}, \citenamefont {Miyahara}, \citenamefont
  {Arita}, \citenamefont {Chainani},\ and\ \citenamefont {Shin}}]{Okazaki12}%
  \BibitemOpen
  \bibfield  {author} {\bibinfo {author} {\bibfnamefont {K.}~\bibnamefont
  {Okazaki}}, \bibinfo {author} {\bibfnamefont {Y.}~\bibnamefont {Ota}},
  \bibinfo {author} {\bibfnamefont {Y.}~\bibnamefont {Kotani}}, \bibinfo
  {author} {\bibfnamefont {W.}~\bibnamefont {Malaeb}}, \bibinfo {author}
  {\bibfnamefont {Y.}~\bibnamefont {Ishida}}, \bibinfo {author} {\bibfnamefont
  {T.}~\bibnamefont {Shimojima}}, \bibinfo {author} {\bibfnamefont
  {T.}~\bibnamefont {Kiss}}, \bibinfo {author} {\bibfnamefont {S.}~\bibnamefont
  {Watanabe}}, \bibinfo {author} {\bibfnamefont {C.-T.}\ \bibnamefont {Chen}},
  \bibinfo {author} {\bibfnamefont {K.}~\bibnamefont {Kihou}}, \bibinfo
  {author} {\bibfnamefont {C.~H.}\ \bibnamefont {Lee}}, \bibinfo {author}
  {\bibfnamefont {A.}~\bibnamefont {Iyo}}, \bibinfo {author} {\bibfnamefont
  {H.}~\bibnamefont {Eisaki}}, \bibinfo {author} {\bibfnamefont
  {T.}~\bibnamefont {Saito}}, \bibinfo {author} {\bibfnamefont
  {H.}~\bibnamefont {Fukazawa}}, \bibinfo {author} {\bibfnamefont
  {Y.}~\bibnamefont {Kohori}}, \bibinfo {author} {\bibfnamefont
  {K.}~\bibnamefont {Hashimoto}}, \bibinfo {author} {\bibfnamefont
  {T.}~\bibnamefont {Shibauchi}}, \bibinfo {author} {\bibfnamefont
  {Y.}~\bibnamefont {Matsuda}}, \bibinfo {author} {\bibfnamefont
  {H.}~\bibnamefont {Ikeda}}, \bibinfo {author} {\bibfnamefont
  {H.}~\bibnamefont {Miyahara}}, \bibinfo {author} {\bibfnamefont
  {R.}~\bibnamefont {Arita}}, \bibinfo {author} {\bibfnamefont
  {A.}~\bibnamefont {Chainani}},\ and\ \bibinfo {author} {\bibfnamefont
  {S.}~\bibnamefont {Shin}},\ }\bibfield  {title} {\bibinfo {title}
  {{Octet-Line Node Structure of Superconducting Order Parameter in
  KFe$_2$As$_2$}},\ }\href {https://doi.org/10.1126/science.1222793} {\bibfield
   {journal} {\bibinfo  {journal} {Science}\ }\textbf {\bibinfo {volume}
  {337}},\ \bibinfo {pages} {1314} (\bibinfo {year} {2012})}\BibitemShut
  {NoStop}%
\bibitem [{\citenamefont {Hirschfeld}(2016)}]{Hirschfeld_review}%
  \BibitemOpen
  \bibfield  {author} {\bibinfo {author} {\bibfnamefont {P.~J.}\ \bibnamefont
  {Hirschfeld}},\ }\bibfield  {title} {\bibinfo {title} {Using gap symmetry and
  structure to reveal the pairing mechanism in fe-based superconductors},\
  }\href {https://doi.org/https://doi.org/10.1016/j.crhy.2015.10.002}
  {\bibfield  {journal} {\bibinfo  {journal} {Comptes Rendus Physique}\
  }\textbf {\bibinfo {volume} {17}},\ \bibinfo {pages} {197} (\bibinfo {year}
  {2016})}\BibitemShut {NoStop}%
\bibitem [{\citenamefont {Lee}\ \emph {et~al.}(2018)\citenamefont {Lee},
  \citenamefont {Chubukov}, \citenamefont {Miao},\ and\ \citenamefont
  {Kotliar}}]{Lee18}%
  \BibitemOpen
  \bibfield  {author} {\bibinfo {author} {\bibfnamefont {T.-H.}\ \bibnamefont
  {Lee}}, \bibinfo {author} {\bibfnamefont {A.~V.}\ \bibnamefont {Chubukov}},
  \bibinfo {author} {\bibfnamefont {H.}~\bibnamefont {Miao}},\ and\ \bibinfo
  {author} {\bibfnamefont {G.}~\bibnamefont {Kotliar}},\ }\bibfield  {title}
  {\bibinfo {title} {{Pairing Mechanism in Hund's Metal Superconductors and the
  Universality of the Superconducting Gap to Critical Temperature Ratio}},\
  }\href {https://doi.org/10.1103/PhysRevLett.121.187003} {\bibfield  {journal}
  {\bibinfo  {journal} {Phys. Rev. Lett.}\ }\textbf {\bibinfo {volume} {121}},\
  \bibinfo {pages} {187003} (\bibinfo {year} {2018})}\BibitemShut {NoStop}%
\bibitem [{\citenamefont {Stanev}\ and\ \citenamefont {Te\ifmmode
  \check{s}\else \v{s}\fi{}anovi\ifmmode~\acute{c}\else
  \'{c}\fi{}}(2010)}]{Stanev10}%
  \BibitemOpen
  \bibfield  {author} {\bibinfo {author} {\bibfnamefont {V.}~\bibnamefont
  {Stanev}}\ and\ \bibinfo {author} {\bibfnamefont {Z.}~\bibnamefont
  {Te\ifmmode \check{s}\else \v{s}\fi{}anovi\ifmmode~\acute{c}\else
  \'{c}\fi{}}},\ }\bibfield  {title} {\bibinfo {title} {Three-band
  superconductivity and the order parameter that breaks time-reversal
  symmetry},\ }\href {https://doi.org/10.1103/PhysRevB.81.134522} {\bibfield
  {journal} {\bibinfo  {journal} {Phys. Rev. B}\ }\textbf {\bibinfo {volume}
  {81}},\ \bibinfo {pages} {134522} (\bibinfo {year} {2010})}\BibitemShut
  {NoStop}%
\bibitem [{\citenamefont {Grinenko}\ \emph {et~al.}(2020)\citenamefont
  {Grinenko}, \citenamefont {Sarkar}, \citenamefont {Kihou}, \citenamefont
  {Lee}, \citenamefont {Morozov}, \citenamefont {Aswartham}, \citenamefont
  {B{\"u}chner}, \citenamefont {Chekhonin}, \citenamefont {Skrotzki},
  \citenamefont {Nenkov}, \citenamefont {Hühne}, \citenamefont {Nielsch},
  \citenamefont {Drechsler}, \citenamefont {Vadimov}, \citenamefont {Silaev},
  \citenamefont {Volkov}, \citenamefont {Eremin}, \citenamefont {Luetkens},\
  and\ \citenamefont {Klauss}}]{Grinenko20}%
  \BibitemOpen
  \bibfield  {author} {\bibinfo {author} {\bibfnamefont {V.}~\bibnamefont
  {Grinenko}}, \bibinfo {author} {\bibfnamefont {R.}~\bibnamefont {Sarkar}},
  \bibinfo {author} {\bibfnamefont {K.}~\bibnamefont {Kihou}}, \bibinfo
  {author} {\bibfnamefont {C.}~\bibnamefont {Lee}}, \bibinfo {author}
  {\bibfnamefont {I.}~\bibnamefont {Morozov}}, \bibinfo {author} {\bibfnamefont
  {S.}~\bibnamefont {Aswartham}}, \bibinfo {author} {\bibfnamefont
  {B.}~\bibnamefont {B{\"u}chner}}, \bibinfo {author} {\bibfnamefont
  {P.}~\bibnamefont {Chekhonin}}, \bibinfo {author} {\bibfnamefont
  {W.}~\bibnamefont {Skrotzki}}, \bibinfo {author} {\bibfnamefont
  {K.}~\bibnamefont {Nenkov}}, \bibinfo {author} {\bibfnamefont
  {R.}~\bibnamefont {Hühne}}, \bibinfo {author} {\bibfnamefont
  {K.}~\bibnamefont {Nielsch}}, \bibinfo {author} {\bibfnamefont {S.~L.}\
  \bibnamefont {Drechsler}}, \bibinfo {author} {\bibfnamefont {V.~L.}\
  \bibnamefont {Vadimov}}, \bibinfo {author} {\bibfnamefont {M.~A.}\
  \bibnamefont {Silaev}}, \bibinfo {author} {\bibfnamefont {P.~A.}\
  \bibnamefont {Volkov}}, \bibinfo {author} {\bibfnamefont {I.}~\bibnamefont
  {Eremin}}, \bibinfo {author} {\bibfnamefont {H.}~\bibnamefont {Luetkens}},\
  and\ \bibinfo {author} {\bibfnamefont {H.~H.}\ \bibnamefont {Klauss}},\
  }\bibfield  {title} {\bibinfo {title} {Superconductivity with broken
  time-reversal symmetry inside a superconducting s-wave state},\ }\href
  {https://doi.org/10.1038/s41567-020-0886-9} {\bibfield  {journal} {\bibinfo
  {journal} {Nature Physics}\ }\textbf {\bibinfo {volume} {16}},\ \bibinfo
  {pages} {789} (\bibinfo {year} {2020})}\BibitemShut {NoStop}%
\bibitem [{\citenamefont {Karahasanovic}\ and\ \citenamefont
  {Schmalian}(2016)}]{Karahasanovic16}%
  \BibitemOpen
  \bibfield  {author} {\bibinfo {author} {\bibfnamefont {U.}~\bibnamefont
  {Karahasanovic}}\ and\ \bibinfo {author} {\bibfnamefont {J.}~\bibnamefont
  {Schmalian}},\ }\bibfield  {title} {\bibinfo {title} {Elastic coupling and
  spin-driven nematicity in iron-based superconductors},\ }\href
  {https://doi.org/10.1103/PhysRevB.93.064520} {\bibfield  {journal} {\bibinfo
  {journal} {Phys. Rev. B}\ }\textbf {\bibinfo {volume} {93}},\ \bibinfo
  {pages} {064520} (\bibinfo {year} {2016})}\BibitemShut {NoStop}%
\bibitem [{\citenamefont {Paul}\ and\ \citenamefont {Garst}(2017)}]{Paul17}%
  \BibitemOpen
  \bibfield  {author} {\bibinfo {author} {\bibfnamefont {I.}~\bibnamefont
  {Paul}}\ and\ \bibinfo {author} {\bibfnamefont {M.}~\bibnamefont {Garst}},\
  }\bibfield  {title} {\bibinfo {title} {{Lattice Effects on Nematic Quantum
  Criticality in Metals}},\ }\href
  {https://doi.org/10.1103/PhysRevLett.118.227601} {\bibfield  {journal}
  {\bibinfo  {journal} {Phys. Rev. Lett.}\ }\textbf {\bibinfo {volume} {118}},\
  \bibinfo {pages} {227601} (\bibinfo {year} {2017})}\BibitemShut {NoStop}%
\bibitem [{\citenamefont {Reiss}\ \emph {et~al.}(2020)\citenamefont {Reiss},
  \citenamefont {Graf}, \citenamefont {Haghighirad}, \citenamefont {Knafo},
  \citenamefont {Drigo}, \citenamefont {Bristow}, \citenamefont {Schofield},\
  and\ \citenamefont {Coldea}}]{Reiss20}%
  \BibitemOpen
  \bibfield  {author} {\bibinfo {author} {\bibfnamefont {P.}~\bibnamefont
  {Reiss}}, \bibinfo {author} {\bibfnamefont {D.}~\bibnamefont {Graf}},
  \bibinfo {author} {\bibfnamefont {A.~A.}\ \bibnamefont {Haghighirad}},
  \bibinfo {author} {\bibfnamefont {W.}~\bibnamefont {Knafo}}, \bibinfo
  {author} {\bibfnamefont {L.}~\bibnamefont {Drigo}}, \bibinfo {author}
  {\bibfnamefont {M.}~\bibnamefont {Bristow}}, \bibinfo {author} {\bibfnamefont
  {A.~J.}\ \bibnamefont {Schofield}},\ and\ \bibinfo {author} {\bibfnamefont
  {A.~I.}\ \bibnamefont {Coldea}},\ }\bibfield  {title} {\bibinfo {title}
  {Quenched nematic criticality and two superconducting domes in an iron-based
  superconductor},\ }\href {https://doi.org/10.1038/s41567-019-0694-2}
  {\bibfield  {journal} {\bibinfo  {journal} {Nature Physics}\ }\textbf
  {\bibinfo {volume} {16}},\ \bibinfo {pages} {89} (\bibinfo {year}
  {2020})}\BibitemShut {NoStop}%
\bibitem [{\citenamefont {Dioguardi}\ \emph {et~al.}(2015)\citenamefont
  {Dioguardi}, \citenamefont {Lawson}, \citenamefont {Bush}, \citenamefont
  {Crocker}, \citenamefont {Shirer}, \citenamefont {Nisson}, \citenamefont
  {Kissikov}, \citenamefont {Ran}, \citenamefont {Bud'ko}, \citenamefont
  {Canfield}, \citenamefont {Yuan}, \citenamefont {Kuhns}, \citenamefont
  {Reyes}, \citenamefont {Grafe},\ and\ \citenamefont {Curro}}]{Curro15}%
  \BibitemOpen
  \bibfield  {author} {\bibinfo {author} {\bibfnamefont {A.~P.}\ \bibnamefont
  {Dioguardi}}, \bibinfo {author} {\bibfnamefont {M.~M.}\ \bibnamefont
  {Lawson}}, \bibinfo {author} {\bibfnamefont {B.~T.}\ \bibnamefont {Bush}},
  \bibinfo {author} {\bibfnamefont {J.}~\bibnamefont {Crocker}}, \bibinfo
  {author} {\bibfnamefont {K.~R.}\ \bibnamefont {Shirer}}, \bibinfo {author}
  {\bibfnamefont {D.~M.}\ \bibnamefont {Nisson}}, \bibinfo {author}
  {\bibfnamefont {T.}~\bibnamefont {Kissikov}}, \bibinfo {author}
  {\bibfnamefont {S.}~\bibnamefont {Ran}}, \bibinfo {author} {\bibfnamefont
  {S.~L.}\ \bibnamefont {Bud'ko}}, \bibinfo {author} {\bibfnamefont {P.~C.}\
  \bibnamefont {Canfield}}, \bibinfo {author} {\bibfnamefont {S.}~\bibnamefont
  {Yuan}}, \bibinfo {author} {\bibfnamefont {P.~L.}\ \bibnamefont {Kuhns}},
  \bibinfo {author} {\bibfnamefont {A.~P.}\ \bibnamefont {Reyes}}, \bibinfo
  {author} {\bibfnamefont {H.-J.}\ \bibnamefont {Grafe}},\ and\ \bibinfo
  {author} {\bibfnamefont {N.~J.}\ \bibnamefont {Curro}},\ }\bibfield  {title}
  {\bibinfo {title} {{NMR evidence for inhomogeneous glassy behavior driven by
  nematic fluctuations in iron arsenide superconductors}},\ }\href
  {https://doi.org/10.1103/PhysRevB.92.165116} {\bibfield  {journal} {\bibinfo
  {journal} {Phys. Rev. B}\ }\textbf {\bibinfo {volume} {92}},\ \bibinfo
  {pages} {165116} (\bibinfo {year} {2015})}\BibitemShut {NoStop}%
\bibitem [{\citenamefont {Frandsen}\ \emph {et~al.}(2019)\citenamefont
  {Frandsen}, \citenamefont {Wang}, \citenamefont {Wu}, \citenamefont {Zhao},\
  and\ \citenamefont {Birgeneau}}]{Frandsen19}%
  \BibitemOpen
  \bibfield  {author} {\bibinfo {author} {\bibfnamefont {B.~A.}\ \bibnamefont
  {Frandsen}}, \bibinfo {author} {\bibfnamefont {Q.}~\bibnamefont {Wang}},
  \bibinfo {author} {\bibfnamefont {S.}~\bibnamefont {Wu}}, \bibinfo {author}
  {\bibfnamefont {J.}~\bibnamefont {Zhao}},\ and\ \bibinfo {author}
  {\bibfnamefont {R.~J.}\ \bibnamefont {Birgeneau}},\ }\bibfield  {title}
  {\bibinfo {title} {{Quantitative characterization of short-range orthorhombic
  fluctuations in FeSe through pair distribution function analysis}},\ }\href
  {https://doi.org/10.1103/PhysRevB.100.020504} {\bibfield  {journal} {\bibinfo
   {journal} {Phys. Rev. B}\ }\textbf {\bibinfo {volume} {100}},\ \bibinfo
  {pages} {020504} (\bibinfo {year} {2019})}\BibitemShut {NoStop}%
\bibitem [{\citenamefont {Koch}\ \emph {et~al.}(2019)\citenamefont {Koch},
  \citenamefont {Konstantinova}, \citenamefont {Abeykoon}, \citenamefont
  {Wang}, \citenamefont {Petrovic}, \citenamefont {Zhu}, \citenamefont
  {Bozin},\ and\ \citenamefont {Billinge}}]{Billinge19}%
  \BibitemOpen
  \bibfield  {author} {\bibinfo {author} {\bibfnamefont {R.~J.}\ \bibnamefont
  {Koch}}, \bibinfo {author} {\bibfnamefont {T.}~\bibnamefont {Konstantinova}},
  \bibinfo {author} {\bibfnamefont {M.}~\bibnamefont {Abeykoon}}, \bibinfo
  {author} {\bibfnamefont {A.}~\bibnamefont {Wang}}, \bibinfo {author}
  {\bibfnamefont {C.}~\bibnamefont {Petrovic}}, \bibinfo {author}
  {\bibfnamefont {Y.}~\bibnamefont {Zhu}}, \bibinfo {author} {\bibfnamefont
  {E.~S.}\ \bibnamefont {Bozin}},\ and\ \bibinfo {author} {\bibfnamefont
  {S.~J.~L.}\ \bibnamefont {Billinge}},\ }\bibfield  {title} {\bibinfo {title}
  {{Room temperature local nematicity in FeSe superconductor}},\ }\href
  {https://doi.org/10.1103/PhysRevB.100.020501} {\bibfield  {journal} {\bibinfo
   {journal} {Phys. Rev. B}\ }\textbf {\bibinfo {volume} {100}},\ \bibinfo
  {pages} {020501} (\bibinfo {year} {2019})}\BibitemShut {NoStop}%
\bibitem [{\citenamefont {Kuo}\ \emph {et~al.}(2016)\citenamefont {Kuo},
  \citenamefont {Chu}, \citenamefont {Palmstrom}, \citenamefont {Kivelson},\
  and\ \citenamefont {Fisher}}]{Kuo16}%
  \BibitemOpen
  \bibfield  {author} {\bibinfo {author} {\bibfnamefont {H.-H.}\ \bibnamefont
  {Kuo}}, \bibinfo {author} {\bibfnamefont {J.-H.}\ \bibnamefont {Chu}},
  \bibinfo {author} {\bibfnamefont {J.~C.}\ \bibnamefont {Palmstrom}}, \bibinfo
  {author} {\bibfnamefont {S.~A.}\ \bibnamefont {Kivelson}},\ and\ \bibinfo
  {author} {\bibfnamefont {I.~R.}\ \bibnamefont {Fisher}},\ }\bibfield  {title}
  {\bibinfo {title} {{Ubiquitous signatures of nematic quantum criticality in
  optimally doped Fe-based superconductors}},\ }\href
  {https://doi.org/10.1126/science.aab0103} {\bibfield  {journal} {\bibinfo
  {journal} {Science}\ }\textbf {\bibinfo {volume} {352}},\ \bibinfo {pages}
  {958} (\bibinfo {year} {2016})}\BibitemShut {NoStop}%
\bibitem [{\citenamefont {Huang}\ and\ \citenamefont
  {Hoffman}(2017)}]{Hoffman}%
  \BibitemOpen
  \bibfield  {author} {\bibinfo {author} {\bibfnamefont {D.}~\bibnamefont
  {Huang}}\ and\ \bibinfo {author} {\bibfnamefont {J.~E.}\ \bibnamefont
  {Hoffman}},\ }\bibfield  {title} {\bibinfo {title} {{Monolayer FeSe on
  SrTiO$_3$}},\ }\href
  {https://doi.org/10.1146/annurev-conmatphys-031016-025242} {\bibfield
  {journal} {\bibinfo  {journal} {Annual Review of Condensed Matter Physics}\
  }\textbf {\bibinfo {volume} {8}},\ \bibinfo {pages} {311} (\bibinfo {year}
  {2017})}\BibitemShut {NoStop}%
\bibitem [{\citenamefont {Hosono}\ \emph {et~al.}(2018)\citenamefont {Hosono},
  \citenamefont {Yamamoto}, \citenamefont {Hiramatsu},\ and\ \citenamefont
  {Ma}}]{Hoson18}%
  \BibitemOpen
  \bibfield  {author} {\bibinfo {author} {\bibfnamefont {H.}~\bibnamefont
  {Hosono}}, \bibinfo {author} {\bibfnamefont {A.}~\bibnamefont {Yamamoto}},
  \bibinfo {author} {\bibfnamefont {H.}~\bibnamefont {Hiramatsu}},\ and\
  \bibinfo {author} {\bibfnamefont {Y.}~\bibnamefont {Ma}},\ }\bibfield
  {title} {\bibinfo {title} {Recent advances in iron-based superconductors
  toward applications},\ }\href
  {https://doi.org/https://doi.org/10.1016/j.mattod.2017.09.006} {\bibfield
  {journal} {\bibinfo  {journal} {Materials Today}\ }\textbf {\bibinfo {volume}
  {21}},\ \bibinfo {pages} {278} (\bibinfo {year} {2018})}\BibitemShut
  {NoStop}%
\bibitem [{\citenamefont {Boeri}\ \emph {et~al.}(2008)\citenamefont {Boeri},
  \citenamefont {Dolgov},\ and\ \citenamefont {Golubov}}]{Boeri08}%
  \BibitemOpen
  \bibfield  {author} {\bibinfo {author} {\bibfnamefont {L.}~\bibnamefont
  {Boeri}}, \bibinfo {author} {\bibfnamefont {O.~V.}\ \bibnamefont {Dolgov}},\
  and\ \bibinfo {author} {\bibfnamefont {A.~A.}\ \bibnamefont {Golubov}},\
  }\bibfield  {title} {\bibinfo {title} {{Is
  ${\mathrm{LaFeAsO}}_{1\ensuremath{-}x}{\mathrm{F}}_{x}$ an Electron-Phonon
  Superconductor?}},\ }\href {https://doi.org/10.1103/PhysRevLett.101.026403}
  {\bibfield  {journal} {\bibinfo  {journal} {Phys. Rev. Lett.}\ }\textbf
  {\bibinfo {volume} {101}},\ \bibinfo {pages} {026403} (\bibinfo {year}
  {2008})}\BibitemShut {NoStop}%
\bibitem [{\citenamefont {Mandal}\ \emph {et~al.}(2014)\citenamefont {Mandal},
  \citenamefont {Cohen},\ and\ \citenamefont {Haule}}]{Mandal14}%
  \BibitemOpen
  \bibfield  {author} {\bibinfo {author} {\bibfnamefont {S.}~\bibnamefont
  {Mandal}}, \bibinfo {author} {\bibfnamefont {R.~E.}\ \bibnamefont {Cohen}},\
  and\ \bibinfo {author} {\bibfnamefont {K.}~\bibnamefont {Haule}},\ }\bibfield
   {title} {\bibinfo {title} {{Strong pressure-dependent electron-phonon
  coupling in FeSe}},\ }\href {https://doi.org/10.1103/PhysRevB.89.220502}
  {\bibfield  {journal} {\bibinfo  {journal} {Phys. Rev. B}\ }\textbf {\bibinfo
  {volume} {89}},\ \bibinfo {pages} {220502} (\bibinfo {year}
  {2014})}\BibitemShut {NoStop}%
\bibitem [{\citenamefont {Lee}\ \emph {et~al.}(2014)\citenamefont {Lee},
  \citenamefont {Schmitt}, \citenamefont {Moore}, \citenamefont {Johnston},
  \citenamefont {Cui}, \citenamefont {Li}, \citenamefont {Yi}, \citenamefont
  {Liu}, \citenamefont {Hashimoto}, \citenamefont {Zhang} \emph
  {et~al.}}]{Lee2014}%
  \BibitemOpen
  \bibfield  {author} {\bibinfo {author} {\bibfnamefont {J.}~\bibnamefont
  {Lee}}, \bibinfo {author} {\bibfnamefont {F.}~\bibnamefont {Schmitt}},
  \bibinfo {author} {\bibfnamefont {R.}~\bibnamefont {Moore}}, \bibinfo
  {author} {\bibfnamefont {S.}~\bibnamefont {Johnston}}, \bibinfo {author}
  {\bibfnamefont {Y.-T.}\ \bibnamefont {Cui}}, \bibinfo {author} {\bibfnamefont
  {W.}~\bibnamefont {Li}}, \bibinfo {author} {\bibfnamefont {M.}~\bibnamefont
  {Yi}}, \bibinfo {author} {\bibfnamefont {Z.}~\bibnamefont {Liu}}, \bibinfo
  {author} {\bibfnamefont {M.}~\bibnamefont {Hashimoto}}, \bibinfo {author}
  {\bibfnamefont {Y.}~\bibnamefont {Zhang}}, \emph {et~al.},\ }\bibfield
  {title} {\bibinfo {title} {{Interfacial mode coupling as the origin of the
  enhancement of $T_c$ in FeSe films on SrTiO$_3$}},\ }\href
  {https://doi.org/10.1038/nature13894} {\bibfield  {journal} {\bibinfo
  {journal} {Nature}\ }\textbf {\bibinfo {volume} {515}},\ \bibinfo {pages}
  {245} (\bibinfo {year} {2014})}\BibitemShut {NoStop}%
\bibitem [{\citenamefont {Platt}\ \emph {et~al.}(2013)\citenamefont {Platt},
  \citenamefont {Hanke},\ and\ \citenamefont {Thomale}}]{Platt13}%
  \BibitemOpen
  \bibfield  {author} {\bibinfo {author} {\bibfnamefont {C.}~\bibnamefont
  {Platt}}, \bibinfo {author} {\bibfnamefont {W.}~\bibnamefont {Hanke}},\ and\
  \bibinfo {author} {\bibfnamefont {R.}~\bibnamefont {Thomale}},\ }\bibfield
  {title} {\bibinfo {title} {Functional renormalization group for multi-orbital
  fermi surface instabilities},\ }\href
  {https://doi.org/10.1080/00018732.2013.862020} {\bibfield  {journal}
  {\bibinfo  {journal} {Advances in Physics}\ }\textbf {\bibinfo {volume}
  {62}},\ \bibinfo {pages} {453} (\bibinfo {year} {2013})}\BibitemShut
  {NoStop}%
\bibitem [{\citenamefont {Kontani}\ and\ \citenamefont
  {Onari}(2010)}]{Kontani10}%
  \BibitemOpen
  \bibfield  {author} {\bibinfo {author} {\bibfnamefont {H.}~\bibnamefont
  {Kontani}}\ and\ \bibinfo {author} {\bibfnamefont {S.}~\bibnamefont
  {Onari}},\ }\bibfield  {title} {\bibinfo {title}
  {{Orbital-Fluctuation-Mediated Superconductivity in Iron Pnictides: Analysis
  of the Five-Orbital Hubbard-Holstein Model}},\ }\href
  {https://doi.org/10.1103/PhysRevLett.104.157001} {\bibfield  {journal}
  {\bibinfo  {journal} {Phys. Rev. Lett.}\ }\textbf {\bibinfo {volume} {104}},\
  \bibinfo {pages} {157001} (\bibinfo {year} {2010})}\BibitemShut {NoStop}%
\bibitem [{\citenamefont {Chen}\ \emph {et~al.}(2010)\citenamefont {Chen},
  \citenamefont {Tsuei}, \citenamefont {Ketchen}, \citenamefont {Ren},\ and\
  \citenamefont {Zhao}}]{Chen10}%
  \BibitemOpen
  \bibfield  {author} {\bibinfo {author} {\bibfnamefont {C.-T.}\ \bibnamefont
  {Chen}}, \bibinfo {author} {\bibfnamefont {C.}~\bibnamefont {Tsuei}},
  \bibinfo {author} {\bibfnamefont {M.}~\bibnamefont {Ketchen}}, \bibinfo
  {author} {\bibfnamefont {Z.-A.}\ \bibnamefont {Ren}},\ and\ \bibinfo {author}
  {\bibfnamefont {Z.}~\bibnamefont {Zhao}},\ }\bibfield  {title} {\bibinfo
  {title} {Integer and half-integer flux-quantum transitions in a niobium-iron
  pnictide loop},\ }\href {https://doi.org/10.1038/nphys1531} {\bibfield
  {journal} {\bibinfo  {journal} {Nature Physics}\ }\textbf {\bibinfo {volume}
  {6}},\ \bibinfo {pages} {260} (\bibinfo {year} {2010})}\BibitemShut {NoStop}%
\bibitem [{\citenamefont {Cho}\ \emph {et~al.}(2018)\citenamefont {Cho},
  \citenamefont {Ko{\'{n}}czykowski}, \citenamefont {Teknowijoyo},
  \citenamefont {Tanatar},\ and\ \citenamefont {Prozorov}}]{Cho18}%
  \BibitemOpen
  \bibfield  {author} {\bibinfo {author} {\bibfnamefont {K.}~\bibnamefont
  {Cho}}, \bibinfo {author} {\bibfnamefont {M.}~\bibnamefont
  {Ko{\'{n}}czykowski}}, \bibinfo {author} {\bibfnamefont {S.}~\bibnamefont
  {Teknowijoyo}}, \bibinfo {author} {\bibfnamefont {M.~A.}\ \bibnamefont
  {Tanatar}},\ and\ \bibinfo {author} {\bibfnamefont {R.}~\bibnamefont
  {Prozorov}},\ }\bibfield  {title} {\bibinfo {title} {Using electron
  irradiation to probe iron-based superconductors},\ }\href
  {https://doi.org/10.1088/1361-6668/aabfa8} {\bibfield  {journal} {\bibinfo
  {journal} {Superconductor Science and Technology}\ }\textbf {\bibinfo
  {volume} {31}},\ \bibinfo {pages} {064002} (\bibinfo {year}
  {2018})}\BibitemShut {NoStop}%
\bibitem [{\citenamefont {Yang}\ \emph {et~al.}(2013)\citenamefont {Yang},
  \citenamefont {Wang}, \citenamefont {Fang}, \citenamefont {Deng},
  \citenamefont {Wang}, \citenamefont {Xiang}, \citenamefont {Yang},\ and\
  \citenamefont {Wen}}]{Yang13}%
  \BibitemOpen
  \bibfield  {author} {\bibinfo {author} {\bibfnamefont {H.}~\bibnamefont
  {Yang}}, \bibinfo {author} {\bibfnamefont {Z.}~\bibnamefont {Wang}}, \bibinfo
  {author} {\bibfnamefont {D.}~\bibnamefont {Fang}}, \bibinfo {author}
  {\bibfnamefont {Q.}~\bibnamefont {Deng}}, \bibinfo {author} {\bibfnamefont
  {Q.-H.}\ \bibnamefont {Wang}}, \bibinfo {author} {\bibfnamefont {Y.-Y.}\
  \bibnamefont {Xiang}}, \bibinfo {author} {\bibfnamefont {Y.}~\bibnamefont
  {Yang}},\ and\ \bibinfo {author} {\bibfnamefont {H.-H.}\ \bibnamefont
  {Wen}},\ }\bibfield  {title} {\bibinfo {title} {{In-gap quasiparticle
  excitations induced by non-magnetic Cu impurities in
  Na(Fe$_{0.96}$Co$_{0.03}$Cu$_{0.01}$)As revealed by scanning tunnelling
  spectroscopy}},\ }\href {https://doi.org/10.1038/ncomms3749} {\bibfield
  {journal} {\bibinfo  {journal} {Nature Communications}\ }\textbf {\bibinfo
  {volume} {4}},\ \bibinfo {pages} {2749} (\bibinfo {year} {2013})}\BibitemShut
  {NoStop}%
\bibitem [{\citenamefont {Kuroki}\ \emph {et~al.}(2009)\citenamefont {Kuroki},
  \citenamefont {Usui}, \citenamefont {Onari}, \citenamefont {Arita},\ and\
  \citenamefont {Aoki}}]{Kuroki09}%
  \BibitemOpen
  \bibfield  {author} {\bibinfo {author} {\bibfnamefont {K.}~\bibnamefont
  {Kuroki}}, \bibinfo {author} {\bibfnamefont {H.}~\bibnamefont {Usui}},
  \bibinfo {author} {\bibfnamefont {S.}~\bibnamefont {Onari}}, \bibinfo
  {author} {\bibfnamefont {R.}~\bibnamefont {Arita}},\ and\ \bibinfo {author}
  {\bibfnamefont {H.}~\bibnamefont {Aoki}},\ }\bibfield  {title} {\bibinfo
  {title} {{Pnictogen height as a possible switch between high-${T}_{c}$
  nodeless and low-${T}_{c}$ nodal pairings in the iron-based
  superconductors}},\ }\href {https://doi.org/10.1103/PhysRevB.79.224511}
  {\bibfield  {journal} {\bibinfo  {journal} {Phys. Rev. B}\ }\textbf {\bibinfo
  {volume} {79}},\ \bibinfo {pages} {224511} (\bibinfo {year}
  {2009})}\BibitemShut {NoStop}%
\bibitem [{\citenamefont {Vafek}\ and\ \citenamefont
  {Chubukov}(2017)}]{Vafek_Chubukov}%
  \BibitemOpen
  \bibfield  {author} {\bibinfo {author} {\bibfnamefont {O.}~\bibnamefont
  {Vafek}}\ and\ \bibinfo {author} {\bibfnamefont {A.~V.}\ \bibnamefont
  {Chubukov}},\ }\bibfield  {title} {\bibinfo {title} {{Hund Interaction,
  Spin-Orbit Coupling, and the Mechanism of Superconductivity in Strongly
  Hole-Doped Iron Pnictides}},\ }\href
  {https://doi.org/10.1103/PhysRevLett.118.087003} {\bibfield  {journal}
  {\bibinfo  {journal} {Phys. Rev. Lett.}\ }\textbf {\bibinfo {volume} {118}},\
  \bibinfo {pages} {087003} (\bibinfo {year} {2017})}\BibitemShut {NoStop}%
\bibitem [{\citenamefont {Lee}\ \emph {et~al.}(2009{\natexlab{b}})\citenamefont
  {Lee}, \citenamefont {Zhang},\ and\ \citenamefont {Wu}}]{Congjun09}%
  \BibitemOpen
  \bibfield  {author} {\bibinfo {author} {\bibfnamefont {W.-C.}\ \bibnamefont
  {Lee}}, \bibinfo {author} {\bibfnamefont {S.-C.}\ \bibnamefont {Zhang}},\
  and\ \bibinfo {author} {\bibfnamefont {C.}~\bibnamefont {Wu}},\ }\bibfield
  {title} {\bibinfo {title} {{Pairing State with a Time-Reversal Symmetry
  Breaking in FeAs-Based Superconductors}},\ }\href
  {https://doi.org/10.1103/PhysRevLett.102.217002} {\bibfield  {journal}
  {\bibinfo  {journal} {Phys. Rev. Lett.}\ }\textbf {\bibinfo {volume} {102}},\
  \bibinfo {pages} {217002} (\bibinfo {year} {2009}{\natexlab{b}})}\BibitemShut
  {NoStop}%
\bibitem [{\citenamefont {Kretzschmar}\ \emph {et~al.}(2013)\citenamefont
  {Kretzschmar}, \citenamefont {Muschler}, \citenamefont {B\"ohm},
  \citenamefont {Baum}, \citenamefont {Hackl}, \citenamefont {Wen},
  \citenamefont {Tsurkan}, \citenamefont {Deisenhofer},\ and\ \citenamefont
  {Loidl}}]{Hackl13}%
  \BibitemOpen
  \bibfield  {author} {\bibinfo {author} {\bibfnamefont {F.}~\bibnamefont
  {Kretzschmar}}, \bibinfo {author} {\bibfnamefont {B.}~\bibnamefont
  {Muschler}}, \bibinfo {author} {\bibfnamefont {T.}~\bibnamefont {B\"ohm}},
  \bibinfo {author} {\bibfnamefont {A.}~\bibnamefont {Baum}}, \bibinfo {author}
  {\bibfnamefont {R.}~\bibnamefont {Hackl}}, \bibinfo {author} {\bibfnamefont
  {H.-H.}\ \bibnamefont {Wen}}, \bibinfo {author} {\bibfnamefont
  {V.}~\bibnamefont {Tsurkan}}, \bibinfo {author} {\bibfnamefont
  {J.}~\bibnamefont {Deisenhofer}},\ and\ \bibinfo {author} {\bibfnamefont
  {A.}~\bibnamefont {Loidl}},\ }\bibfield  {title} {\bibinfo {title}
  {{Raman-Scattering Detection of Nearly Degenerate $s$-Wave and $d$-Wave
  Pairing Channels in Iron-Based
  ${\mathrm{Ba}}_{0.6}{\mathrm{K}}_{0.4}{\mathrm{Fe}}_{2}{\mathrm{As}}_{2}$ and
  ${\mathrm{Rb}}_{0.8}{\mathrm{Fe}}_{1.6}{\mathrm{Se}}_{2}$ Superconductors}},\
  }\href {https://doi.org/10.1103/PhysRevLett.110.187002} {\bibfield  {journal}
  {\bibinfo  {journal} {Phys. Rev. Lett.}\ }\textbf {\bibinfo {volume} {110}},\
  \bibinfo {pages} {187002} (\bibinfo {year} {2013})}\BibitemShut {NoStop}%
\bibitem [{\citenamefont {Thorsm\o{}lle}\ \emph {et~al.}(2016)\citenamefont
  {Thorsm\o{}lle}, \citenamefont {Khodas}, \citenamefont {Yin}, \citenamefont
  {Zhang}, \citenamefont {Carr}, \citenamefont {Dai},\ and\ \citenamefont
  {Blumberg}}]{Blumberg16}%
  \BibitemOpen
  \bibfield  {author} {\bibinfo {author} {\bibfnamefont {V.~K.}\ \bibnamefont
  {Thorsm\o{}lle}}, \bibinfo {author} {\bibfnamefont {M.}~\bibnamefont
  {Khodas}}, \bibinfo {author} {\bibfnamefont {Z.~P.}\ \bibnamefont {Yin}},
  \bibinfo {author} {\bibfnamefont {C.}~\bibnamefont {Zhang}}, \bibinfo
  {author} {\bibfnamefont {S.~V.}\ \bibnamefont {Carr}}, \bibinfo {author}
  {\bibfnamefont {P.}~\bibnamefont {Dai}},\ and\ \bibinfo {author}
  {\bibfnamefont {G.}~\bibnamefont {Blumberg}},\ }\bibfield  {title} {\bibinfo
  {title} {Critical quadrupole fluctuations and collective modes in iron
  pnictide superconductors},\ }\href
  {https://doi.org/10.1103/PhysRevB.93.054515} {\bibfield  {journal} {\bibinfo
  {journal} {Phys. Rev. B}\ }\textbf {\bibinfo {volume} {93}},\ \bibinfo
  {pages} {054515} (\bibinfo {year} {2016})}\BibitemShut {NoStop}%
\bibitem [{\citenamefont {Gallais}\ \emph {et~al.}(2016)\citenamefont
  {Gallais}, \citenamefont {Paul}, \citenamefont {Chauvi\`ere},\ and\
  \citenamefont {Schmalian}}]{Gallais16}%
  \BibitemOpen
  \bibfield  {author} {\bibinfo {author} {\bibfnamefont {Y.}~\bibnamefont
  {Gallais}}, \bibinfo {author} {\bibfnamefont {I.}~\bibnamefont {Paul}},
  \bibinfo {author} {\bibfnamefont {L.}~\bibnamefont {Chauvi\`ere}},\ and\
  \bibinfo {author} {\bibfnamefont {J.}~\bibnamefont {Schmalian}},\ }\bibfield
  {title} {\bibinfo {title} {{Nematic Resonance in the Raman Response of
  Iron-Based Superconductors}},\ }\href
  {https://doi.org/10.1103/PhysRevLett.116.017001} {\bibfield  {journal}
  {\bibinfo  {journal} {Phys. Rev. Lett.}\ }\textbf {\bibinfo {volume} {116}},\
  \bibinfo {pages} {017001} (\bibinfo {year} {2016})}\BibitemShut {NoStop}%
\bibitem [{\citenamefont {Tafti}\ \emph {et~al.}(2013)\citenamefont {Tafti},
  \citenamefont {Juneau-Fecteau}, \citenamefont {Delage}, \citenamefont
  {De~Cotret}, \citenamefont {Reid}, \citenamefont {Wang}, \citenamefont {Luo},
  \citenamefont {Chen}, \citenamefont {Doiron-Leyraud},\ and\ \citenamefont
  {Taillefer}}]{Tafti13}%
  \BibitemOpen
  \bibfield  {author} {\bibinfo {author} {\bibfnamefont {F.}~\bibnamefont
  {Tafti}}, \bibinfo {author} {\bibfnamefont {A.}~\bibnamefont
  {Juneau-Fecteau}}, \bibinfo {author} {\bibfnamefont {M.-E.}\ \bibnamefont
  {Delage}}, \bibinfo {author} {\bibfnamefont {S.~R.}\ \bibnamefont
  {De~Cotret}}, \bibinfo {author} {\bibfnamefont {J.-P.}\ \bibnamefont {Reid}},
  \bibinfo {author} {\bibfnamefont {A.}~\bibnamefont {Wang}}, \bibinfo {author}
  {\bibfnamefont {X.}~\bibnamefont {Luo}}, \bibinfo {author} {\bibfnamefont
  {X.}~\bibnamefont {Chen}}, \bibinfo {author} {\bibfnamefont {N.}~\bibnamefont
  {Doiron-Leyraud}},\ and\ \bibinfo {author} {\bibfnamefont {L.}~\bibnamefont
  {Taillefer}},\ }\bibfield  {title} {\bibinfo {title} {{Sudden reversal in the
  pressure dependence of $T_c$ in the iron-based superconductor
  KFe$_2$As$_2$}},\ }\href {https://doi.org/10.1038/nphys2617} {\bibfield
  {journal} {\bibinfo  {journal} {Nature Physics}\ }\textbf {\bibinfo {volume}
  {9}},\ \bibinfo {pages} {349} (\bibinfo {year} {2013})}\BibitemShut {NoStop}%
\bibitem [{\citenamefont {Rinott}\ \emph {et~al.}(2017)\citenamefont {Rinott},
  \citenamefont {Chashka}, \citenamefont {Ribak}, \citenamefont {Rienks},
  \citenamefont {Taleb-Ibrahimi}, \citenamefont {Le~Fevre}, \citenamefont
  {Bertran}, \citenamefont {Randeria},\ and\ \citenamefont
  {Kanigel}}]{Kanigel17}%
  \BibitemOpen
  \bibfield  {author} {\bibinfo {author} {\bibfnamefont {S.}~\bibnamefont
  {Rinott}}, \bibinfo {author} {\bibfnamefont {K.~B.}\ \bibnamefont {Chashka}},
  \bibinfo {author} {\bibfnamefont {A.}~\bibnamefont {Ribak}}, \bibinfo
  {author} {\bibfnamefont {E.~D.~L.}\ \bibnamefont {Rienks}}, \bibinfo {author}
  {\bibfnamefont {A.}~\bibnamefont {Taleb-Ibrahimi}}, \bibinfo {author}
  {\bibfnamefont {P.}~\bibnamefont {Le~Fevre}}, \bibinfo {author}
  {\bibfnamefont {F.}~\bibnamefont {Bertran}}, \bibinfo {author} {\bibfnamefont
  {M.}~\bibnamefont {Randeria}},\ and\ \bibinfo {author} {\bibfnamefont
  {A.}~\bibnamefont {Kanigel}},\ }\bibfield  {title} {\bibinfo {title} {{Tuning
  across the BCS-BEC crossover in the multiband superconductor
  Fe$_{1+y}$Se$_x$Te$_{1-x}$: An angle-resolved photoemission study}},\ }\href
  {https://doi.org/10.1126/sciadv.1602372} {\bibfield  {journal} {\bibinfo
  {journal} {Science Advances}\ }\textbf {\bibinfo {volume} {3}},\ \bibinfo
  {pages} {e1602372} (\bibinfo {year} {2017})}\BibitemShut {NoStop}%
\bibitem [{\citenamefont {Kong}\ \emph {et~al.}(2019)\citenamefont {Kong},
  \citenamefont {Zhu}, \citenamefont {Papaj}, \citenamefont {Chen},
  \citenamefont {Cao}, \citenamefont {Isobe}, \citenamefont {Xing},
  \citenamefont {Liu}, \citenamefont {Wang}, \citenamefont {Fan} \emph
  {et~al.}}]{Kong19}%
  \BibitemOpen
  \bibfield  {author} {\bibinfo {author} {\bibfnamefont {L.}~\bibnamefont
  {Kong}}, \bibinfo {author} {\bibfnamefont {S.}~\bibnamefont {Zhu}}, \bibinfo
  {author} {\bibfnamefont {M.}~\bibnamefont {Papaj}}, \bibinfo {author}
  {\bibfnamefont {H.}~\bibnamefont {Chen}}, \bibinfo {author} {\bibfnamefont
  {L.}~\bibnamefont {Cao}}, \bibinfo {author} {\bibfnamefont {H.}~\bibnamefont
  {Isobe}}, \bibinfo {author} {\bibfnamefont {Y.}~\bibnamefont {Xing}},
  \bibinfo {author} {\bibfnamefont {W.}~\bibnamefont {Liu}}, \bibinfo {author}
  {\bibfnamefont {D.}~\bibnamefont {Wang}}, \bibinfo {author} {\bibfnamefont
  {P.}~\bibnamefont {Fan}}, \emph {et~al.},\ }\bibfield  {title} {\bibinfo
  {title} {Half-integer level shift of vortex bound states in an iron-based
  superconductor},\ }\href {https://doi.org/10.1038/s41567-019-0630-5}
  {\bibfield  {journal} {\bibinfo  {journal} {Nature Physics}\ }\textbf
  {\bibinfo {volume} {15}},\ \bibinfo {pages} {1181} (\bibinfo {year}
  {2019})}\BibitemShut {NoStop}%
\bibitem [{\citenamefont {Wu}\ \emph {et~al.}(2020{\natexlab{a}})\citenamefont
  {Wu}, \citenamefont {Zhang}, \citenamefont {Xu}, \citenamefont {Hu},\ and\
  \citenamefont {Liu}}]{Wu_review}%
  \BibitemOpen
  \bibfield  {author} {\bibinfo {author} {\bibfnamefont {X.}~\bibnamefont
  {Wu}}, \bibinfo {author} {\bibfnamefont {R.-X.}\ \bibnamefont {Zhang}},
  \bibinfo {author} {\bibfnamefont {G.}~\bibnamefont {Xu}}, \bibinfo {author}
  {\bibfnamefont {J.}~\bibnamefont {Hu}},\ and\ \bibinfo {author}
  {\bibfnamefont {C.-X.}\ \bibnamefont {Liu}},\ }\bibfield  {title} {\bibinfo
  {title} {{In the Pursuit of Majorana Modes in Iron-based High-$T_c$
  Superconductors}},\ }\href@noop {} {\bibfield  {journal} {\bibinfo  {journal}
  {arXiv:2005.03603}\ } (\bibinfo {year} {2020}{\natexlab{a}})}\BibitemShut
  {NoStop}%
\bibitem [{\citenamefont {Kong}\ and\ \citenamefont
  {Ding}(2020)}]{Kong_review}%
  \BibitemOpen
  \bibfield  {author} {\bibinfo {author} {\bibfnamefont {L.-Y.}\ \bibnamefont
  {Kong}}\ and\ \bibinfo {author} {\bibfnamefont {H.}~\bibnamefont {Ding}},\
  }\bibfield  {title} {\bibinfo {title} {{Emergent vortex Majorana zero mode in
  iron-based superconductors}},\ }\href@noop {} {\bibfield  {journal} {\bibinfo
   {journal} {Acta Physica Sinica}\ }\textbf {\bibinfo {volume} {69}},\
  \bibinfo {pages} {110301} (\bibinfo {year} {2020})}\BibitemShut {NoStop}%
\bibitem [{\citenamefont {Lohani}\ \emph {et~al.}(2020)\citenamefont {Lohani},
  \citenamefont {Hazra}, \citenamefont {Ribak}, \citenamefont {Nitzav},
  \citenamefont {Fu}, \citenamefont {Yan}, \citenamefont {Randeria},\ and\
  \citenamefont {Kanigel}}]{Kanigel20}%
  \BibitemOpen
  \bibfield  {author} {\bibinfo {author} {\bibfnamefont {H.}~\bibnamefont
  {Lohani}}, \bibinfo {author} {\bibfnamefont {T.}~\bibnamefont {Hazra}},
  \bibinfo {author} {\bibfnamefont {A.}~\bibnamefont {Ribak}}, \bibinfo
  {author} {\bibfnamefont {Y.}~\bibnamefont {Nitzav}}, \bibinfo {author}
  {\bibfnamefont {H.}~\bibnamefont {Fu}}, \bibinfo {author} {\bibfnamefont
  {B.}~\bibnamefont {Yan}}, \bibinfo {author} {\bibfnamefont {M.}~\bibnamefont
  {Randeria}},\ and\ \bibinfo {author} {\bibfnamefont {A.}~\bibnamefont
  {Kanigel}},\ }\bibfield  {title} {\bibinfo {title} {{Band inversion and
  topology of the bulk electronic structure in
  ${\mathrm{FeSe}}_{0.45}{\mathrm{Te}}_{0.55}$}},\ }\href
  {https://doi.org/10.1103/PhysRevB.101.245146} {\bibfield  {journal} {\bibinfo
   {journal} {Phys. Rev. B}\ }\textbf {\bibinfo {volume} {101}},\ \bibinfo
  {pages} {245146} (\bibinfo {year} {2020})}\BibitemShut {NoStop}%
\bibitem [{\citenamefont {Zhang}\ \emph
  {et~al.}(2019{\natexlab{a}})\citenamefont {Zhang}, \citenamefont {Wang},
  \citenamefont {Wu}, \citenamefont {Yaji}, \citenamefont {Ishida},
  \citenamefont {Kohama}, \citenamefont {Dai}, \citenamefont {Sun},
  \citenamefont {Bareille}, \citenamefont {Kuroda} \emph {et~al.}}]{Zhang19}%
  \BibitemOpen
  \bibfield  {author} {\bibinfo {author} {\bibfnamefont {P.}~\bibnamefont
  {Zhang}}, \bibinfo {author} {\bibfnamefont {Z.}~\bibnamefont {Wang}},
  \bibinfo {author} {\bibfnamefont {X.}~\bibnamefont {Wu}}, \bibinfo {author}
  {\bibfnamefont {K.}~\bibnamefont {Yaji}}, \bibinfo {author} {\bibfnamefont
  {Y.}~\bibnamefont {Ishida}}, \bibinfo {author} {\bibfnamefont
  {Y.}~\bibnamefont {Kohama}}, \bibinfo {author} {\bibfnamefont
  {G.}~\bibnamefont {Dai}}, \bibinfo {author} {\bibfnamefont {Y.}~\bibnamefont
  {Sun}}, \bibinfo {author} {\bibfnamefont {C.}~\bibnamefont {Bareille}},
  \bibinfo {author} {\bibfnamefont {K.}~\bibnamefont {Kuroda}}, \emph
  {et~al.},\ }\bibfield  {title} {\bibinfo {title} {Multiple topological states
  in iron-based superconductors},\ }\href
  {https://doi.org/10.1038/s41567-018-0280-z} {\bibfield  {journal} {\bibinfo
  {journal} {Nature Physics}\ }\textbf {\bibinfo {volume} {15}},\ \bibinfo
  {pages} {41} (\bibinfo {year} {2019}{\natexlab{a}})}\BibitemShut {NoStop}%
\bibitem [{\citenamefont {Chen}\ \emph {et~al.}(2018)\citenamefont {Chen},
  \citenamefont {Chen}, \citenamefont {Yang}, \citenamefont {Du}, \citenamefont
  {Zhu}, \citenamefont {Wang},\ and\ \citenamefont {Wen}}]{Chen18}%
  \BibitemOpen
  \bibfield  {author} {\bibinfo {author} {\bibfnamefont {M.}~\bibnamefont
  {Chen}}, \bibinfo {author} {\bibfnamefont {X.}~\bibnamefont {Chen}}, \bibinfo
  {author} {\bibfnamefont {H.}~\bibnamefont {Yang}}, \bibinfo {author}
  {\bibfnamefont {Z.}~\bibnamefont {Du}}, \bibinfo {author} {\bibfnamefont
  {X.}~\bibnamefont {Zhu}}, \bibinfo {author} {\bibfnamefont {E.}~\bibnamefont
  {Wang}},\ and\ \bibinfo {author} {\bibfnamefont {H.-H.}\ \bibnamefont
  {Wen}},\ }\bibfield  {title} {\bibinfo {title} {{Discrete energy levels of
  Caroli-de Gennes-Matricon states in quantum limit in
  FeTe$_{0.55}$Se$_{0.45}$}},\ }\href
  {https://doi.org/10.1038/s41467-018-03404-8} {\bibfield  {journal} {\bibinfo
  {journal} {Nature communications}\ }\textbf {\bibinfo {volume} {9}},\
  \bibinfo {pages} {970} (\bibinfo {year} {2018})}\BibitemShut {NoStop}%
\bibitem [{\citenamefont {Wang}\ \emph {et~al.}(2018)\citenamefont {Wang},
  \citenamefont {Kong}, \citenamefont {Fan}, \citenamefont {Chen},
  \citenamefont {Zhu}, \citenamefont {Liu}, \citenamefont {Cao}, \citenamefont
  {Sun}, \citenamefont {Du}, \citenamefont {Schneeloch}, \citenamefont {Zhong},
  \citenamefont {Gu}, \citenamefont {Fu}, \citenamefont {Ding},\ and\
  \citenamefont {Gao}}]{Gao18}%
  \BibitemOpen
  \bibfield  {author} {\bibinfo {author} {\bibfnamefont {D.}~\bibnamefont
  {Wang}}, \bibinfo {author} {\bibfnamefont {L.}~\bibnamefont {Kong}}, \bibinfo
  {author} {\bibfnamefont {P.}~\bibnamefont {Fan}}, \bibinfo {author}
  {\bibfnamefont {H.}~\bibnamefont {Chen}}, \bibinfo {author} {\bibfnamefont
  {S.}~\bibnamefont {Zhu}}, \bibinfo {author} {\bibfnamefont {W.}~\bibnamefont
  {Liu}}, \bibinfo {author} {\bibfnamefont {L.}~\bibnamefont {Cao}}, \bibinfo
  {author} {\bibfnamefont {Y.}~\bibnamefont {Sun}}, \bibinfo {author}
  {\bibfnamefont {S.}~\bibnamefont {Du}}, \bibinfo {author} {\bibfnamefont
  {J.}~\bibnamefont {Schneeloch}}, \bibinfo {author} {\bibfnamefont
  {R.}~\bibnamefont {Zhong}}, \bibinfo {author} {\bibfnamefont
  {G.}~\bibnamefont {Gu}}, \bibinfo {author} {\bibfnamefont {L.}~\bibnamefont
  {Fu}}, \bibinfo {author} {\bibfnamefont {H.}~\bibnamefont {Ding}},\ and\
  \bibinfo {author} {\bibfnamefont {H.-J.}\ \bibnamefont {Gao}},\ }\bibfield
  {title} {\bibinfo {title} {{Evidence for Majorana bound states in an
  iron-based superconductor}},\ }\href
  {https://doi.org/10.1126/science.aao1797} {\bibfield  {journal} {\bibinfo
  {journal} {Science}\ }\textbf {\bibinfo {volume} {362}},\ \bibinfo {pages}
  {333} (\bibinfo {year} {2018})}\BibitemShut {NoStop}%
\bibitem [{\citenamefont {Machida}\ \emph {et~al.}(2019)\citenamefont
  {Machida}, \citenamefont {Sun}, \citenamefont {Pyon}, \citenamefont {Takeda},
  \citenamefont {Kohsaka}, \citenamefont {Hanaguri}, \citenamefont {Sasagawa},\
  and\ \citenamefont {Tamegai}}]{Machida19}%
  \BibitemOpen
  \bibfield  {author} {\bibinfo {author} {\bibfnamefont {T.}~\bibnamefont
  {Machida}}, \bibinfo {author} {\bibfnamefont {Y.}~\bibnamefont {Sun}},
  \bibinfo {author} {\bibfnamefont {S.}~\bibnamefont {Pyon}}, \bibinfo {author}
  {\bibfnamefont {S.}~\bibnamefont {Takeda}}, \bibinfo {author} {\bibfnamefont
  {Y.}~\bibnamefont {Kohsaka}}, \bibinfo {author} {\bibfnamefont
  {T.}~\bibnamefont {Hanaguri}}, \bibinfo {author} {\bibfnamefont
  {T.}~\bibnamefont {Sasagawa}},\ and\ \bibinfo {author} {\bibfnamefont
  {T.}~\bibnamefont {Tamegai}},\ }\bibfield  {title} {\bibinfo {title}
  {{Zero-energy vortex bound state in the superconducting topological surface
  state of Fe(Se,Te)}},\ }\href {https://doi.org/10.1038/s41563-019-0397-1}
  {\bibfield  {journal} {\bibinfo  {journal} {Nature Materials}\ }\textbf
  {\bibinfo {volume} {18}},\ \bibinfo {pages} {811} (\bibinfo {year}
  {2019})}\BibitemShut {NoStop}%
\bibitem [{\citenamefont {Yin}\ \emph {et~al.}(2015)\citenamefont {Yin},
  \citenamefont {Wu}, \citenamefont {Wang}, \citenamefont {Ye}, \citenamefont
  {Gong}, \citenamefont {Hou}, \citenamefont {Shan}, \citenamefont {Li},
  \citenamefont {Liang}, \citenamefont {Wu} \emph {et~al.}}]{Yin15}%
  \BibitemOpen
  \bibfield  {author} {\bibinfo {author} {\bibfnamefont {J.-X.}\ \bibnamefont
  {Yin}}, \bibinfo {author} {\bibfnamefont {Z.}~\bibnamefont {Wu}}, \bibinfo
  {author} {\bibfnamefont {J.}~\bibnamefont {Wang}}, \bibinfo {author}
  {\bibfnamefont {Z.}~\bibnamefont {Ye}}, \bibinfo {author} {\bibfnamefont
  {J.}~\bibnamefont {Gong}}, \bibinfo {author} {\bibfnamefont {X.}~\bibnamefont
  {Hou}}, \bibinfo {author} {\bibfnamefont {L.}~\bibnamefont {Shan}}, \bibinfo
  {author} {\bibfnamefont {A.}~\bibnamefont {Li}}, \bibinfo {author}
  {\bibfnamefont {X.}~\bibnamefont {Liang}}, \bibinfo {author} {\bibfnamefont
  {X.}~\bibnamefont {Wu}}, \emph {et~al.},\ }\bibfield  {title} {\bibinfo
  {title} {{Observation of a robust zero-energy bound state in iron-based
  superconductor Fe(Te,Se)}},\ }\href {https://doi.org/10.1038/nphys3371}
  {\bibfield  {journal} {\bibinfo  {journal} {Nature Physics}\ }\textbf
  {\bibinfo {volume} {11}},\ \bibinfo {pages} {543} (\bibinfo {year}
  {2015})}\BibitemShut {NoStop}%
\bibitem [{\citenamefont {Chen}\ \emph {et~al.}(2020)\citenamefont {Chen},
  \citenamefont {Jiang}, \citenamefont {Zhang}, \citenamefont {Liu},
  \citenamefont {Liu}, \citenamefont {Wang},\ and\ \citenamefont
  {Wang}}]{Chen20}%
  \BibitemOpen
  \bibfield  {author} {\bibinfo {author} {\bibfnamefont {C.}~\bibnamefont
  {Chen}}, \bibinfo {author} {\bibfnamefont {K.}~\bibnamefont {Jiang}},
  \bibinfo {author} {\bibfnamefont {Y.}~\bibnamefont {Zhang}}, \bibinfo
  {author} {\bibfnamefont {C.}~\bibnamefont {Liu}}, \bibinfo {author}
  {\bibfnamefont {Y.}~\bibnamefont {Liu}}, \bibinfo {author} {\bibfnamefont
  {Z.}~\bibnamefont {Wang}},\ and\ \bibinfo {author} {\bibfnamefont
  {J.}~\bibnamefont {Wang}},\ }\bibfield  {title} {\bibinfo {title} {{Atomic
  line defects and zero-energy end states in monolayer Fe(Te,Se)
  high-temperature superconductors}},\ }\href
  {https://doi.org/10.1038/s41567-020-0813-0} {\bibfield  {journal} {\bibinfo
  {journal} {Nature Physics}\ }\textbf {\bibinfo {volume} {16}},\ \bibinfo
  {pages} {536} (\bibinfo {year} {2020})}\BibitemShut {NoStop}%
\bibitem [{\citenamefont {Wang}\ \emph {et~al.}(2020)\citenamefont {Wang},
  \citenamefont {Rodriguez}, \citenamefont {Jiao}, \citenamefont {Howard},
  \citenamefont {Graham}, \citenamefont {Gu}, \citenamefont {Hughes},
  \citenamefont {Morr},\ and\ \citenamefont {Madhavan}}]{Madhavan20}%
  \BibitemOpen
  \bibfield  {author} {\bibinfo {author} {\bibfnamefont {Z.}~\bibnamefont
  {Wang}}, \bibinfo {author} {\bibfnamefont {J.~O.}\ \bibnamefont {Rodriguez}},
  \bibinfo {author} {\bibfnamefont {L.}~\bibnamefont {Jiao}}, \bibinfo {author}
  {\bibfnamefont {S.}~\bibnamefont {Howard}}, \bibinfo {author} {\bibfnamefont
  {M.}~\bibnamefont {Graham}}, \bibinfo {author} {\bibfnamefont {G.~D.}\
  \bibnamefont {Gu}}, \bibinfo {author} {\bibfnamefont {T.~L.}\ \bibnamefont
  {Hughes}}, \bibinfo {author} {\bibfnamefont {D.~K.}\ \bibnamefont {Morr}},\
  and\ \bibinfo {author} {\bibfnamefont {V.}~\bibnamefont {Madhavan}},\
  }\bibfield  {title} {\bibinfo {title} {{Evidence for dispersing 1D Majorana
  channels in an iron-based superconductor}},\ }\href
  {https://doi.org/10.1126/science.aaw8419} {\bibfield  {journal} {\bibinfo
  {journal} {Science}\ }\textbf {\bibinfo {volume} {367}},\ \bibinfo {pages}
  {104} (\bibinfo {year} {2020})}\BibitemShut {NoStop}%
\bibitem [{\citenamefont {Li}\ \emph {et~al.}(2020)\citenamefont {Li},
  \citenamefont {Zaki}, \citenamefont {Garlea}, \citenamefont {Savici},
  \citenamefont {Fobes}, \citenamefont {Xu}, \citenamefont {Camino},
  \citenamefont {Petrovic}, \citenamefont {Gu}, \citenamefont {Johnson},
  \citenamefont {Tranquada},\ and\ \citenamefont {Zaliznyak}}]{Yangmu20}%
  \BibitemOpen
  \bibfield  {author} {\bibinfo {author} {\bibfnamefont {Y.}~\bibnamefont
  {Li}}, \bibinfo {author} {\bibfnamefont {N.}~\bibnamefont {Zaki}}, \bibinfo
  {author} {\bibfnamefont {V.~O.}\ \bibnamefont {Garlea}}, \bibinfo {author}
  {\bibfnamefont {A.~T.}\ \bibnamefont {Savici}}, \bibinfo {author}
  {\bibfnamefont {D.}~\bibnamefont {Fobes}}, \bibinfo {author} {\bibfnamefont
  {Z.}~\bibnamefont {Xu}}, \bibinfo {author} {\bibfnamefont {F.}~\bibnamefont
  {Camino}}, \bibinfo {author} {\bibfnamefont {C.}~\bibnamefont {Petrovic}},
  \bibinfo {author} {\bibfnamefont {G.}~\bibnamefont {Gu}}, \bibinfo {author}
  {\bibfnamefont {P.~D.}\ \bibnamefont {Johnson}}, \bibinfo {author}
  {\bibfnamefont {J.~M.}\ \bibnamefont {Tranquada}},\ and\ \bibinfo {author}
  {\bibfnamefont {I.~A.}\ \bibnamefont {Zaliznyak}},\ }\bibfield  {title}
  {\bibinfo {title} {{Magnetic, superconducting, and topological surface states
  on Fe$_{1+y}$Te$_{1-x}$Se$_ {x}$}},\ }\href@noop {} {\bibfield  {journal}
  {\bibinfo  {journal} {arXiv:2012.07893}\ } (\bibinfo {year}
  {2020})}\BibitemShut {NoStop}%
\bibitem [{\citenamefont {K\"onig}\ and\ \citenamefont
  {Coleman}(2019)}]{Konig_Majorana}%
  \BibitemOpen
  \bibfield  {author} {\bibinfo {author} {\bibfnamefont {E.~J.}\ \bibnamefont
  {K\"onig}}\ and\ \bibinfo {author} {\bibfnamefont {P.}~\bibnamefont
  {Coleman}},\ }\bibfield  {title} {\bibinfo {title}
  {{Crystalline-Symmetry-Protected Helical Majorana Modes in the Iron
  Pnictides}},\ }\href {https://doi.org/10.1103/PhysRevLett.122.207001}
  {\bibfield  {journal} {\bibinfo  {journal} {Phys. Rev. Lett.}\ }\textbf
  {\bibinfo {volume} {122}},\ \bibinfo {pages} {207001} (\bibinfo {year}
  {2019})}\BibitemShut {NoStop}%
\bibitem [{\citenamefont {Zhang}\ \emph
  {et~al.}(2019{\natexlab{b}})\citenamefont {Zhang}, \citenamefont {Cole},\
  and\ \citenamefont {Das~Sarma}}]{DasSarma19}%
  \BibitemOpen
  \bibfield  {author} {\bibinfo {author} {\bibfnamefont {R.-X.}\ \bibnamefont
  {Zhang}}, \bibinfo {author} {\bibfnamefont {W.~S.}\ \bibnamefont {Cole}},\
  and\ \bibinfo {author} {\bibfnamefont {S.}~\bibnamefont {Das~Sarma}},\
  }\bibfield  {title} {\bibinfo {title} {{Helical Hinge Majorana Modes in
  Iron-Based Superconductors}},\ }\href
  {https://doi.org/10.1103/PhysRevLett.122.187001} {\bibfield  {journal}
  {\bibinfo  {journal} {Phys. Rev. Lett.}\ }\textbf {\bibinfo {volume} {122}},\
  \bibinfo {pages} {187001} (\bibinfo {year} {2019}{\natexlab{b}})}\BibitemShut
  {NoStop}%
\bibitem [{\citenamefont {Heinsdorf}\ \emph {et~al.}(2021)\citenamefont
  {Heinsdorf}, \citenamefont {Christensen}, \citenamefont {Iraola},
  \citenamefont {Zhang}, \citenamefont {Yang}, \citenamefont {Birol},
  \citenamefont {Batista}, \citenamefont {Valent{\'\i}},\ and\ \citenamefont
  {Fernandes}}]{Heinsdorf21}%
  \BibitemOpen
  \bibfield  {author} {\bibinfo {author} {\bibfnamefont {N.}~\bibnamefont
  {Heinsdorf}}, \bibinfo {author} {\bibfnamefont {M.~H.}\ \bibnamefont
  {Christensen}}, \bibinfo {author} {\bibfnamefont {M.}~\bibnamefont {Iraola}},
  \bibinfo {author} {\bibfnamefont {S.}~\bibnamefont {Zhang}}, \bibinfo
  {author} {\bibfnamefont {F.}~\bibnamefont {Yang}}, \bibinfo {author}
  {\bibfnamefont {T.}~\bibnamefont {Birol}}, \bibinfo {author} {\bibfnamefont
  {C.~D.}\ \bibnamefont {Batista}}, \bibinfo {author} {\bibfnamefont
  {R.}~\bibnamefont {Valent{\'\i}}},\ and\ \bibinfo {author} {\bibfnamefont
  {R.~M.}\ \bibnamefont {Fernandes}},\ }\bibfield  {title} {\bibinfo {title}
  {{Prediction of Double-Weyl Points in the Iron-Based Superconductor
  CaKFe$_4$As$_4$}},\ }\href@noop {} {\bibfield  {journal} {\bibinfo  {journal}
  {arXiv:2101.05301}\ } (\bibinfo {year} {2021})}\BibitemShut {NoStop}%
\bibitem [{\citenamefont {Kong}\ \emph {et~al.}(2020)\citenamefont {Kong},
  \citenamefont {Cao}, \citenamefont {Zhu}, \citenamefont {Papaj},
  \citenamefont {Dai}, \citenamefont {Li}, \citenamefont {Fan}, \citenamefont
  {Liu}, \citenamefont {Yang}, \citenamefont {Wang} \emph {et~al.}}]{Kong20}%
  \BibitemOpen
  \bibfield  {author} {\bibinfo {author} {\bibfnamefont {L.}~\bibnamefont
  {Kong}}, \bibinfo {author} {\bibfnamefont {L.}~\bibnamefont {Cao}}, \bibinfo
  {author} {\bibfnamefont {S.}~\bibnamefont {Zhu}}, \bibinfo {author}
  {\bibfnamefont {M.}~\bibnamefont {Papaj}}, \bibinfo {author} {\bibfnamefont
  {G.}~\bibnamefont {Dai}}, \bibinfo {author} {\bibfnamefont {G.}~\bibnamefont
  {Li}}, \bibinfo {author} {\bibfnamefont {P.}~\bibnamefont {Fan}}, \bibinfo
  {author} {\bibfnamefont {W.}~\bibnamefont {Liu}}, \bibinfo {author}
  {\bibfnamefont {F.}~\bibnamefont {Yang}}, \bibinfo {author} {\bibfnamefont
  {X.}~\bibnamefont {Wang}}, \emph {et~al.},\ }\bibfield  {title} {\bibinfo
  {title} {{Tunable vortex Majorana zero modes in LiFeAs superconductor}},\
  }\href@noop {} {\bibfield  {journal} {\bibinfo  {journal} {arXiv:2010.04735}\
  } (\bibinfo {year} {2020})}\BibitemShut {NoStop}%
\bibitem [{\citenamefont {Katayama}\ \emph {et~al.}(2013)\citenamefont
  {Katayama}, \citenamefont {Kudo}, \citenamefont {Onari}, \citenamefont
  {Mizukami}, \citenamefont {Sugawara}, \citenamefont {Sugiyama}, \citenamefont
  {Kitahama}, \citenamefont {Iba}, \citenamefont {Fujimura}, \citenamefont
  {Nishimoto}, \citenamefont {Nohara},\ and\ \citenamefont
  {Sawa}}]{Katayama13}%
  \BibitemOpen
  \bibfield  {author} {\bibinfo {author} {\bibfnamefont {N.}~\bibnamefont
  {Katayama}}, \bibinfo {author} {\bibfnamefont {K.}~\bibnamefont {Kudo}},
  \bibinfo {author} {\bibfnamefont {S.}~\bibnamefont {Onari}}, \bibinfo
  {author} {\bibfnamefont {T.}~\bibnamefont {Mizukami}}, \bibinfo {author}
  {\bibfnamefont {K.}~\bibnamefont {Sugawara}}, \bibinfo {author}
  {\bibfnamefont {Y.}~\bibnamefont {Sugiyama}}, \bibinfo {author}
  {\bibfnamefont {Y.}~\bibnamefont {Kitahama}}, \bibinfo {author}
  {\bibfnamefont {K.}~\bibnamefont {Iba}}, \bibinfo {author} {\bibfnamefont
  {K.}~\bibnamefont {Fujimura}}, \bibinfo {author} {\bibfnamefont
  {N.}~\bibnamefont {Nishimoto}}, \bibinfo {author} {\bibfnamefont
  {M.}~\bibnamefont {Nohara}},\ and\ \bibinfo {author} {\bibfnamefont
  {H.}~\bibnamefont {Sawa}},\ }\bibfield  {title} {\bibinfo {title}
  {{Superconductivity in Ca$_{1-x}$La$_x$FeAs$_2$: A Novel 112-Type Iron
  Pnictide with Arsenic Zigzag Bonds}},\ }\href
  {https://doi.org/10.7566/JPSJ.82.123702} {\bibfield  {journal} {\bibinfo
  {journal} {Journal of the Physical Society of Japan}\ }\textbf {\bibinfo
  {volume} {82}},\ \bibinfo {pages} {123702} (\bibinfo {year}
  {2013})}\BibitemShut {NoStop}%
\bibitem [{\citenamefont {Dagotto}(2013)}]{Dagotto_selenides}%
  \BibitemOpen
  \bibfield  {author} {\bibinfo {author} {\bibfnamefont {E.}~\bibnamefont
  {Dagotto}},\ }\bibfield  {title} {\bibinfo {title} {Colloquium: The
  unexpected properties of alkali metal iron selenide superconductors},\ }\href
  {https://doi.org/10.1103/RevModPhys.85.849} {\bibfield  {journal} {\bibinfo
  {journal} {Rev. Mod. Phys.}\ }\textbf {\bibinfo {volume} {85}},\ \bibinfo
  {pages} {849} (\bibinfo {year} {2013})}\BibitemShut {NoStop}%
\bibitem [{\citenamefont {Wu}\ \emph {et~al.}(2020{\natexlab{b}})\citenamefont
  {Wu}, \citenamefont {Frandsen}, \citenamefont {Wang}, \citenamefont {Yi},\
  and\ \citenamefont {Birgeneau}}]{Birgeneau20}%
  \BibitemOpen
  \bibfield  {author} {\bibinfo {author} {\bibfnamefont {S.}~\bibnamefont
  {Wu}}, \bibinfo {author} {\bibfnamefont {B.~A.}\ \bibnamefont {Frandsen}},
  \bibinfo {author} {\bibfnamefont {M.}~\bibnamefont {Wang}}, \bibinfo {author}
  {\bibfnamefont {M.}~\bibnamefont {Yi}},\ and\ \bibinfo {author}
  {\bibfnamefont {R.}~\bibnamefont {Birgeneau}},\ }\bibfield  {title} {\bibinfo
  {title} {{Iron-Based Chalcogenide Spin Ladder BaFe$_2 X_3$ ($X=$ Se, S)}},\
  }\href {https://doi.org/10.1007/s10948-019-05304-4} {\bibfield  {journal}
  {\bibinfo  {journal} {Journal of Superconductivity and Novel Magnetism}\
  }\textbf {\bibinfo {volume} {33}},\ \bibinfo {pages} {143} (\bibinfo {year}
  {2020}{\natexlab{b}})}\BibitemShut {NoStop}%
\bibitem [{\citenamefont {Maharaj}\ \emph {et~al.}(2017)\citenamefont
  {Maharaj}, \citenamefont {Rosenberg}, \citenamefont {Hristov}, \citenamefont
  {Berg}, \citenamefont {Fernandes}, \citenamefont {Fisher},\ and\
  \citenamefont {Kivelson}}]{Maharaj17}%
  \BibitemOpen
  \bibfield  {author} {\bibinfo {author} {\bibfnamefont {A.~V.}\ \bibnamefont
  {Maharaj}}, \bibinfo {author} {\bibfnamefont {E.~W.}\ \bibnamefont
  {Rosenberg}}, \bibinfo {author} {\bibfnamefont {A.~T.}\ \bibnamefont
  {Hristov}}, \bibinfo {author} {\bibfnamefont {E.}~\bibnamefont {Berg}},
  \bibinfo {author} {\bibfnamefont {R.~M.}\ \bibnamefont {Fernandes}}, \bibinfo
  {author} {\bibfnamefont {I.~R.}\ \bibnamefont {Fisher}},\ and\ \bibinfo
  {author} {\bibfnamefont {S.~A.}\ \bibnamefont {Kivelson}},\ }\bibfield
  {title} {\bibinfo {title} {{Transverse fields to tune an Ising-nematic
  quantum phase transition}},\ }\href {https://doi.org/10.1073/pnas.1712533114}
  {\bibfield  {journal} {\bibinfo  {journal} {Proceedings of the National
  Academy of Sciences}\ }\textbf {\bibinfo {volume} {114}},\ \bibinfo {pages}
  {13430} (\bibinfo {year} {2017})}\BibitemShut {NoStop}%
\end{thebibliography}%

\end{document}